\documentclass[fleqn,usenatbib]{mnras}
\usepackage{float}
\usepackage{amsmath}
\usepackage{url}
\usepackage{newtxtext,newtxmath}

\usepackage[T1]{fontenc}
\usepackage[normalem]{ulem} 
\DeclareRobustCommand{\VAN}[3]{#2}
\let\VANthebibliography\thebibliography
\def\thebibliography{\DeclareRobustCommand{\VAN}[3]{##3}\VANthebibliography}

\usepackage{graphicx}	
\usepackage{amsmath}	
\usepackage{siunitx}

\usepackage{orcidlink}

\title[Helium Reionization from Empirical QLF before and after \texttt{JWST}]{Helium Reionization from Empirical Quasar Luminosity Functions before and after \texttt{JWST}}

\author[Basu et al.]{
Arghyadeep Basu \orcidlink{0000-0001-8104-9751},$^{1,2}$\thanks{E-mail: basu@mpa-garching.mpg.de}
Enrico Garaldi \orcidlink{0000-0002-6021-7020},$^{3,1}$
Benedetta Ciardi \orcidlink{0000-0002-5037-310X}$^{1}$
\\
$^{1}$Max-planck-Institut f$\ddot{u}$r Astrophysik, Karl-Schwarzschild-Strasse 1, D-85741, Garching, Germany\\
$^{2}$Ludwig-Maximilians-Universität München (LMU), Geschwister-Scholl-Platz 1, 80539 München, Germany\\
$^{3}$Institute for Fundamental Physics of the Universe, via Beirut 2, 34151 Trieste, Italy\\
}

\date{Accepted 2024 June 12. Received 2024 June 11; in original form 2024 April 09}

\pubyear{2024}

\begin{document}
\label{firstpage}
\pagerange{\pageref{firstpage}--\pageref{lastpage}}
\maketitle

\begin{abstract}
Recently, models of the quasar luminosity function (QLF) rooted on large observational compilations have been produced that, unlike their predecessors, feature a smooth evolution with time. This  bypasses the need to assume an ionizing emissivity evolution when simulating helium reionization with observations-based QLF, thus yielding more robust constraints. We combine one such QLF with a cosmological hydrodynamical simulation and 3D multi-frequency radiative transfer.  
The simulated reionization history is consistently delayed in comparison to most other models in the literature. The predicted intergalactic medium temperature is larger than the observed one at $z \lesssim 3$. Through forward modeling of the  He~{\sc ii} Lyman-$\alpha$ forest, we show that our model produces an extended helium reionization and successfully matches the bulk of the observed effective optical depth distribution, although it over-ionizes the Universe at $z\lesssim2.8$ as the effect of small-scale Lyman Limit Systems not being resolved. We thoroughly characterize transmission regions and dark gaps in He~{\sc ii} Lyman-$\alpha$ forest sightlines. We quantify their sensitivity to the helium reionization, opening a new avenue for further observational studies of this epoch. 
Finally, we explore the implications for helium reionization of the large number of active galactic nuclei revealed at $z\gtrsim5$ by \texttt{JWST}. We find that such modifications do not affect any observable at $z\leq4$, except in our most extreme model, indicating that the observed abundance of high-$z$ AGNs does not bear consequences for helium reionization.

\end{abstract}

\begin{keywords}
radiative transfer -- (galaxies:) intergalactic medium -- (galaxies:) quasars: absorption lines -- cosmology: theory -- (galaxies:) quasars: general
\end{keywords}



\section{Introduction}

The first stars and galaxies produced photons capable of ionizing hydrogen (i.e. with energy $E_\gamma > 13.6 \rm{eV}$) and singly ionizing helium (i.e. $E_\gamma > 24.6 \rm{eV}$). However, the ionization of intergalactic He~{\sc ii} into He~{\sc iii}, commonly known as helium reionization, requires more energetic photons ($E_\gamma >54.4 \rm{eV}$) that were not produced in sufficient number by these first sources. Instead, the increasing abundance of quasars (QSOs) in the redshift range 2 < \textit{z} < 4 enabled the full ionization of the intergalactic helium.

Recent large-scale spectroscopic surveys such as \texttt{SDSS} \footnote{The Sloan Digital Sky Survey, \url{https://www.sdss.org/}} and \texttt{DES} \footnote{The Dark Energy Survey, \url{https://www.darkenergysurvey.org/}} have significantly expanded the sample size of known quasars across cosmic epochs. This extensive dataset has enabled the derivation of robust constraints on the quasar luminosity function (QLF). Remarkably, \citet{kulkarni2019} and \citet{Shen2020} recently introduced QLF smoothly evolving with redshift, based on compilations of observational data in the redshift range $0 \leq z \lesssim 7$. The former focuses on 1450\AA\ magnitude, while the latter employs a wider range of spectra, including rest-frame Infrared (IR), B band, Ultraviolet (UV), soft and hard X-rays. The QLF, the spatial clustering of QSOs \citep[which  encodes information concerning the spatial distribution of these objects, e.g.][]{Porciani2004,White2012,Eftekharzadeh2015,Greiner2021} and quasar properties such as lifetimes and spectral energy distribution (SED), rule the timing and features of the cosmic helium reionization 
\citep{Sokasian2002,Furnaletto&Lidz2011,Khrykin2016,Khrykin2017,Khrykin2019,laplante2017}.
Their impact on structure formation is modulated by the distribution of gaseous structures dense enough to self-shield from the ionizing radiation and therefore acting as photon sinks \citep[e.g.][]{Kapahtia2024}. 

Reliably constraining helium reionization is very challenging. The most direct avenue for observing this transformation is via the He~{\sc ii} Lyman-$\alpha$ forest in the spectra of high-$z$ quasars.
For instance, \citet{Dixon2009} interpreted the rapid decrease in the He~{\sc ii} optical depth at \textit{z} $\sim$ 2.7 as indicative of the end of helium reionization. 
Moreover, the fluctuations in opacity observed along multiple lines of sight offer valuable insights into the concluding stages of this epoch \citep{Anderson1999,Heap2000,Smette2002,Reimers2005}. \citet{Furlanetto2011} suggested that the large fluctuations observed at \textit{z} $\sim$ 2.8 were more likely the product of ongoing reionization. The fact that the process of helium reionization is extended is confirmed by sightline-to-sightline variation of optical depth in more recent observations \citep{worseck2016,worseck2019,makan2021,makan2022}, but these are limited by the low number of sightlines due to the intrinsic difficulties of observing the He~{\sc ii} Lyman-$\alpha$ forest. In the near future,  \texttt{WEAVE} \footnote{The WHT Enhanced Area Velocity Explorer, \url{https://ingconfluence.ing.iac.es/confluence//display/WEAV}} is expected to more than double the number of sightlines at $z>2$ over several thousand square degrees\footnote{Although the wavelength coverage of WEAVE does not include the He~{\sc ii} Lyman-$\alpha$ line, this survey can} serve as target selection of clean sightlines where the He~{\sc ii} Lyman-$\alpha$ forest can be detected by subsequent observations.. Additionally, WEAVE-QSO \citep{pieri2016} will constrain the IGM temperature at $z \sim 3$, providing new tighter constraints on the helium reionization. 
Overall, despite the difficulties, a consensus has emerged that helium reionization occurred in the redshift range $2.7 \lesssim z \lesssim 4.5$ \citep{Miralda1993,Giroux1995,Croft1997,Fardal1998,Schaye2000,Theuns2002,Riccoti2000,Bolton2006,Bolton2009,Meiksin2012}.

A detailed understanding of helium reionization is crucial not only to test our structure formation framework, but also to assess its impact on galaxies and the intergalactic medium (IGM). In fact, the excess energy from the He~{\sc ii} to He~{\sc iii} transition significantly heats up the IGM \citep{hui&gnedin1997,Schaye2000,McQuinn2009,Garzili2012}. 
However, modelling helium reionization is computationally very challenging. It requires to simultaneously capture the quasars clustering properties on scales of hundreds of Mpc and the galaxy-scale gas physics. To cope with such requirements, many numerical simulations of this epoch focus on scales $\le 100$ $\rm{cMpc}$ \citep{Sokasian2002,Meiksin2012,compostella2013,compostella2014}, at the cost of missing the large ionized bubbles expected during helium reionization and failing to include the rarest sources. Alternatively, several studies employ various combinations of analytic and semi-analytic models \citep{Tom2000,Gleser2005,upton2020}, that however result in less solid predictions (e.g. concerning the morphology of ionized helium bubbles, heating of the IGM, He-ionizing background). 

Until now, the modeling of He~{\sc ii} reionization has typically either relayed on the simulated (instead of observed) QSOs population, or on selecting the observed QLF at a specific redshift and extrapolating it based on some emissivity evolution \citep{compostella2013,compostella2014,garaldi2019}. Here, we explore the prediction of the recent smoothly-time-evolving QLF from \citet{Shen2020} on helium reionization, its morphology and the properties of the $2<z<4$ IGM. To this end, we post-process the large-scale hydrodynamical simulation \texttt{TNG300} \citep[e.g.][]{pillepich2018,nelson2018} using the 3D multi-frequency radiative transfer code \texttt{CRASH} \citep[e.g.][]{ciardi2001,glatzle2022} which follows the formation and evolution of He~{\sc iii} regions produced by a population of quasars extracted from the \citet{Shen2020} QLF. We forward model these simulations to produce synthetic Lyman-$\alpha$ forest spectra and compare them to available data. Finally, we investigate the impact of the recent \textit{James Webb Space Telescope} (\texttt{JWST}) observations of a large number of active galactic nuclei (AGNs) at $z>5$ \citep[e.g.][]{Fudamoto2022,Harikane2023,Maiolinoa2023,Maiolino2023,Goulding2023,Larson2023,Juodzbalis2023,Greene2023}. We introduce the simulations in Section \ref{method}, whereas our results are discussed in Section \ref{results}. We summarize the conclusion and the future prospects in Section \ref{conclusions}.

\section{Methodology}
\label{method}
In order to provide a faithful picture of helium reionization, we have combined the outputs of a hydrodynamical simulation (section \ref{tng300}) with a multi-frequency radiative transfer code (section \ref{crash}). The latter is sourced by a population of quasars following the QLF from \citet[][in particular their Model 2]{Shen2020}, that are placed in the simulations volume as described in Section \ref{tng+crash}. This procedure ensures that our results are as genuine predictions of the observed QLF as possible.

\subsection{The \texttt{TNG300} hydrodynamical simulation}
\label{tng300}

In this work we leverage the \texttt{TNG300} simulation, which is a part of the \texttt{Illustris TNG project} \citep{volker2018,naiman2018,marinacci2018,pillepich2018,nelson2018}, to model the formation and evolution of structures in the Universe. It has been performed with the \texttt{AREPO} code \citep{springel2010}, which is used to solve the idealized magneto-hydrodynamicals equations \citep{pakmor2011} describing the non-gravitational interactions of baryonic matter, as well as the gravitational interaction of all matter. The simulation employs the recent \texttt{TNG} galaxy formation model \citep{weinberger2017,pillepich2018}, and star formation is incorporated by converting gas cells into star particles above a density threshold of $\textit{n}\rm{_{H}}$ $\rm{\sim 0.1 cm^{-3}}$, following the Kennicutt-Schmidt relation \citep{springel2003}. Stellar populations are self-consistently evolved, and inject metals, energy and mass into the  interstellar medium (ISM) throughout their lifetime, including their supernova (SN) explosions. AGN feedback has two-modes: a more efficient kinetic channel at low Eddington ratio (`radio mode') and a less efficient thermal channel at high Eddington ratio (`quasar mode') \citep{weinberger2018}.

\texttt{TNG300} has been run in a comoving box of length $L_\mathrm{box} = 205 \,h^{-1} {\rm cMpc}$, with (initially) $2\times2500^{3}$ gas and dark matter (DM) particles. The average gas particle mass is $\bar{m}\rm{_{gas}}=7.44 \times 10^{6}\,M_{\odot}$, while the DM particle mass is constant and amounts to $m\rm{_{DM}}=3.98 \times 10^{7}\,M_{\odot}$. Haloes are identified on-the-fly using a friends-of-friends algorithm with a linking length of $0.2$ times the mean inter-particle separation. \texttt{TNG300} adopts a \citep{planck2016}-consistent cosmology with $\Omega_{m}=0.3089$, $\Omega_{\Lambda}=0.6911$, $\Omega_{b}=0.0486$, $h=0.6774$, $\sigma_{8}=0.8159$ and $n_{s}=0.9667$, where the symbols have their usual meaning.

In particular, from \texttt{TNG300} we have used 19 outputs covering the redshift range $5.53 \leq z \leq 2.32$. For the additional simulations presented in Section \ref{jwst}, we employ additional outputs at $z = 6$ and $z=5.85$. These serve as basis for the radiative transfer, that we describe next.

\subsection{The \texttt{CRASH} radiative transfer code}
\label{crash}
We have implemented the radiative transfer (RT) of ionizing photons through the IGM by post-processing the outputs from the hydrodynamical simulation with the code \texttt{CRASH} \citep{ciardi2001,maselli2003,maselli2009,Maselli2005,partl2011},
which self-consistently calculates the evolution of the hydrogen and helium ionization state and the gas temperature. \texttt{CRASH} uses a Monte-Carlo-based ray tracing scheme, where the ionizing radiation and its time varying distribution in space is represented by multi-frequency photon packets travelling through the simulation volume. The latest version of CRASH features a self-consistent treatment of UV and soft X-ray photons, in which X-ray ionization, heating as well as detailed secondary electron physics \citep{graziani2013,graziani2018} and dust absorption \citep{glatzle2019,glatzle2022} are included. We refer the reader to the original papers for more details on \texttt{CRASH}. 

\texttt{CRASH} performs the RT on grids of gas density and temperature. In our setup, these correspond to snapshots from the \texttt{TNG300} simulation (see Section \ref{tng300}). In order to account for the expansion of the Universe between the $i$-th and $(i+1)$-th snapshots, the gas number density is evolved as $n(\mathbf{x},z) = n(\mathbf{x}, z_{i})(1 + z)^{3}/(1 + z_{i})^{3}$, where $\mathbf{x} \equiv (x_{c}, y_{c}, z_{c})$ are the coordinates of each cell c. 

Since we are only interested in helium reionization, we follow radiation covering the energy range $h_p \nu \in [54.4 \ \rm{eV}, 2 \ \rm{keV}]$, where $h_p$ is the Planck constant, and we fix the IGM H~{\sc I} and He~{\sc II} ionization fractions to $x_{\rm HI} = 10^{-4}$ and $x_{\rm HeI}=0$. This corresponds to a fully-completed hydrogen reionization.

We generate five RT outputs at regular time intervals in between each pair of adjacent \texttt{TNG300} snapshots. 

\begin{figure*}
    \includegraphics[width=175mm]{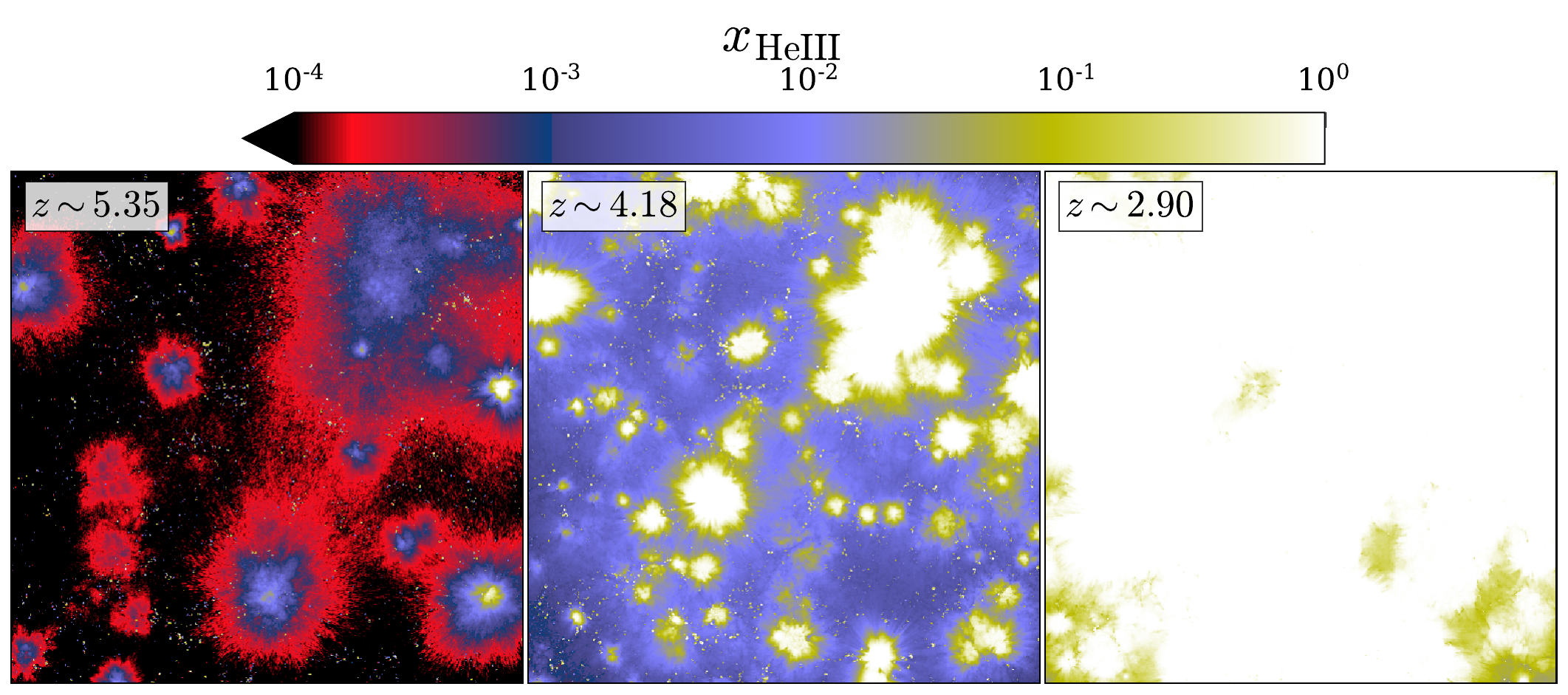}
    \includegraphics[width=175mm]{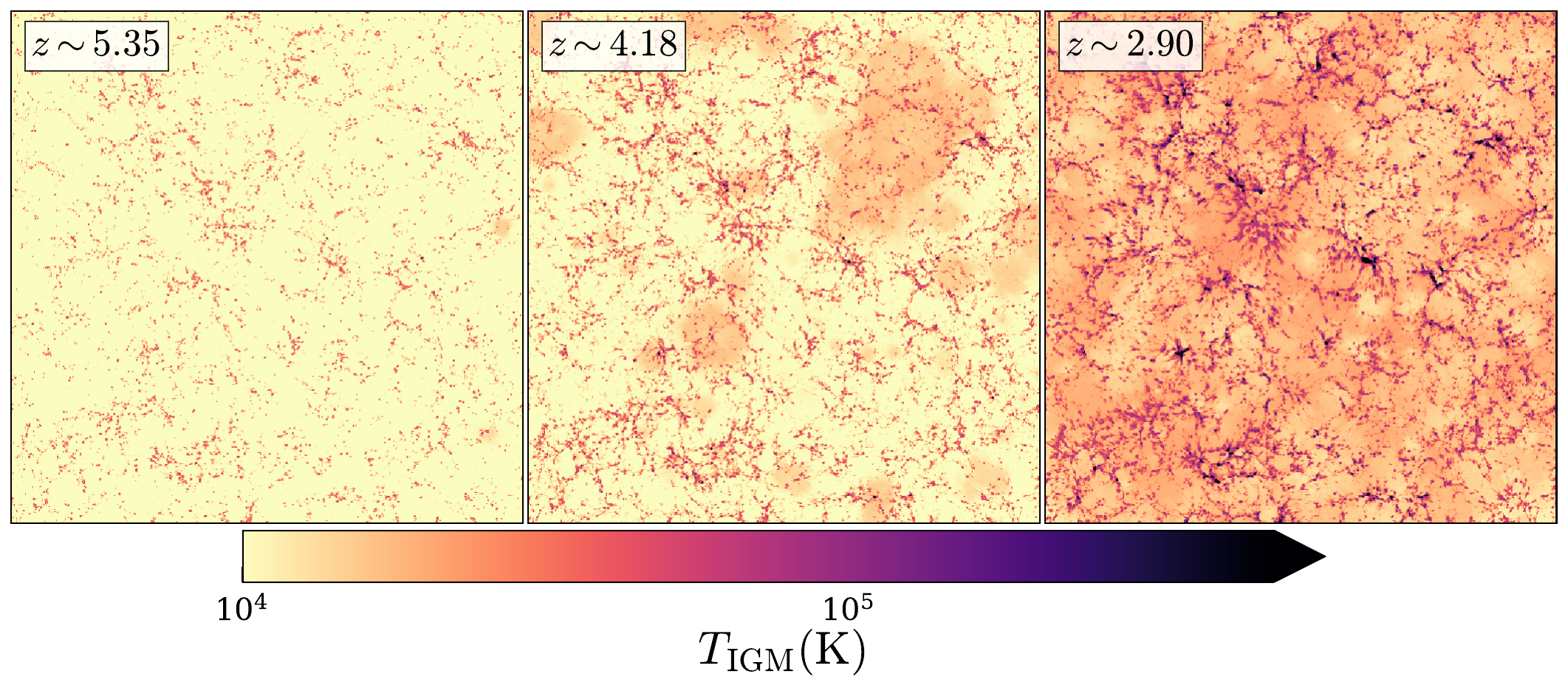}
    
    \caption{Slice across the simulation box of \texttt{N512-ph1e6} at \textit{z} =5.35, 4.18, and 2.90 (from left to right). Top and bottom panels show the He~{\sc iii} fraction and the gas temperature, respectively. The maps are $205 h^{-1} \rm{cMpc}$ wide and $400 h^{-1}$ ckpc thick.}
    \label{fig:reion_history}
\end{figure*}

\subsection{Combining \texttt{CRASH} and \texttt{TNG300}}
\label{tng+crash}

\subsubsection{Initial Conditions and Outputs}
\label{ics}
To incorporate the hydrodynamical simulation outputs into \texttt{CRASH}, the gas density and temperature extracted from the snapshots need to be deposited onto 3D uniform grids. We adopt a smoothing length for each Voronoi gas cell equal to the radius of the sphere enclosing its 64 nearest neighbours, and use the usual SPH cubic-spline kernel for its volume distribution. We initialize the He~{\sc iii} fraction to $x_{\rm HeIII}=10^{-4}$. 
To account for a fully-completed hydrogen reionization in the initial conditions, we set a gas temperature floor of $10^4$~K (which is approximately the average of the two available observations \citep{walther2019,Gaikwad2020} of IGM temperature at mean density ($T_{0}$) at $z \sim 5.4$, see figure \ref{fig:temp_evol} for a reference). This is consistent with our assumption of having  H~{\sc ii} fraction equal to unity in all cells.
Additionally, we assume that the fraction of H~{\sc i} and He~{\sc i} remain fixed for the entire range of redshifts.

\subsubsection{Spectral Energy Distribution}
\label{sed}
To derive the SED of quasars, we employ the model developed by \citet{Marius2018, Marius2020}. In the range 13.6 $\rm{eV}$--200 $\rm{eV}$, this model explicitly calculates the average over a sample of 108,104 QSO SEDs from \citet{Krawczyk2013} in the interval  $0.064 < z < 5.46$. Beyond 200 $\rm{eV}$, the spectral shape is modeled as a power law with an index of -1. We assume no evolution of the SED with redshift (refer to Section 2 in \citealt{Marius2018} for more details). For all RT simulations, the spectra of ionizing sources extend to a maximum frequency of $\sim 2 \ \rm{keV}$, covering the UV and soft X-ray regime. The discretization of the source spectrum is finer around the ionization thresholds of He~{\sc ii} ($\sim$ 54.4 $\rm{eV}$).

\subsubsection{Quasar distribution}
\label{qso_assignment}

In order to ensure that the simulated helium reionization is a genuine prediction of the QLF from \citet[][Model 2]{Shen2020}, we disregard the black holes within \texttt{TNG300}, since their location and properties critically depend on the black hole seeding prescription, accretion physics and feedback model implemented in the simulation. Instead, we follow a more agnostic approach and populate the simulated haloes with quasars following the prescription described below. 

While the QLF offers predictions for the number of quasars of a given luminosity, it lacks information on their spatial distribution. Thus, to associate quasars to halos we adopt an abundance matching approach which establishes a one-to-one correspondence between the quasar bolometric luminosity ($L\rm{_{bol}}$) and the mass of DM haloes ($M\rm{_{DM}}$).
In practice, this amounts to placing the $i$-th brightest quasar at the center of the $i$-th most massive halo in the simulation. We refer to this approach as `direct'. 
In order to account for randomness in the $L\rm{_{bol}}$ -- $M\rm{_{DM}}$ correspondence, we design a slightly modified approach. In this `fiducial' method we split the simulated halo mass function and the QLF into $N_\mathrm{bins} = 20$ equal bins in logarithmic space. We then place the quasars as follows:
(i) we first select a quasar from the QLF at a given redshift, accounting for the finite simulation volume; 
(ii) then we identify the bin of the QLF in which the quasar resides (e.g. the $i$-th brightest bin of QLF); (iii) we choose a random halo from the corresponding $i$-th most massive bin of the halo mass function;
(iv) we place the quasar at the centre of such halo;
(v) finally, we introduce a 1\% random noise on the $L\rm{_{bol}}$ of the quasar;
(vi) we repeat this step for all the quasars in the simulation volume.
Note that the QLF computed self-consistently from the \texttt{TNG300} simulation is not in agreement with the \citet{Shen2020} model. Therefore, we opted to re-assign QSO luminosities as described in order to ensure our results are as genuine predictions of the observed QLF as possible.
Additionally, we did not incorporate any correction to the simulated AGN feedback, which is likely to leave an imprint on the properties of nearby gas. The reason is that the impact of feedback is highly non-linear and coupled to other simulated processes, and it is therefore impossible to remove that directly in post-processing. However, if we were hypothetically able to undo the heat injection associated with AGN feedback, it would likely improve the comparison with observations (specifically, by raising the effective optical depths in the low-$z$ regime) which we discuss later in the section \ref{forest_tau_eff}. 

In both approaches, the quasar-halo pairing is repeated for each snapshot of the hydrodynamical simulation, and haloes that hosted a QSO in the previous snapshot are temporarily excluded from the procedure. This approximately accounts for the quasar duty cycle, since the interval between two snapshots is comparable (although somewhat larger) than the expected quasar lifetime \citep{Morey2021,Khrykin2021,Soltinsky2023}. 
In the remainder of the paper we show results from our fiducial approach. In Appendix \ref{appendix:BH_to_halo} we demonstrate that these two approaches give nearly identical results, with only $\mathcal{O}(10\%)$ differences in the initial stages of helium reionization, as reionization fronts takes more time to escape the largest haloes because of their higher densities.

\setlength{\tabcolsep}{5pt}
\begin{table}
\centering
    \begin{tabular}{lccccc}
        \hline
        Simulation Name   &$N\rm{_{grid}}$ &$N_{\gamma}$ &PBC  &QSO assignment  &$z_{\rm{final}}$\\
        \hline
        \hline
        \texttt{N256-ph1e5}   &$256^{3}$   &$10^{5}$  &No  &fiducial &2.32\\
        \texttt{N256-ph5e5}   &$256^{3}$   &$5 \times 10^{5}$  &No  &fiducial &2.32\\
        \texttt{N256-ph1e6}   &$256^{3}$   &$10^{6}$  &No  &fiducial &2.58\\
        \texttt{N256-ph1e5-PBC}   &$256^{3}$   &$10^{5}$  &Yes  &fiducial &2.32\\
        \hline
        \texttt{N512-ph1e5}   &$512^{3}$   &$10^{5}$  &No  &fiducial &2.32\\
        \textbf{\texttt{N512-ph5e5}}   &$\boldsymbol{512^{3}}$   &$\boldsymbol{5 \times 10^{5}}$  &\textbf{No}  &\textbf{fiducial} &\textbf{2.32}\\
        \texttt{N512-ph1e6}   &$512^{3}$   &$10^{6}$  &No  &fiducial &2.73\\
        \texttt{N512-ph1e5-PBC}   &$512^{3}$   &$10^{5}$  &Yes  &fiducial &3.71\\
        \texttt{N512-ph5e5-DIR}   &$512^{3}$   &$5 \times 10^{5}$  &No  &direct &2.44\\
        \hline
        \texttt{N768-ph5e5}   &$768^{3}$   &$5 \times 10^{5}$  &No  &fiducial &2.73\\
        \texttt{N768-ph1e6}   &$768^{3}$   &$10^{6}$  &No  &fiducial &3.82\\
        \hline
    \end{tabular}
    \caption{List of simulations and associated parameters. From left to right: name of the simulation, grid size ($N\rm{_{grid}}$), number of photon packets emitted per source per timestep ($N_{\gamma}$), presence of periodic boundary conditions (PBC), the method used to assign quasars to the DM haloes and the final redshift until which the simulation has run ($z_{\rm{final}}$). The reference simulation is shown in bold.}
    \label{tab:table-sim}
\end{table}

\subsubsection{Simulation Setup}
\label{sim_setup}

We have run a suite of simulations with different resolutions for the gas and radiation fields. The former is quantified as the dimension of the Cartesian grid ($N_{\rm{grid}}$) used for the deposition of the \texttt{TNG300} particle data. The latter is controlled by the number of photon packets emitted by each source for every output of the hydrodynamical simulation\footnote{The RT time step is computed as the ratio of the time between two hydrodynamical outputs and $N_{\gamma}$, so that at each RT time step, every source emits one packet.}. We have explored $N_{\rm{grid}} = 256^{3}$, $512^{3}$ and $768^{3}$, corresponding to spatial resolutions of $\delta x = L_\mathrm{box} / N_{\rm{grid}}^{1/3} \approx 800$, $400$ and $267$ $h^{-1}\,\mathrm{ckpc}$, respectively, and $N_{\gamma} = 10^{5}$, $5 \times 10^{5}$ and $10^{6}$.
We refer the reader to Appendix \ref{appendix:convergence_Ngamma} and \ref{appendix:convergence_Ngrid} for a quantitative discussion of the convergence tests performed with respect to photon packet sampling and grid dimension, respectively. Details of the simulations have been summarized in Table \ref{tab:table-sim}. Notice that computational constraints prevented us from running the highest-resolution boxes to low redshift, and therefore they are used only for convergence tests. We designate simulation \texttt{N512-ph5e5} as our reference and present only results from this run unless stated otherwise.

In all simulations, the photon packets are lost once they leave the simulation box, i.e. no periodic boundary conditions (PBC) have been applied. This choice is dictated by the otherwise steep rise in the computational cost once the vast majority of the volume is ionized, since in the presence of PBC photons can cross the simulation box multiple times before being absorbed. We have checked that this does not significantly affect our results (see Appendix \ref{appendix:convergence_PBC}). Nevertheless, we take the conservative approach of removing in our analysis all cells within $5 \times \delta x$ from each side of the box.

\section{Results}
\label{results}
The main question we want to address here is: \textit{What are the implications for helium reionization of the recent QLF constraints?}
In order to answer it, in the following we discuss the outcome of our fiducial simulation and compare it with available data. Then, we explore the implications of the high AGN number density reported by recent \texttt{JWST} observations.

\subsection{Reionization history}
\label{reion_history}
For a visual examination of the progress of helium reionization, we show in figure \ref{fig:reion_history} the He~{\sc iii} fraction (top panels) and IGM temperature (bottom panels) at \textit{z} = 5.35, 4.18 and 2.90 in a slice extracted from simulation \texttt{N512-ph1e6}. From the figure, it is evident how He~{\sc ii} reionization proceeds in an inhomogeneous manner, with ionized regions initially forming around quasars, evolving rapidly, and eventually merging \citep[e.g.][]{compostella2013,laplante2017}. The partially-ionized gas surrounding ionized regions primarily results from soft X-ray and hard UV photons, as already discussed in prior research \citep{koki2017,Marius2018,Marius2020}. The bottom panels show how the ionization of He~{\sc ii} coincides with heating of the IGM.

\begin{figure}
    \includegraphics[width=\columnwidth]{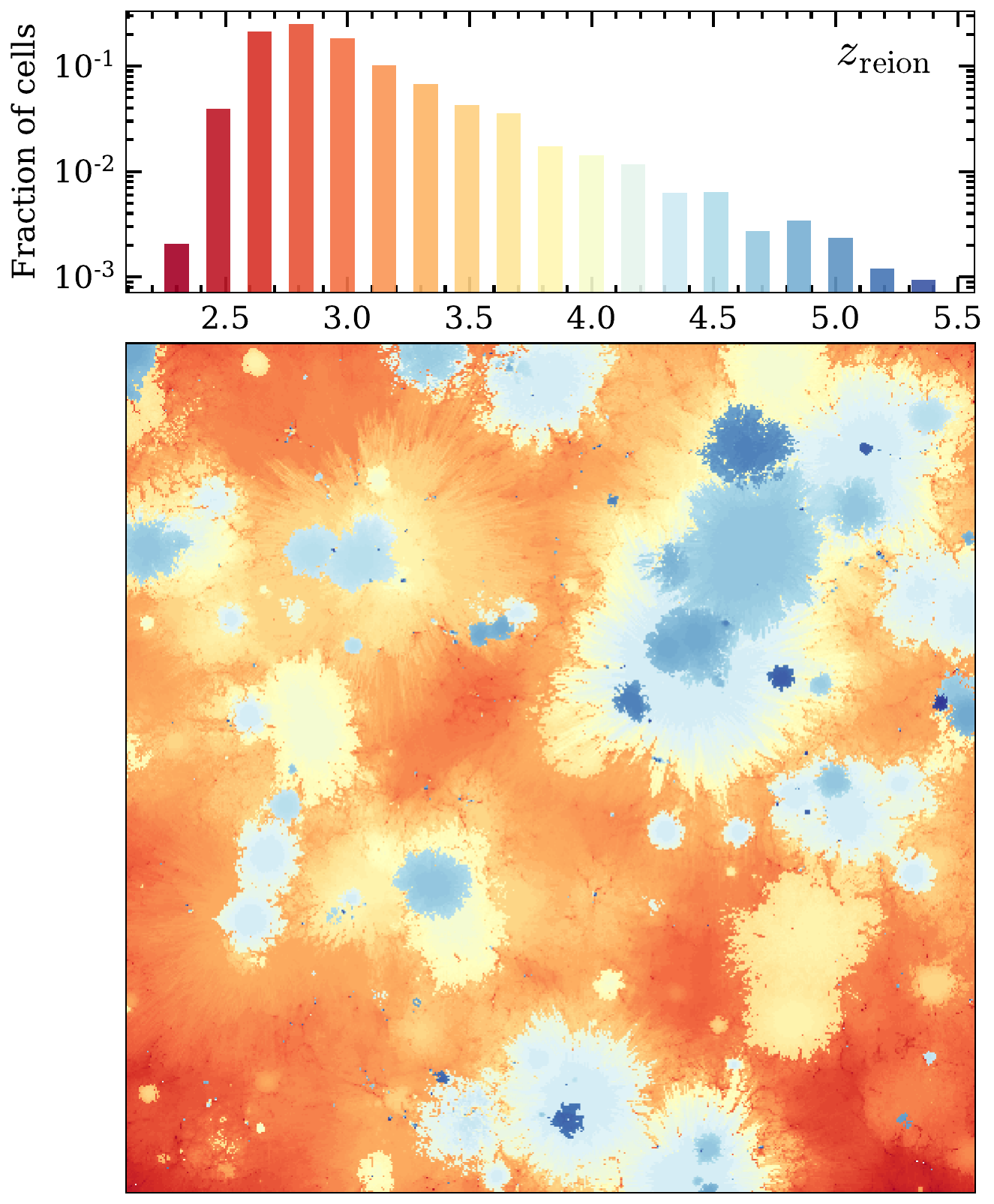}
    
    \caption{Fraction of cells (top panel) in the entire simulation volume that underwent reionization at a given redshift ($z\rm{_{reion}}$, see text for the definition). The color associated to each individual bin is used in the bottom panel to color code  cells in a slice of \texttt{N512-ph1e6} (the same slice as figure \ref{fig:reion_history}).}
    \label{fig:reion_redshift}
\end{figure}

\begin{figure}
    \includegraphics[width=\columnwidth]{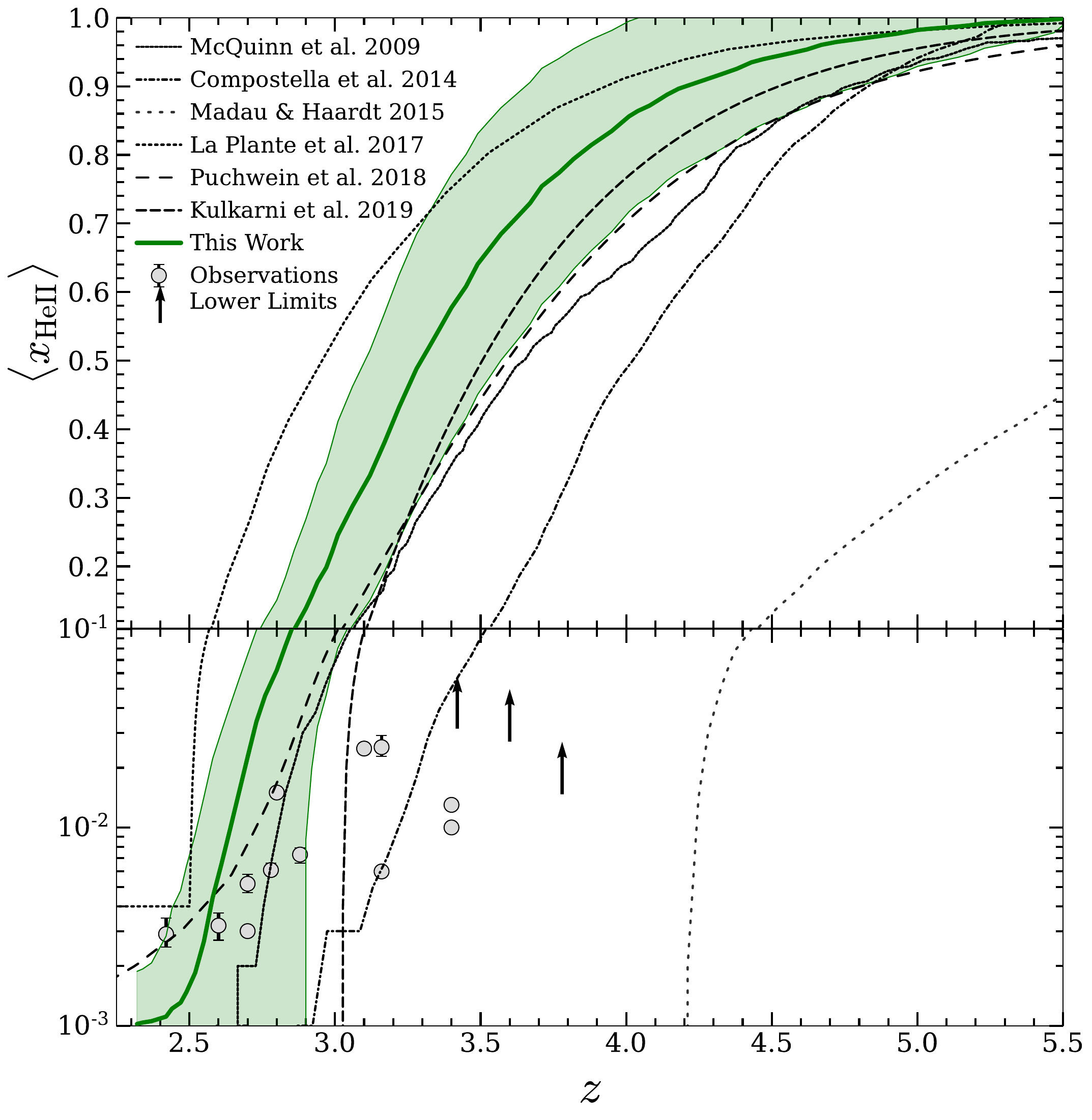}
    
    \caption{Redshift evolution of the volume averaged He~{\sc ii} fraction for  \texttt{N512-ph5e5} (solid green line). Other theoretical models \citep{McQuinn2009,compostella2014,madau2015,laplante2017,puchwein2019,kulkarni2019} are shown in black thin lines with different linestyles as mentioned in the legend. The symbols denote a collection of observational constraints from \citet{worseck2016}, \citet{davies2017}, \citet{worseck2019}, \citet{makan2021}, and \citet{makan2022}. The shaded region displays the central 68$\%$ of the data.}
    \label{fig:xheii}
\end{figure}

\begin{figure}
    \includegraphics[width=\columnwidth]{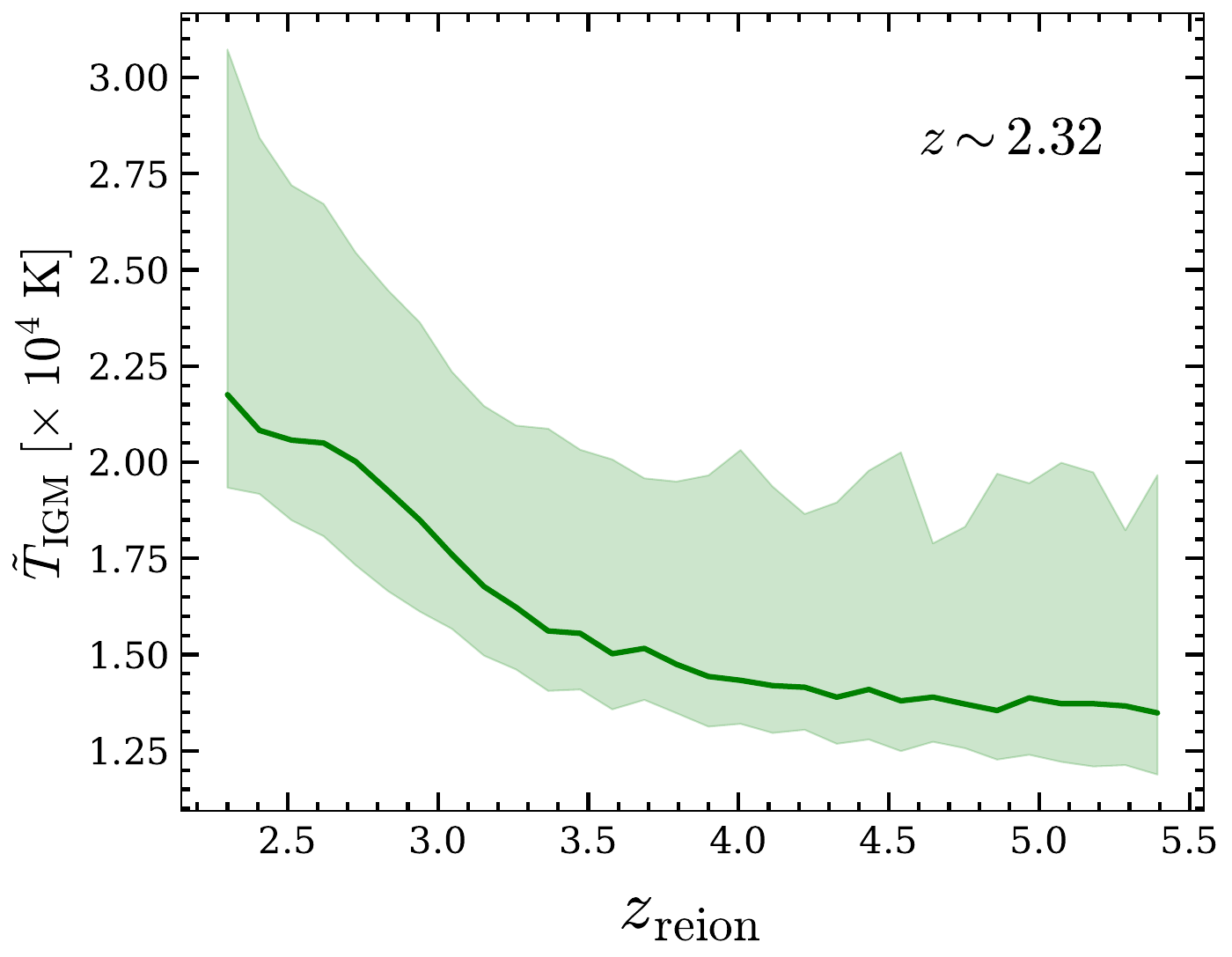}
    
    \caption{Correlation between the median IGM temperature (solid line) and the reionization redshift at $z\sim2.32$.  Only cells from \texttt{N512-ph5e5} with $T$ $\leq 40,000$~K are considered. The shaded region displays the central $68\%$ of the data.}
    \label{fig:reion_redshift_temp}
\end{figure}

\begin{figure}
    \includegraphics[width=\columnwidth]{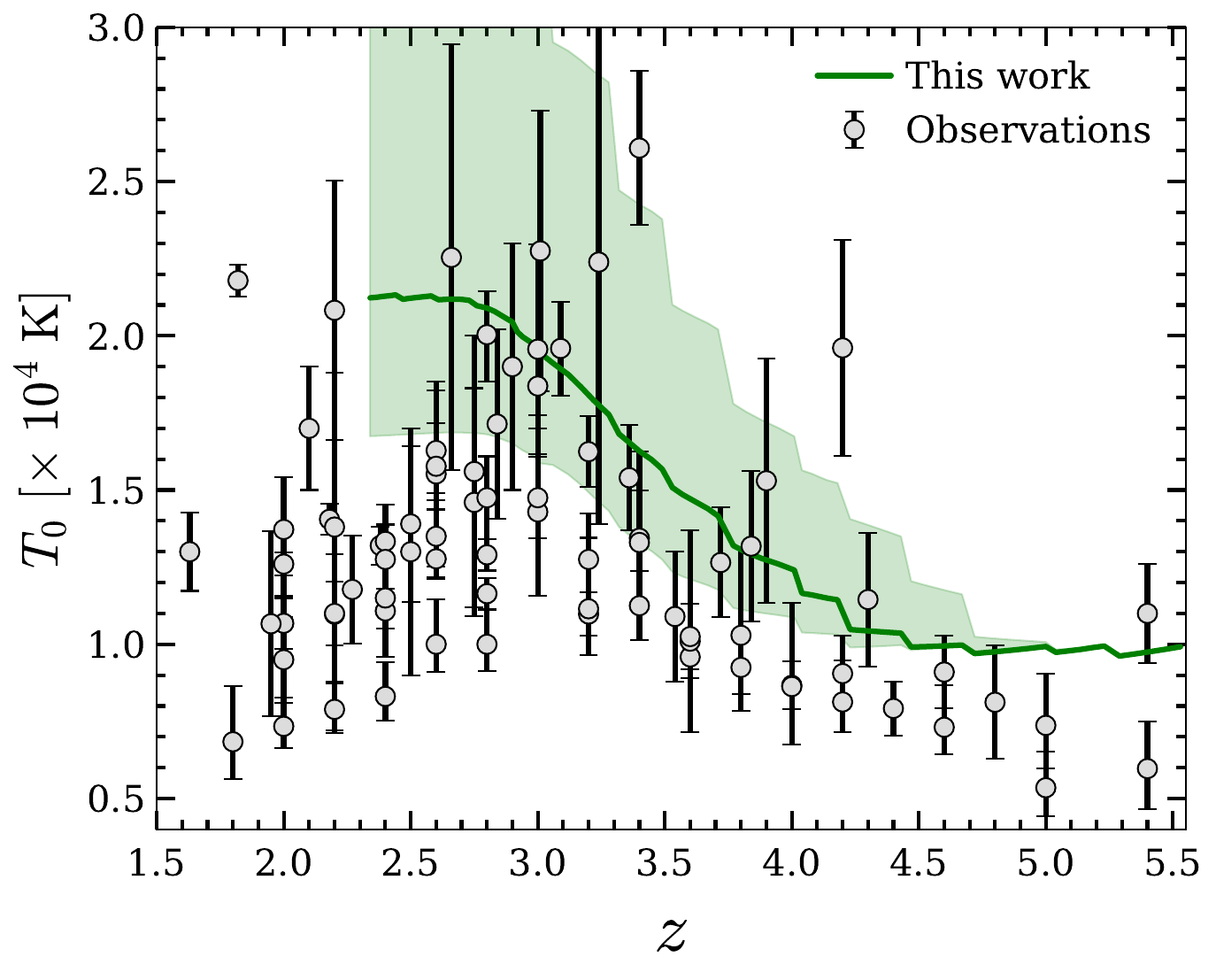}
    \caption{Redshift evolution of the median IGM temperature at mean density (solid line) for \texttt{N512-ph5e5}. The symbols denote a collection of observational constraints from \citet{Schaye2000}, \citet{Bolton2010}, \citet{Lidz2010}, \citet{becker2011}, \citet{Bolton2012}, \citet{Garzili2012}, \citet{bolton2014}, \citet{boera2014}, \citet{rorai2018}, \citet{hiss2018}, \citet{boera2019}, \citet{walther2019}, \citet{Gaikwad2020}, and \citet{gaikwad2021}. The shaded region displays the central $68\%$ of the data.}
    \label{fig:temp_evol}
\end{figure}

To further demonstrate the highly-inhomogeneous nature of helium reionization, we show in figure \ref{fig:reion_redshift} the redshift $z\rm{_{reion}}$ at which each cell in the same slice as figure \ref{fig:reion_history} is reionized. We define $z\rm{_{reion}}$ as the redshift at which the local He~{\sc iii} fraction exceeds 0.99 for the first time. Consistently with the anticipated inside-out scenario, the process initially occurs in proximity to the quasars. As the process unfolds, the majority of cells in the simulation volume experience ionization at lower redshifts, specifically $z < 3$, revealing a prominent peak around $z \sim 2.7$ in the histogram of $z\rm{_{reion}}$. This is reflected in the redshift evolution of the volume-averaged He~{\sc ii} fraction, that is shown in figure \ref{fig:xheii} for our fiducial run. Symbols denote observational constraints from \cite{worseck2016}, \cite{davies2017}, \cite{worseck2019}, \cite{makan2021} and \cite{makan2022}.
Our model completes reionization too late to accommodate the reported He~{\sc ii} fraction at $3 \lesssim z \lesssim 3.5$. However, an observational determination of this quantity is notoriously difficult, especially at higher redshift, due to the small number of sightlines suitable for such measurement. 
Additionally, the local nature of such constraints  combined with the inhomogeneous nature of reionization enables multiple values to coexist at the same time. Finally, the steeply-declining sensitivity of our probes to large neutral fractions biases observational results. Our reionization history, though, appears delayed also compared to other theoretical models \citep{madau2015, laplante2017, puchwein2019, McQuinn2009, compostella2014, kulkarni2019}, with the exception of \citet{laplante2017}, which employ a fully coupled radiation-hydrodynamical approach and model quasar activity as a lightbulb with luminosity-dependent lifetimes reproducing QLF constraints from SDSS and \texttt{COSMOS}\footnote{The Cosmic Evolution Survey, \url{https://cosmos.astro.caltech.edu/}}\citep{Masters2012,Mcgreer2013,Ross2013}.
It is essential to note that all theoretical models differ in their methodologies and modeling of source properties, with many lacking a RT scheme, so that a detailed comparison is not straightforward and outside the scope of this work. 

Once helium reionization is completed, the residual He~{\sc ii} fraction in our model is somewhat lower than the one inferred from observations. This is confirmed by the analysis performed in Section \ref{forest_tau_eff}. This discrepancy is indicative of the fact that our simulations do not fully capture the surviving sinks of radiation in the post-helium-reionization Universe. 
This is due to the fact that the resolution of the \texttt{TNG300} simulation is not high enough to capture the Lyman limit systems scales, as well as that by gridding the density field we smooth  out the largest gas overdensities that can self-shield from radiation and therefore maintain He~{\sc ii} reservoirs until late time. 

Finally, we demonstrate the impact of helium reionization onto the IGM properties in figure \ref{fig:reion_redshift_temp}, where we show the dependence of the median IGM temperature ($\tilde{T}\rm{_{IGM}}$) at $z=2.32$  as a function of $z\rm{_{reion}}$ for our reference simulation \texttt{N512-ph5e5}. Note that here we only considered cells with $ T \leq 40,000 \ \rm{K}$ filter out cells affected by feedback precesses other than photo-ionization, which typically have higher temperatures. From the figure it is clear that an early He~{\sc ii} reionization results in a longer period of adiabatic cooling and thus a lower IGM temperature at the specific redshift investigated (the same trend is found at other redshifts).

\begin{figure*}
    \includegraphics[width=175mm]{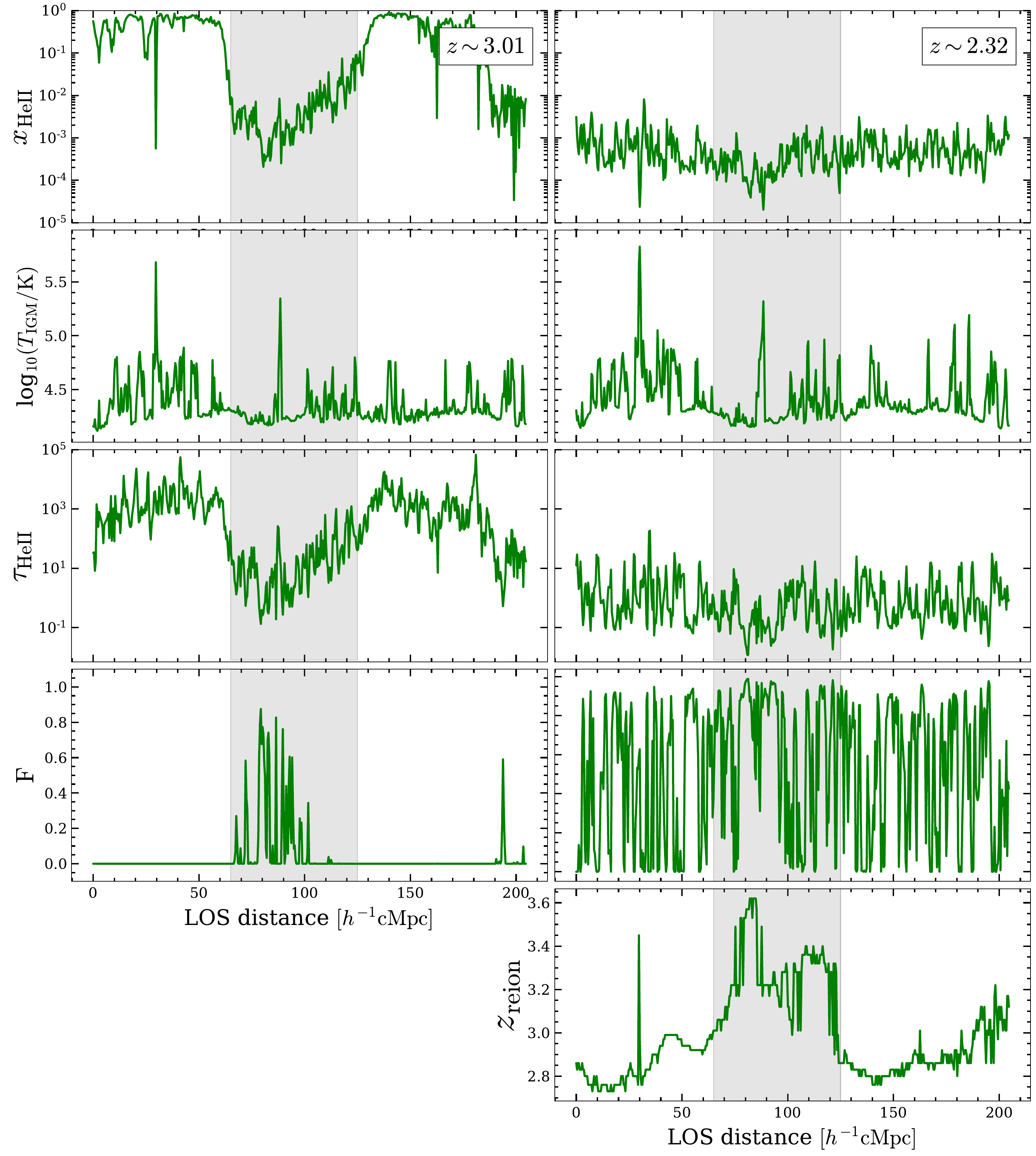}
    
    \caption{Line of sight IGM properties at \textit{z} = 3.01 and 2.32 extracted from our reference simulation \texttt{N512-ph5e5}. The four panels from top to bottom display the He~{\sc ii} fraction, the IGM temperature, the He~{\sc ii} optical depth, and the normalized transmission flux. The local He~{\sc ii} reionization redshift along the line of sight extracted at $z=2.32$ is shown in the right bottom panel. The shaded region highlights cells that undergo an early reionization.}
    \label{fig:los}
\end{figure*}

\begin{figure}
    \includegraphics[width=\columnwidth]{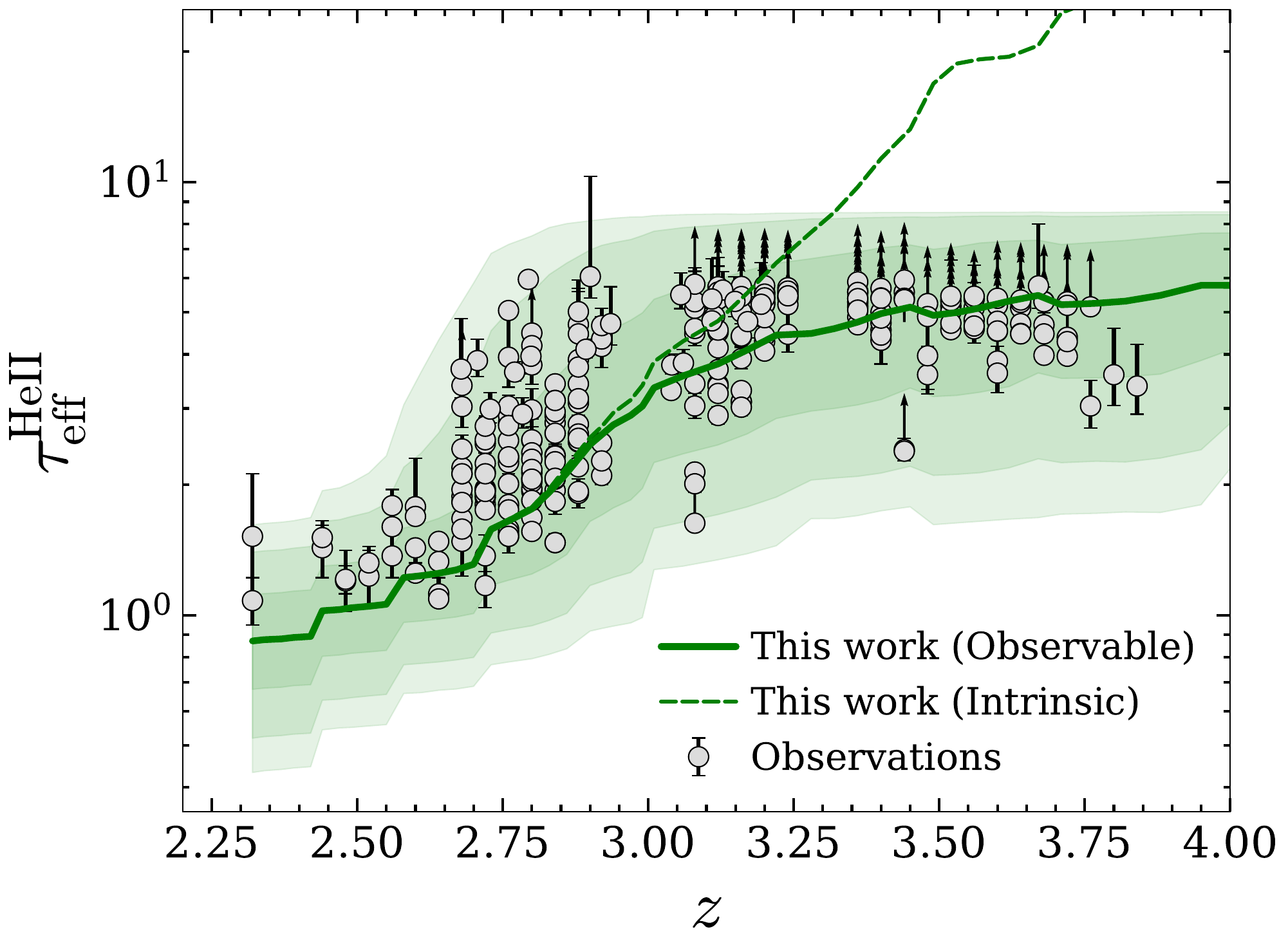}
    
  \caption{Redshift evolution of the median He~{\sc ii} effective optical depth computed from the synthetic spectra in our reference simulation \texttt{N512-ph5e5} (solid green curve). The shaded regions indicate 68\%, 95\%, and 99\% confidence intervals. Observational estimates from \citet{worseck2016}, \citet{worseck2019}, \citet{makan2021} and \citet{makan2022} are shown as symbols. 
  For a consistent comparison with observations, in the evaluation of the median we include only spectral chunks with an effective optical depth below 8.56 (i.e. `observable', see text for more details), while the dashed line refers to the median calculated for the `intrinsic' distribution.}
    \label{fig:taueff}
\end{figure}

\begin{figure}
    \centering
    \includegraphics[width=0.9\columnwidth]{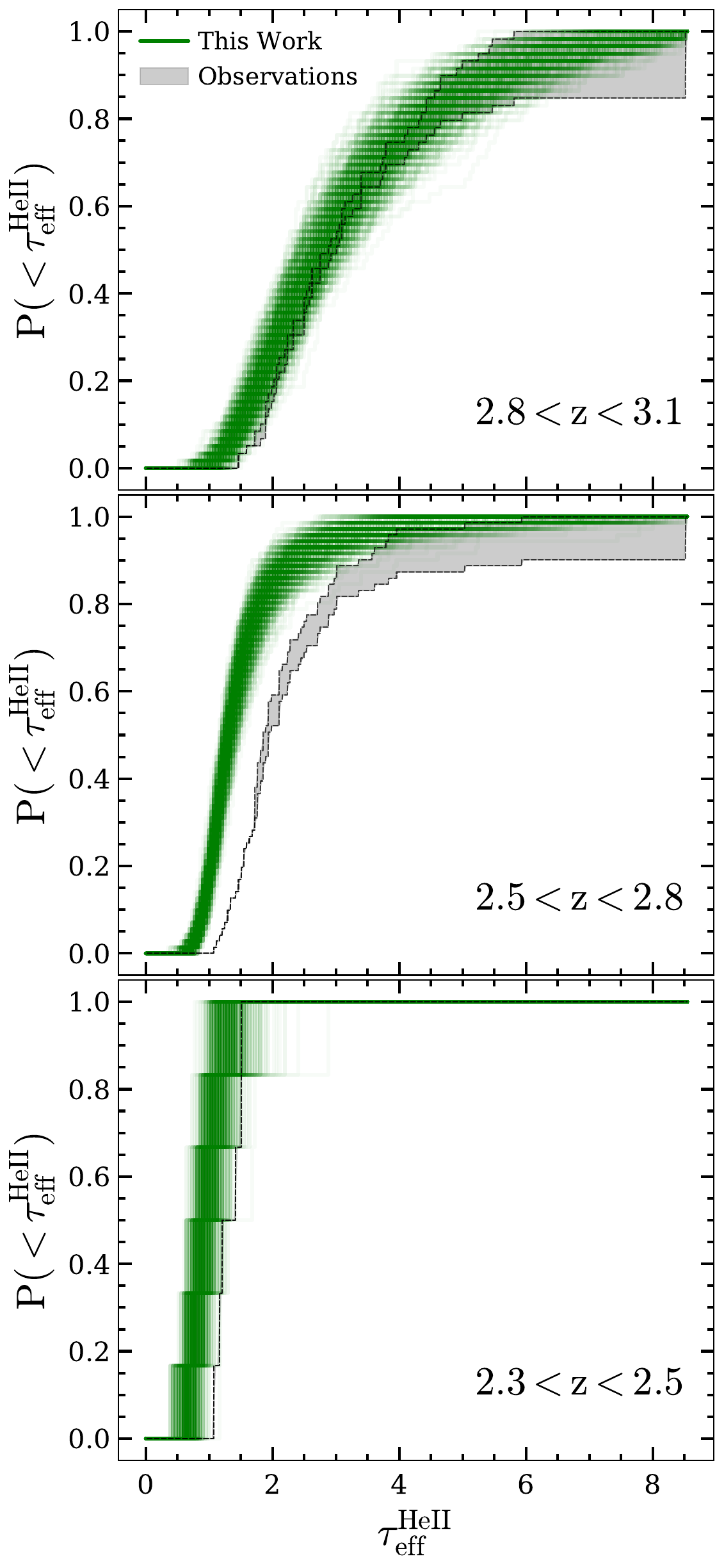}
    
    \caption{Cumulative distribution function of He~{\sc ii} effective optical depth in different redshift bins, averaged over spectral chunks of length $50 h^{-1} {\rm cMpc}$ extracted from simulation \texttt{N512-ph5e5}. All the curves are generated by sampling the true distribution in the simulation with the same number of observational data points in respective redshift bins (see text for more details). The grey shaded regions are observational limits adopting the "optimistic" and "pessimistic" approach used by \citet{bosman2018}.}
    \label{fig:cum_taueff}
\end{figure}
\subsection{Thermal history}
\label{thermal_history}
The ionization of helium injects energy in the IGM through photo-heating, rising its temperature. The evolution of IGM temperature is not only a test of reionization models, but also the main physical mechanism linking helium reionization to the evolution of structures, since it  hanges the dynamics of gas, its cooling rate and its accretion onto galaxies. We compute the IGM temperature at mean density by selecting all the cells in the simulation volume having gas density within $1\%$ of mean gas density. In figure \ref{fig:temp_evol}, we show the redshift evolution of the median of this quantity (solid curve) alongside the central 68$\%$ of the data (shaded region). As anticipated, $T_{0}$ increases rapidly in the first half of helium reionization through photo-heating. Once the majority of the simulation volume has been reionized at $z \sim 3$, the rate of heat injection due to photo-heating diminishes, resulting in a flattening at $z \sim 2.7$, when $\gtrsim 99\%$ of cells are ionized and the He~{\sc ii} reionization process is basically complete. 

In figure \ref{fig:temp_evol} we also plot a collection of observational data by \citet[computed using Lyman-$\alpha$ line width distribution]{Schaye2000}, \citet[from analysing QSO proximity zone]{Bolton2010}, \citet[using Morlet wave filter analysis]{Lidz2010}, \citet[using curvature statistics]{becker2011}, \citet[from Doppler line width of Lyman-$\alpha$ absorption]{Bolton2012}, \citet[via wavelet filtering analysis]{Garzili2012}, \citet[from line-width distributions]{bolton2014}, \citet[from curvature statistics]{boera2014}, \citet[from distribution of H~{\sc i} column density and Doppler parameters]{rorai2018}, \citet[from distribution of Doppler parameters]{hiss2018}, \citet[via Lyman-$\alpha$ forest flux power spectra]{boera2019}, \citet[via Lyman-$\alpha$ forest flux power spectra]{walther2019} and \citet[via flux power spectra, Doppler parameter distribution, wavelet and curvature statistics]{Gaikwad2020,gaikwad2021}.
Despite a large scatter among them, these observations yield a coherent picture in which the IGM undergoes a phase of rapid heating at $3 \lesssim z \lesssim 4$, and then rapidly cools down. 
In the initial phase and until $z \sim 4$, our fiducial simulation predicts somewhat higher temperature compared to observations. This mismatch is a consequence of introducing a temperature floor at $10^{4}$ K (see Section~\ref{ics}). This does not affect our conclusions, since our entire analysis relies on later times, where the self-consistently calculated IGM temperature significantly exceeds this value.

Interestingly, The IGM temperature at mean density does not decrease at all at $z \lesssim 3$, but rather stays constant. There are multiple possible reasons for this. First, if the simulated heating provided by feedback processes during structure formation is too large, it might artificially compensate for the adiabatic cooling of the IGM, maintaining its temperature approximately constant. We have checked in our simulation that this plays only a minor role by excluding from the computation of $T_0$ all cells affected by feedback in {\tt TNG300} and finding that this only changes the IGM temperature at mean density by approximately 5$\%$. Alternatively, the quasar SED employed might be responsible for this offset. In fact, if our SED overestimates the number of photons with $E_\gamma \gg 54.4$ eV, it might artificially boost the IGM photo-heating. However, we remind the reader that such SEDs are based on a  compilation of observations covering the redshift range investigated (see section~\ref{sed}). Finally, it might be the case that the simulated end of helium reionization is not rapid enough, producing a broad peak, of which we miss the late-time part. This explanation is in broad agreement with the somewhat late end of helium reionization in our model discussed above. 

It should be noted that the majority of the observational constraints on $T_{0}$ are obtained by calibrating observable quantities with simulations, therefore introducing model-dependencies in the inferred physical quantities. Additionally, many of the simulations employed for this task assume a spatially-uniform time-varying UV-background, thus missing the effect of a fluctuating He~{\sc ii} ionizing background. Nevertheless, the evidence for a peak at $z\sim3$ appears strong, at least from a qualitative perspective.

Interestingly, the flattening point in the simulated curve aligns well with the peak in $T_{0}$ for the majority of observations, indicating that our predicted end of the helium ionization process is similar to the observed one. However, the shallower evolution of $T_{0}$ around this ending phase (i.e. turn-over point) suggests a more extended period of helium reionization. We will explore this in more detail in the next sections by extracting synthetic He~{\sc ii} Lyman-$\alpha$ forest properties.

\subsection{He~{\sc ii} Lyman-$\alpha$ forest}
\label{forest}
To directly compare our model with observations of He~{\sc ii} Lyman-$\alpha$ forest, we have generated synthetic spectra by extracting $16384$ sightlines at each RT snapshot, each spanning the full box length in the z direction.
For each pixel $i$ in a sightline we evaluate
the normalized transmission flux $F(i) = \exp[-\tau_{\rm{HeII}}(i)]$, where
\begin{equation}
    \tau_{\rm{HeII}}(i) = \frac{c \sigma_{\alpha} \delta R}{\sqrt{\pi}} \sum_{j=1}^{N} n\mathrm{_{HeII}}(j) \ \tilde{V}(i,j),
\end{equation}
is the He~{\sc ii} Lyman-$\alpha$ optical depth. 
Here $N$ is the number of pixels in a the sightline, $\sigma_{\alpha}$ is the Lyman-$\alpha$ scattering cross-section, $c$ is the speed of light, $\delta R$ is the size of a pixel in proper distance units, $n_\mathrm{{HeII}}(j)$ is the He~{\sc ii} number density at the position of pixel $j$ and $\tilde{V}$ is the convoluted Voigt profile approximation provided by \citet{tepper2006}. The latter depends on the IGM temperature and the peculiar velocity. Throughout this paper, we ignore the contamination from other lines into the wavelength window of the He~{\sc ii} Lyman-$\alpha$ forest. This is not expected to have any impact on our results. In fact, such contamination is a major obstacle to observations but the data actually employed to constrain the helium reionization are typically free of this issue, making our choice reasonable.

\subsubsection{Individual line of sight}
\label{forest_1los}
To demonstrate the results of our forward modeling procedure, we show in figure \ref{fig:los} (from top to bottom) the IGM ionization state, gas temperature, transmitted flux and optical depth along one individual line of sight at $z \sim 3.01$ (left column) and 2.32 (right column). For the sake of clarity, here (and for all the following plots) we omit the $i$ when referring to values of the physical quantities associated to a single pixel. As anticipated, at $z \sim 3.01$  $x_{\textrm{HeII}}$ (top row) is notably higher than at $z \sim 2.32$, when most helium is fully ionized. Correspondingly, as redshift decreases, $T_{\rm IGM}$ (second row) rises alongside He~{\sc iii} fractions, leading to lower values of $\tau_{\rm HeII}$ (third row) and more areas of transmitted radiation (bottom). Notice that the IGM temperature is set by the combined effect of photo-heating and other feedback processes associated to structure formation (such as shock heating, AGN feedback, etc.). Therefore it can locally reach values significantly higher than those expected from pure photo-heating and, simultaneously, drive localised collisional ionization of He~{\sc ii} into He~{\sc iii}. This can be seen in the narrow temperature spikes in the figure. We also show the local He~{\sc ii} reionization redshift ($z\rm{_{reion}}$) extracted at the end of the simulation (i.e. at $z\sim2.32$). As discussed in section \ref{reion_history}, regions undergoing an earlier reionization (highlighted by the shaded region)
are characterized by a slightly lower temperature than the rest.

While the spatial resolution of our synthetic spectra prevents us from probing the smallest individual absorption features of the He~{\sc ii} Lyman-$\alpha$ forest, they reliably capture the average IGM opacity (see Appendix \ref{appendix:convergence_Ngrid}), which we compare to observations in the next section.

\subsubsection{Effective optical depth}
\label{forest_tau_eff}
 
To facilitate direct comparison with observations, we evaluate the He~{\sc ii} effective optical depth $(\tau_{\rm{eff}}^{\rm{HeII}})$, a widely used characterization of the Lyman-$\alpha$ forest. To achieve this, we partition each synthetic spectrum into chunks of length 50 $ h^{-1}{\rm cMpc}$ (corresponding to $\Delta z \approx 0.04$ at the relevant redshifts) following \citet{worseck2016}, resulting in a sample of over 60,000 chunks at each redshift. The He{\sc ii} effective optical depth is then computed as:
\begin{equation}
\tau_{\rm{eff}}^{\rm{HeII}} = - \ln (\langle F \rangle),
\label{columndensity}
\end{equation}
where $\langle F \rangle$ is the mean transmitted flux of all the pixels in each chunk. In order to approximately mimic the fact that observations are not able to distinguish values of the optical depth above a certain threshold $\tau_{\rm{eff,th}}^{\rm{HeII}}$, we consider observable those spectral chunks with $\tau_{\rm{eff}}^{\rm{HeII}}<\tau_{\rm{eff,th}}^{\rm{HeII}}$. We set $\tau_{\rm{eff,th}}^{\rm{HeII}}=8.56$ as representative of the typical maximum optical depth detected in the observations at $z>3$ \citep{worseck2016,worseck2019,makan2021,makan2022}. 

Figure \ref{fig:taueff} displays the redshift evolution of the median observable He~{\sc ii} effective optical depth $\tau_{\rm{eff}}^{\rm{HeII}}$ (solid line) along with the central 68\%, 95\% and 99\% of the data (shaded regions of increasing transparency), as well as aformentioned observational data points. 
As a reference, we also show the median of the intrinsic distribution (i.e. the one without any optical depth threshold) using a dashed line. The difference between the intrinsic and the observable distribution is crucial for investigating the highest-redshift observations available, that are only able to sample the low-$\tau_{\rm{eff,th}}^{\rm{HeII}}$ part of the intrinsic distribution. We also predict that at $z\lesssim 3$ observations are able to effectively sample the entire distribution. 

In fact, while the intrinsic median $\tau_{\rm{eff,th}}^{\rm{HeII}}$ keeps increasing with increasing redshift, the observable one flattens out at $z \gtrsim 3.5$, with the bulk of effective optical depths becoming less and less accessible to observations, which are only sensitive to the most transparent regions of the IGM.
Following the completion of the reionization process at $z \sim 2.7$, the effective optical depth shows a residual non-zero value, due to residual He~{\sc ii} in self-shielded systems and recombinations in the IGM. The mild time evolution seen in our simulations develops as a consequence of (i) the evolving thermal state of the IGM mainly through adiabatic cooling and heat injection due to structure formation, and (ii) the reduced recombination rate stemming from the expansion of the Universe lowering the density in the IGM. The noticeable scatter around the median curve reflects the patchy nature of the He~{\sc ii} reionization process, and indeed it decreases once this process is over. 
We find an agreement with most data points within the $95\%$ confidence intervals, demonstrating the effectiveness of our simulation in reproducing the observed behavior of He~{\sc ii} effective optical depth. This success is indicative of the fact that the QLF from \citet{Shen2020} provides a reasonable helium reionization history and topology, when combined with a accurate hydrodynamical and RT simulation. 

When looking at the details, however, we find some discrepancies, which are explored in the following. From figure \ref{fig:taueff}, it can be seen that at $z \lesssim 3$ most of the observed effective optical depths lie above the median value derived from our simulations. In order to better compare the distribution of effective optical depths at fixed redshift, we compute its cumulative distribution function (CDF) in three redshift bins spanning the range $z$ = 2.3 - 3.1. 
In order to provide a fair comparison with observations, for each redshift bin investigated we create 500 realization of the predicted optical depth distribution. We do so by randomly selecting a number of synthetic $\tau_{\rm{eff}}^{\rm{HeII}}$ equal to the number of observed values within the bin. We show these realization in figure \ref{fig:cum_taueff} as thin green lines, along with the observed CDF (obtained employing the same observations shown in figure \ref{fig:taueff}). For the latter, we adopt the "optimistic/pessimistic" approach of \citet{bosman2018} to deal with pixels having flux below the detection threshold, thus resulting in a range of possible observed values (gray shading).  It should be noted that our simulated optical depths are not calibrated to match any observed CDF nor mean transmitted flux. 
Despite this, in the highest redshift bin (corresponding to an approximately 90\% ionised IGM) the simulated CDF reproduces well the observed one. As reionization progresses, the CDF slope becomes as expected progressively steeper. For $z\lesssim 2.8$, though, the simulated CDF is systematically shifted to lower optical depths in comparison to observations, although the shape is still well reproduced. In the lowest redshift bin, the CDF gets saturated at $\tau_{\rm{eff}}^{\rm{HeII}} \sim 1.5$, confirming that the last stage of the reionization process has been reached. Here as well, the simulated CDF is somewhat shifted to lower optical depths. These discrepancies can be revealing of two different phenomena, namely (i) that in our model helium reionization is completed slightly too early with respect to the observed data and (ii) that the limited resolution of our simulations prevents us from fully resolving the residual sinks of radiation in the post-reionization Universe that would increase the IGM effective optical depth. From the discussion in Section \ref{reion_history} and the resolution test we have performed (described in Appendix \ref{appendix:convergence_Ngrid}) we conclude that these differences are probably due to a combination of both effects.
Unfortunately, the resolution needed to properly capture the small scale Lyman-limit systems would make our simulations prohibitively expensive. Therefore, for a more accurate modeling of the final phases of the process, a sub-grid prescription is required (e.g. \citealt{Mao2020,Bianco2021,Cain2021}),
which we plan to include in future investigations.

\subsubsection{Characterisation of transmission regions}
\label{forest_spikes}
Here we move to a characterization of individual features in the He~{\sc ii} Lyman-$\alpha$ forest. These carry information on the sources of reionization \citep[e.g. ][]{garaldi2019, Gaikwad2020}, but are also much more dependent on the simulation resolution than the average quantities discussed so far. We note that the resolution requirements in the case of helium reionization are less stringent than in the case of hydrogen reionization, since the larger bias of the sources implies that the ionized regions are typically much larger, and so are the features in the forest. Therefore, the resolution necessary to capture their existence, if not their details, is lower. In order to maintain a conservative approach, in the following we only investigate quantities that we have found to be resilient to changes in resolution. For instance, we found that the transmission peak height is more robust than its width against resolution changes, since the former only depends on the most ionized regions (which are typically close to a very bright source and therefore fully ionized regardless of the resolution), while the latter depends on the details of the ionized region edges, which are much more sensitive to the employed resolution. Consequently, we elect to investigate only the former. 
Nevertheless, the results that follow need to be taken as qualitative more than quantitative, since an improved spatial resolution will still affect their details. 

\begin{figure}
    \includegraphics[width=1.1\columnwidth]{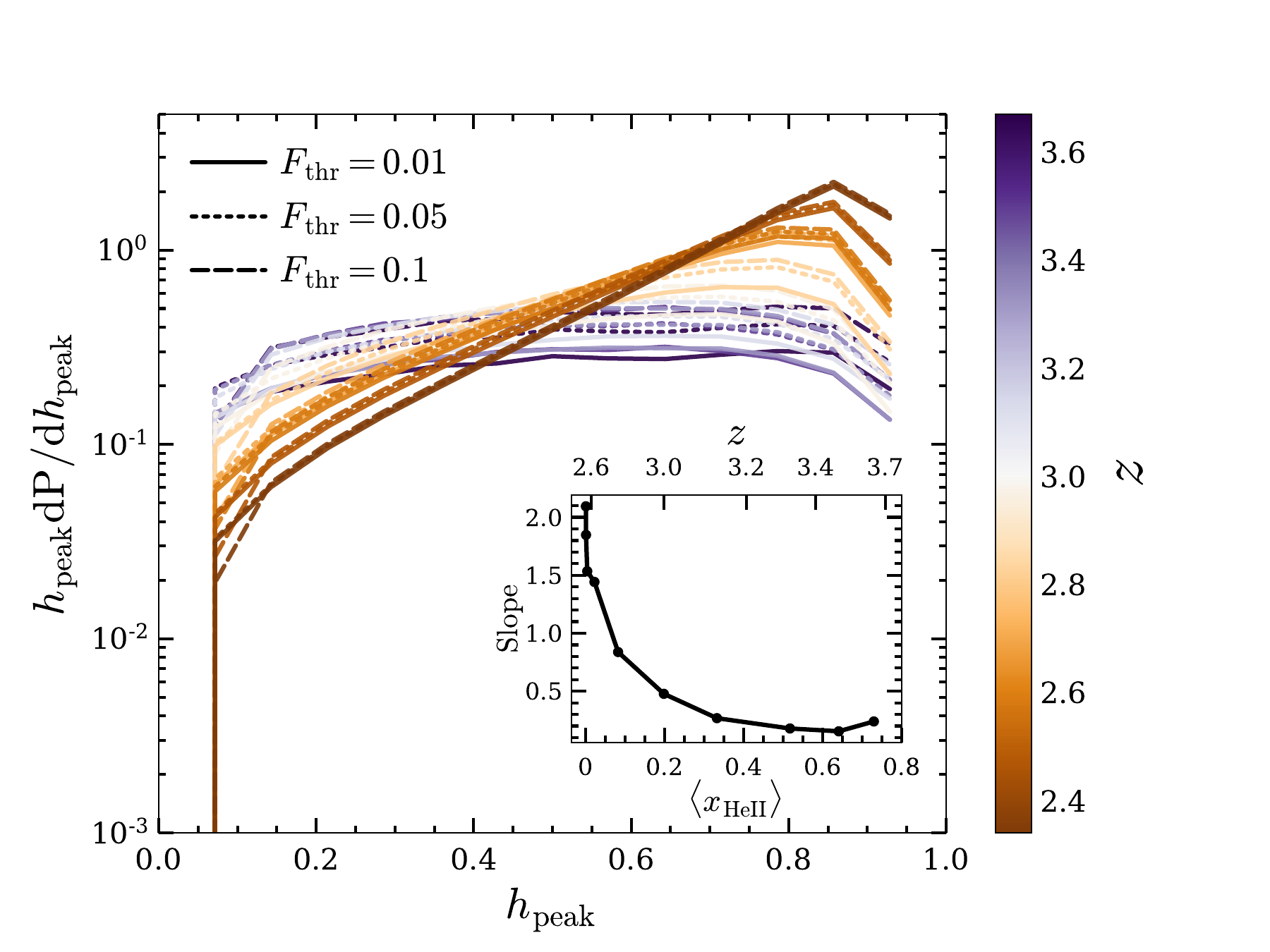}

    \caption{Distribution of  peak heights ($h\mathrm{_{peak}}$) computed over the sample of He~{\sc ii} Lyman-$\alpha$ forest spectra extracted from simulation \texttt{N512-ph5e5} at different redshifts (indicated by the color bar). Different line styles refer to different values of the adopted flux threshold, $F_{\rm thr}$. In the inset, the relation between the slope of the distribution (see text for more details) and $\langle x_\mathrm{HeII}\rangle$ for $F_{\rm thr}=0.01$ is shown as black solid curve.}
    \label{fig:transmitted_region_prop_hpeak}
\end{figure}

We start by identifying transmission regions in the synthetic spectra. We follow the procedure developed in \citet{Garaldi2019b,garaldi2022}, adapted from \citet{Gnedin2017}, and consider a pixel as part of a transmission spike if its normalised flux is higher than a threshold value, $F\mathrm{_{thr}}$. We then characterize the spike through its height, $h_\mathrm{_{peak}}$, defined as the maximum transmitted normalized flux among all pixels associated to it. 
In figure \ref{fig:transmitted_region_prop_hpeak} (main panel) we show the probability density of $h\mathrm{_{peak}}$ for different redshift and $F\mathrm{_{thr}}$. We observe a strong dependence on $z$, with an almost flat curve at $z>3$, indicating a prevalence of small transmission spikes over large ones (notice that the vertical axis is multiplied by $h\mathrm{_{peak}}$). This can easily be understood as at this redshift ionized regions are still relatively small around the first quasars active in the simulated volume, so part of the transmitted flux is absorbed while traveling through the nearby neutral regions. As redshift decreases, more sources turn on and the ionized regions grow in size. This is reflected in a tilt of the peak height distribution towards larger values. Interestingly, this change in the slope does not break the linearity of the relation (for the chosen variables and axis scaling). This allows us to derive an empirical relation between the slope of this distribution and the volume-averaged He~{\sc ii} fraction in the simulation. 
To obtain the former we perform a least-square fit of $\log(h_\mathrm{_{peak}} \mathrm{d}P/\mathrm{d}h_\mathrm{_{peak}})$ in the range $0.2 \leq h\mathrm{_{peak}} \leq 0.8$. We show the co-evolution of these two quantities in the inset of figure \ref{fig:transmitted_region_prop_hpeak}. When $\langle x_\mathrm{HeII}\rangle $ is well above 40$\%$ (or $z>3$), the slope is $\sim $ 0, consistently with the flat distribution discussed above. Towards lower redshift, instead, we observe a steep rise of the slope, confirming the sensitivity of this probe to the last phases of helium reionization with percolation of ionized bubbles. 
Finally, we note that the rightmost bin of the $h_\mathrm{_{peak}}$ distribution shows a decline. This is due to the fact that in our simulations the IGM is not yet sufficiently ionized to allow a complete transmission of the incoming flux. The qualitative behaviour described is independent of the value adopted for $F\mathrm{_{thr}}$, although there are some minor quantitative differences at the highest redshifts. 

\begin{figure}
    \includegraphics[width=1.1\columnwidth]{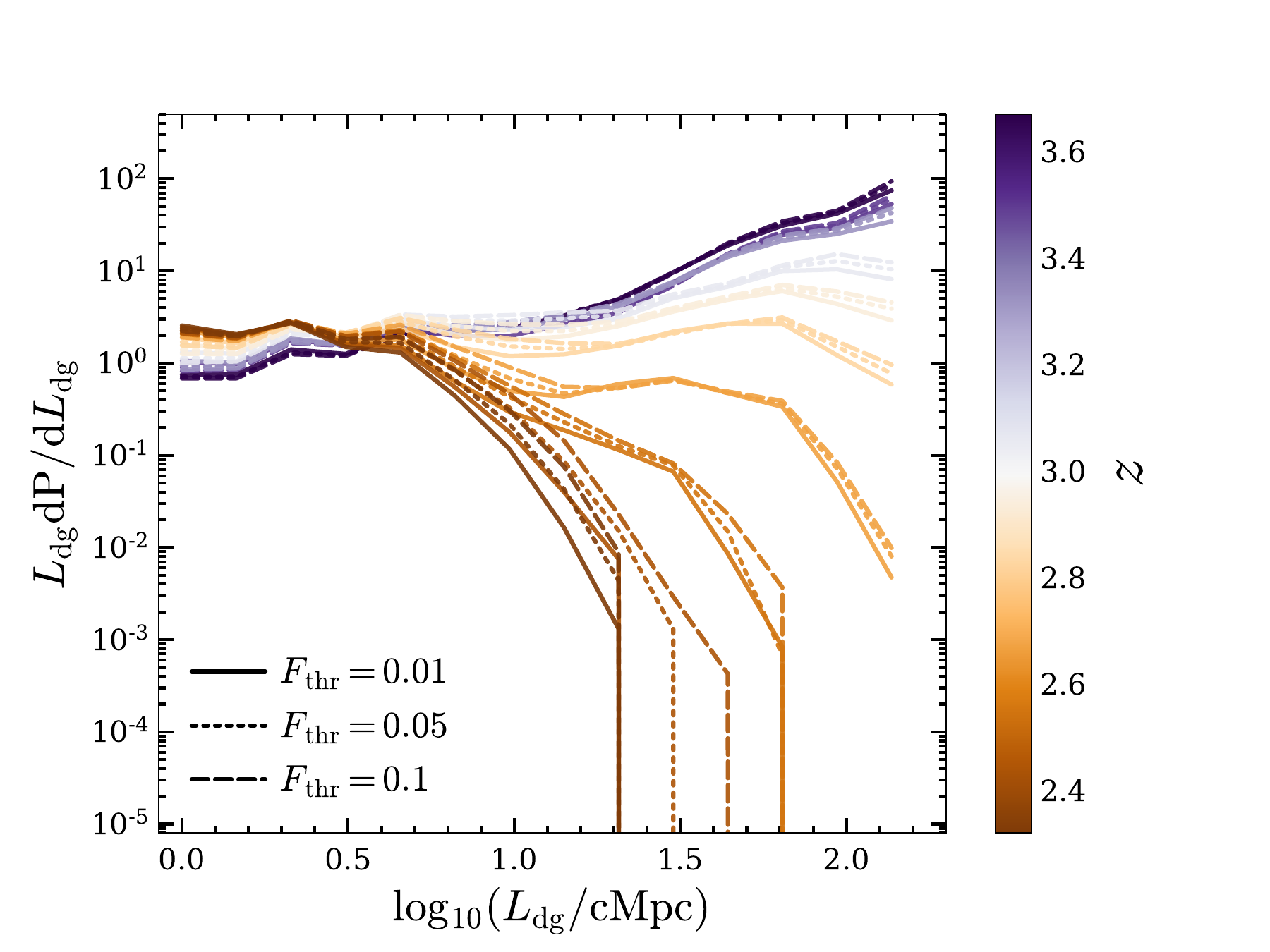}

    \caption{Distribution of  dark gaps lengths ($L\mathrm{_{dg}}$) computed over the sample of He~{\sc ii} Lyman-$\alpha$ forest spectra extracted from simulation \texttt{N512-ph5e5} at different redshifts (indicated by the color bar). Different line styles refer to different values of the adopted flux threshold, $F_{\rm thr}$.}
    \label{fig:transmitted_region_prop_Ldark}
\end{figure}

To complement the peak height distribution analysis, we compute the distribution of the absorbed regions in the spectrum using the dark gap (DG) statistics \citep{paschos2005,fan2006,gallerani2006,gallerani2008}, where a DG is defined as a continuous region with normalised transmission flux below $F\mathrm{_{thr}}$, and it is characterized by its length, $L\mathrm{_{dg}}$. In figure \ref{fig:transmitted_region_prop_Ldark} we show the distribution of $L\mathrm{_{dg}}$ for various redshifts and $F\mathrm{_{thr}}$ values. 
It should be noted that $L\mathrm{_{dg}}$ is limited by construction to the length of each synthetic spectra, which in turn is constrained by the simulation box length (since we do not employ periodic boundary conditions). 
The distribution of $L\mathrm{_{dg}}$ exhibits a clear trend throughout the entire redshift range investigated. The occurrence of the longest gaps ($L\mathrm{_{dg}} \gtrsim 10\, \rm{cMpc}$) monotonically decreases with cosmic time, as a consequence of the increasing number of ionized regions towards the lower redshifts. This drop is significantly more rapid between $z\sim2.7 \lesssim z \lesssim 3$. Conversely, the occurrence of shorter gaps increases with the development of helium reionization. The distribution of $L\mathrm{_{dg}}$ is also practically insensitive to the adopted threshold value. We have confirmed that this statistics is fairly independent of the simulation resolution in the redshift range explored. 

\begin{figure}
    \includegraphics[width=\columnwidth]{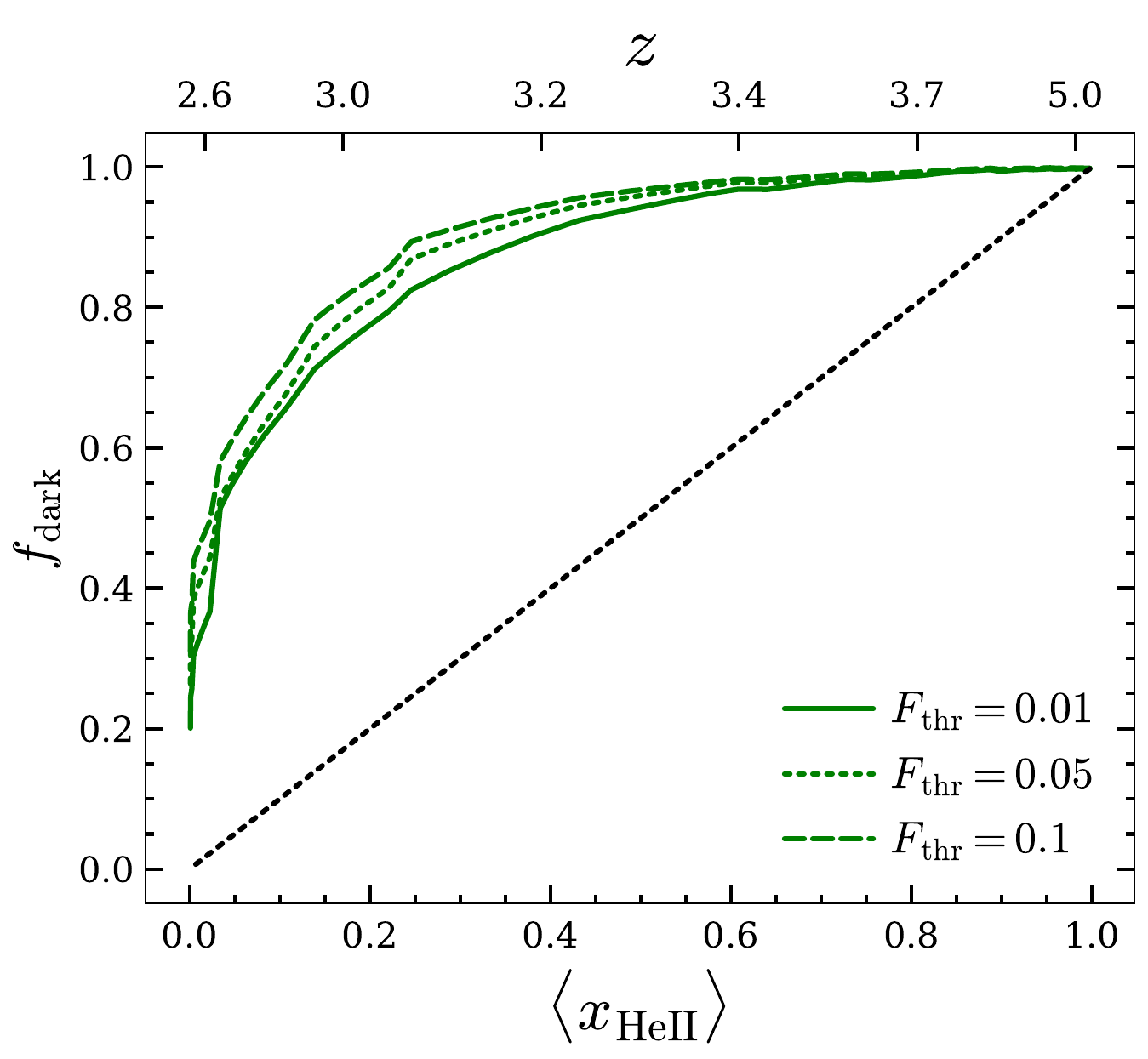}
    
    \caption{Evolution of the dark pixel fraction ($f_\mathrm{dark}$) as a function of the volume-averaged He~{\sc ii} fraction ($\langle x_\mathrm{HeII}\rangle$) in synthetic spectra of the He~{\sc ii} Lyman-$\alpha$ forest extracted from our fiducial simulation. Different line styles (green curves) refer to different values of the adopted flux threshold, $F_{\rm thr}$, in the definition of dark pixels. The black dotted curve shows one-to-one correspondence between these two quantities. In the top axis, the corresponding redshifts are displayed.}
    \label{fig:darkfraction_evol}
\end{figure}

Finally, we compress the information provided by the dark gaps distribution into a single number by computing the fraction of pixels ($f\mathrm{_{dark}}$) in all spectra at a a given redshift that have normalized flux below a threshold value $F\mathrm{_{thr}}$. 
In the context of hydrogen reionization, this quantity is often employed to derive model-independent upper limits on the neutral fraction as $x_\mathrm{HI} \leq f_\mathrm{dark, HI}$ \citep{Mesinger2010,Mcgreer2011,Mcgreer2015}. 
In figure \ref{fig:darkfraction_evol} we show the co-evolution of this dark fraction with volume-averaged He~{\sc ii} fraction in our fiducial simulation. 
A distinct drop of $f\mathrm{_{dark}}$ is visible at $z\sim3.3$, where $\sim$ 50$\%$ of Helium is fully ionized. As expected, $f\mathrm{_{dark}}$ increases with increasing $F\mathrm{_{thr}}$. 
The fact that throughout our simulation $f\mathrm{_{dark}}$ is higher than $\langle x_\mathrm{HeII}\rangle$ (i.e. all green curves are well above the black dotted curve) confirms that this approach works also in the context of helium reionization. 

In general, with enough data available, all these statistics characterizing transmission regions can be used to constrain the timing of He~{\sc ii} reionization, as well as the source properties, as discussed e.g. by \citet{garaldi2022}. 

\begin{figure} 
    \includegraphics[width=\columnwidth]{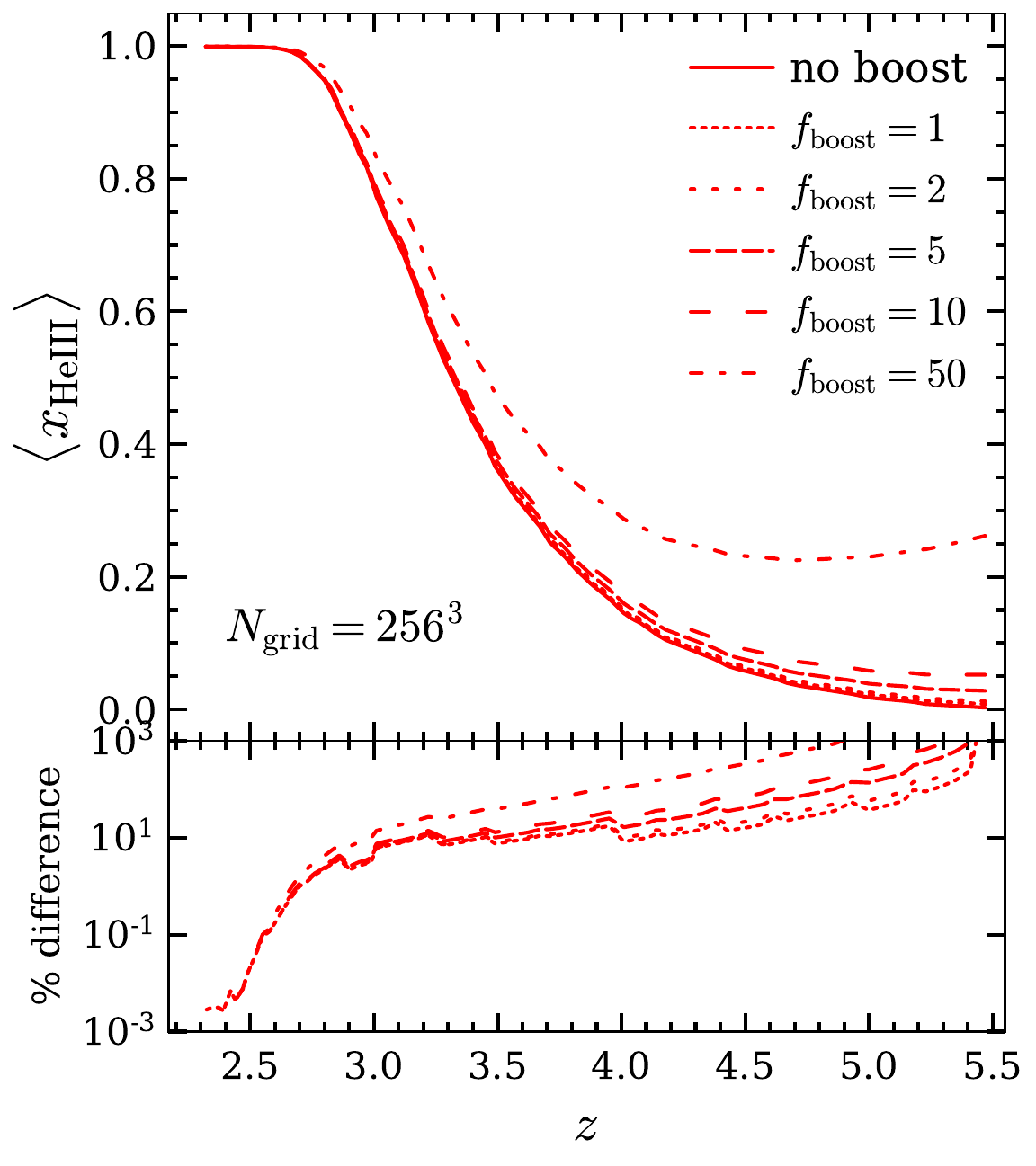}
    \caption{{\it Top panel}: evolution of the volume-averaged He~{\sc ii} fraction for simulations featuring a boosted QLF at $z>5.53$. The boosting factor is reported in the legend. Notice that we assume a step-like transition at $z=5.53$ between the boosted and fiducial QLF.
    The solid curve ($f_\mathrm{boost}=1$) correspond to \texttt{N256-ph1e5}, i.e. a lower-resolution version of our fiducial simulation, which is started at $z_\mathrm{in}=5.53$ with the intergalactic helium singly ionised. {\it Bottom panel}: relative differences of each run with respect to the fiducial \texttt{N256-ph1e5}.}
    \label{fig:jwst}
\end{figure}

\subsection{Impact of \texttt{JWST} detection of $z>5$ QSOs}
\label{jwst}

The \texttt{JWST} has recently detected numerous AGNs at $z > 5$, although in relatively small survey areas. If confirmed to be representative of the entire population, these would imply a significantly boosted QLF (up to approximately a factor of 30 in the faint end of the QLFs) with respect to pre-\texttt{JWST} estimations \citep[e.g.][]{Harikane2023,Maiolinoa2023,Maiolino2023,Goulding2023,Larson2023,Juodzbalis2023,Greene2023}.
Under the assumption that the characteristics of such objects are similar to those of lower redshift ones (in particular, their UV spectrum and their near-unity escape fraction of ionizing photons, but see \citealt{Christiani2016}), such population could have an impact on both hydrogen and helium reionization. Here we assess their contribution to the latter. We caution, however, that our approach is relatively simple, and as such its results should be taken as qualitative. 

Given the large differences in the QLF derived by different studies, and in order to minimize the number of parameters in our study while maintaining flexibility, we choose to model the population of $z>5$ QSOs by boosting our fiducial QLF \citep[i.e Model 2 of ][]{Shen2020} by a factor $f_\mathrm{boost} $\footnote{Note that the $f_\mathrm{boost}$ mentioned here differs from the one used in \citet{pacucii2023}, where it is employed to quantify a boost in black hole masses observed by \texttt{JWST} to account for non-detectability. Effectively, this will cause a uniform shift in the QLF towards more massive luminous bins while we instead introduce a boost across all luminosity bins.} = 1, 2, 5, 10, 50  at $z>5.53$, while maintaining its value at lower redshift, where observations of the QLF are more robust. We then run simulations identical in setup to those discussed so far, but starting from $z_\mathrm{init} = 6$. To reduce the computational requirement, these additional simulations are run only with $N_{\rm grid}=256^3$.

Figure \ref{fig:jwst} shows the simulated helium reionization histories (top panel) and their relative difference with respect to our fiducial QLF (bottom) for the different $f_\mathrm{boost}$ simulated. 
We also show a run labeled `no boost', which we take as reference in the bottom panel. This is a lower-resolution version (to match the other runs discussed in this section) of our fiducial simulation. We use this to show the impact of different initial redshift for the simulation, since `no boost' starts at $z_\mathrm{init} = 5.53$ (like our fiducial run and other runs previously discussed), while the simulation labeled `$f_\mathrm{boost} = 1$' begins at $z_\mathrm{init}=6$. From an inspection of the bottom panel it can be seen that such differece is generally negligible with respect to modification to the QLF. 
For most $f_\mathrm{boost}$ values, the He~{\sc ii} fraction at $z=5.53$ is only mildly affected and the difference between the various runs is quickly erased once helium reionization is ongoing. The exception to this is our most extreme model ($f_\mathrm{boost} = 50$), that maintains some small difference all the way to the completion of helium reionization. Although not shown here for the sake of brevity, we find similar qualitative and quantitative results with respect to the He~{\sc ii} effective optical depth and associated statistics.
The findings unequivocally indicate that an increased (up to an order of magnitude) number of quasars at $z\gtrsim5$ (or more precisely a higher emissivity from quasars at such earlier times), as suggested by \texttt{JWST}, does not necessarily impact the ionization and thermal state of the IGM during the ending phase of this epoch. Should the number density of quasars be larger, their characteristics (chiefly, the ionizing photons escape fraction) be different than lower-$z$ QSOs, or should the boost persist to later times, our results would have to be revised through a tailored, more accurate model.

\section{Discussion and Conclusions}
\label{conclusions}

Recent observational advancements have pushed our ability to probe helium reionization past its tail end. At the same time, new observations-based models of the quasar luminosity function (QLF) have become available. Most notably, some of these models enforce a smooth evolution through time of the QLF, enabling us to bypass the need to assume an ionising emissivity evolution to rescale the observed QLF at a given redshift, as often done in the past \citep{compostella2013, compostella2014, Garaldi2019b}. In this paper, we present the first simulations of helium reionization directly implementing one such QLF \citep[namely the Model 2 from][]{Shen2020}. We carry out a thorough exploration of its predictions concerning helium reionization, including a comparison to the latest observations and a set of predictions for the features of the He~{\sc ii} Lyman-$\alpha$ forest that can be measured in current and future data. To this end, we combine the \citet{Shen2020} QLF with the \texttt{TNG300} cosmological hydrodynamical simulation and the 3D radiative transfer code \texttt{CRASH} (see section \ref{tng+crash}). Another distinguishing feature of our simulations is the inclusion of observationally-derived quasar spectra (see section \ref{sed}). This makes our prediction entirely independent from the (often very approximate) physics of black hole seeding, formation and evolution implemented in large-volume numerical simulations. Employing this setup, we simulate the evolution of the IGM from $z_\mathrm{init} = 5.53$ to the completion of helium reionization and forward model our runs to provide faithful predictions of the He~{\sc ii} Lyman-$\alpha$ forest. Finally, inspired by recent \texttt{JWST} observations, we develop a simple numerical model to assess the impact of high-redshift QSOs on helium reionization. For that, we assume that (i) the QLF is unchanged at $z\leq 5.53$ with a step-like transition, (ii) that the large number of high-redshift AGNs observed results in a rigid shift of our fiducial QLF towards larger number densities, and (iii) that the spectrum of high-redshift quasars is identical to their low-redshift counterparts. While simplistic, this approach captures the main features relevant for helium reionization, and therefore our results are expected to be qualitatively robust.

Our main results can be summarized as follows: 
\begin{itemize}
    \item We predict a later helium reionization history with respect to the majority of models available in the literature. While many details of the modeling differ, the largest impact likely comes from the different QLF employed.
    
    \item Our fiducial simulation predicts an IGM temperature at mean density which is somewhat higher than most of the observational data points. This is likely due to the fact that the latter are obtained by calibrating with respect to simulations employing a uniform UV background, thus missing the impact of inhomogeneous helium reionization.

    \item The simulated IGM temperature at mean density remains constant even after completing the reionization process, although the point of flattening of the curve aligns well with the observational turn-over point.
    
    \item The observed and simulated He~{\sc ii} effective optical depth are in good agreement once observational limitations are properly taken into account. They also show a comparable amount of scatter throughout the simulated history. Therefore, we expect our model to properly capture the bias of the sources responsible for helium reionization and the resulting patchiness of such process. 

   \item We found residual differences between the simulated and observed distribution of He~{\sc ii} effective optical depth at fixed redshift, in particular at $z \le 2.8$. This can be ascribed to an early end of reionization in our simulations. 

    \item We present a comprehensive analysis of transmission regions and dark gaps in simulated He~{\sc ii} Lyman-$\alpha$ forest spectra. We demonstrate how the shape of their distribution is linked to the underlying IGM ionization state, opening up the possibility to use such measure to provide additional constraints from observed quasar spectra. 

    \item We find that, in our simplified approach to estimate the impact on helium reionization of the large number of AGNs observed by \texttt{JWST}, their effect is negligible unless the QLF is boosted by more than an order of magnitude at $z\gtrsim5.5$. 

\end{itemize}

In summary, we have presented a detailed characterization of helium reionization in light of the latest constraints on the QLF and the latest measurements of the IGM optical depth, combining a state-of-the-art hydrodynamical simulation with accurate radiative transfer. Our results show an overall convergence of predicted and observed properties in the redshift range $2.3 \le z\le 3.5$. 

Nevertheless, the discrepancies we notice between our fiducial simulation and observations are likely indicative of a lack of self-consistency between current observations of helium reionization and of the QLF itself. Unfortunately, our simulations are unable at present to point toward on of these two as culprit. Additionally, it is possible that assumptions in our modeling contribute (or even generate) the discrepancies discussed before. Among these, we would like to highlight the spectral energy distributions of QSOs and the quasar ionizing escape fraction. The latter, in particular, is virtually always assumed to be unity (in simulations like the one presented in this work), but it has been shown vary significantly \citep{Christiani2016}. Testing and overcoming these assumptions are as essential as improving observational datasets (in particular concerning helium reionization) in order to assess whether our understanding of quasar assembly and evolution is consistent with that of intergalactic helium reionization. We plan to follow this route in a future dedicated paper.
Investigation like ours demonstrate the power of combining different domains of knowledge to achieve a deeper understanding of structure formation throughout cosmic history.

\section*{Acknowledgements}
All simulations were carried out on the machines of Max Planck Institute for Astrophysics (MPA) and Max Planck Computing and Data Facility (MPCDF). We thank the anonymous reviewer for the useful comments which helped to improve the manuscript. We thank Frederick Davies for useful discussions. AB thanks the entire EoR research group of MPA for all the encouraging comments for this project. EG acknowledges support from the CANON foundation Europe through part of this project. This work made extensive use of publicly available software packages : \texttt{numpy} \citep{vander2011}, \texttt{matplotlib} \citep{Hunter2007}, \texttt{scipy} \citep{Jones2001} and \texttt{CoReCon} \citep{Garaldi2023}. Authors thank the developers of these packages.

\section*{Data Availability}
The final data products from this study will be shared on reasonable request to the authors.



\bibliographystyle{mnras}
\bibliography{mnras_template} 

\begin{thebibliography}{}
\makeatletter
\relax
\def\mn@urlcharsother{\let\do\@makeother \do\$\do\&\do\#\do\^\do\_\do\%\do\~}
\def\mn@doi{\begingroup\mn@urlcharsother \@ifnextchar [ {\mn@doi@}
  {\mn@doi@[]}}
\def\mn@doi@[#1]#2{\def\@tempa{#1}\ifx\@tempa\@empty \href
  {http://dx.doi.org/#2} {doi:#2}\else \href {http://dx.doi.org/#2} {#1}\fi
  \endgroup}
\def\mn@eprint#1#2{\mn@eprint@#1:#2::\@nil}
\def\mn@eprint@arXiv#1{\href {http://arxiv.org/abs/#1} {{\tt arXiv:#1}}}
\def\mn@eprint@dblp#1{\href {http://dblp.uni-trier.de/rec/bibtex/#1.xml}
  {dblp:#1}}
\def\mn@eprint@#1:#2:#3:#4\@nil{\def\@tempa {#1}\def\@tempb {#2}\def\@tempc
  {#3}\ifx \@tempc \@empty \let \@tempc \@tempb \let \@tempb \@tempa \fi \ifx
  \@tempb \@empty \def\@tempb {arXiv}\fi \@ifundefined
  {mn@eprint@\@tempb}{\@tempb:\@tempc}{\expandafter \expandafter \csname
  mn@eprint@\@tempb\endcsname \expandafter{\@tempc}}}

\bibitem[\protect\citeauthoryear{{Anderson}, {Hogan}, {Williams}  \&
  {Carswell}}{{Anderson} et~al.}{1999}]{Anderson1999}
{Anderson} S.~F.,  {Hogan} C.~J.,  {Williams} B.~F.,   {Carswell} R.~F.,  1999,
  \mn@doi [\aj] {10.1086/300698}, \href
  {https://ui.adsabs.harvard.edu/abs/1999AJ....117...56A} {117, 56}

\bibitem[\protect\citeauthoryear{{Becker}, {Bolton}, {Haehnelt}  \&
  {Sargent}}{{Becker} et~al.}{2011}]{becker2011}
{Becker} G.~D.,  {Bolton} J.~S.,  {Haehnelt} M.~G.,   {Sargent} W. L.~W.,
  2011, \mn@doi [\mnras] {10.1111/j.1365-2966.2010.17507.x}, \href
  {https://ui.adsabs.harvard.edu/abs/2011MNRAS.410.1096B} {410, 1096}

\bibitem[\protect\citeauthoryear{{Bianco}, {Iliev}, {Ahn}, {Giri}, {Mao},
  {Park}  \& {Shapiro}}{{Bianco} et~al.}{2021}]{Bianco2021}
{Bianco} M.,  {Iliev} I.~T.,  {Ahn} K.,  {Giri} S.~K.,  {Mao} Y.,  {Park} H.,
  {Shapiro} P.~R.,  2021, \mn@doi [\mnras] {10.1093/mnras/stab787}, \href
  {https://ui.adsabs.harvard.edu/abs/2021MNRAS.504.2443B} {504, 2443}

\bibitem[\protect\citeauthoryear{{Boera}, {Murphy}, {Becker}  \&
  {Bolton}}{{Boera} et~al.}{2014}]{boera2014}
{Boera} E.,  {Murphy} M.~T.,  {Becker} G.~D.,   {Bolton} J.~S.,  2014, \mn@doi
  [\mnras] {10.1093/mnras/stu660}, \href
  {https://ui.adsabs.harvard.edu/abs/2014MNRAS.441.1916B} {441, 1916}

\bibitem[\protect\citeauthoryear{{Boera}, {Becker}, {Bolton}  \&
  {Nasir}}{{Boera} et~al.}{2019}]{boera2019}
{Boera} E.,  {Becker} G.~D.,  {Bolton} J.~S.,   {Nasir} F.,  2019, \mn@doi
  [\apj] {10.3847/1538-4357/aafee4}, \href
  {https://ui.adsabs.harvard.edu/abs/2019ApJ...872..101B} {872, 101}

\bibitem[\protect\citeauthoryear{{Bolton}, {Haehnelt}, {Viel}  \&
  {Carswell}}{{Bolton} et~al.}{2006}]{Bolton2006}
{Bolton} J.~S.,  {Haehnelt} M.~G.,  {Viel} M.,   {Carswell} R.~F.,  2006,
  \mn@doi [\mnras] {10.1111/j.1365-2966.2006.09927.x}, \href
  {https://ui.adsabs.harvard.edu/abs/2006MNRAS.366.1378B} {366, 1378}

\bibitem[\protect\citeauthoryear{{Bolton}, {Oh}  \& {Furlanetto}}{{Bolton}
  et~al.}{2009}]{Bolton2009}
{Bolton} J.~S.,  {Oh} S.~P.,   {Furlanetto} S.~R.,  2009, \mn@doi [\mnras]
  {10.1111/j.1365-2966.2009.14914.x}, \href
  {https://ui.adsabs.harvard.edu/abs/2009MNRAS.396.2405B} {396, 2405}

\bibitem[\protect\citeauthoryear{{Bolton}, {Becker}, {Wyithe}, {Haehnelt}  \&
  {Sargent}}{{Bolton} et~al.}{2010}]{Bolton2010}
{Bolton} J.~S.,  {Becker} G.~D.,  {Wyithe} J. S.~B.,  {Haehnelt} M.~G.,
  {Sargent} W. L.~W.,  2010, \mn@doi [\mnras]
  {10.1111/j.1365-2966.2010.16701.x}, \href
  {https://ui.adsabs.harvard.edu/abs/2010MNRAS.406..612B} {406, 612}

\bibitem[\protect\citeauthoryear{{Bolton}, {Becker}, {Raskutti}, {Wyithe},
  {Haehnelt}  \& {Sargent}}{{Bolton} et~al.}{2012}]{Bolton2012}
{Bolton} J.~S.,  {Becker} G.~D.,  {Raskutti} S.,  {Wyithe} J. S.~B.,
  {Haehnelt} M.~G.,   {Sargent} W. L.~W.,  2012, \mn@doi [\mnras]
  {10.1111/j.1365-2966.2011.19929.x}, \href
  {https://ui.adsabs.harvard.edu/abs/2012MNRAS.419.2880B} {419, 2880}

\bibitem[\protect\citeauthoryear{{Bolton}, {Becker}, {Haehnelt}  \&
  {Viel}}{{Bolton} et~al.}{2014}]{bolton2014}
{Bolton} J.~S.,  {Becker} G.~D.,  {Haehnelt} M.~G.,   {Viel} M.,  2014, \mn@doi
  [\mnras] {10.1093/mnras/stt2374}, \href
  {https://ui.adsabs.harvard.edu/abs/2014MNRAS.438.2499B} {438, 2499}

\bibitem[\protect\citeauthoryear{{Bosman}, {Fan}, {Jiang}, {Reed}, {Matsuoka},
  {Becker}  \& {Haehnelt}}{{Bosman} et~al.}{2018}]{bosman2018}
{Bosman} S. E.~I.,  {Fan} X.,  {Jiang} L.,  {Reed} S.,  {Matsuoka} Y.,
  {Becker} G.,   {Haehnelt} M.,  2018, \mn@doi [\mnras]
  {10.1093/mnras/sty1344}, \href
  {https://ui.adsabs.harvard.edu/abs/2018MNRAS.479.1055B} {479, 1055}

\bibitem[\protect\citeauthoryear{{Boyle}, {Shanks}, {Croom}, {Smith}, {Miller},
  {Loaring}  \& {Heymans}}{{Boyle} et~al.}{2000}]{Boyle2000}
{Boyle} B.~J.,  {Shanks} T.,  {Croom} S.~M.,  {Smith} R.~J.,  {Miller} L.,
  {Loaring} N.,   {Heymans} C.,  2000, \mn@doi [\mnras]
  {10.1046/j.1365-8711.2000.03730.x}, \href
  {https://ui.adsabs.harvard.edu/abs/2000MNRAS.317.1014B} {317, 1014}

\bibitem[\protect\citeauthoryear{{Cain}, {D'Aloisio}, {Gangolli}  \&
  {Becker}}{{Cain} et~al.}{2021}]{Cain2021}
{Cain} C.,  {D'Aloisio} A.,  {Gangolli} N.,   {Becker} G.~D.,  2021, \mn@doi
  [\apjl] {10.3847/2041-8213/ac1ace}, \href
  {https://ui.adsabs.harvard.edu/abs/2021ApJ...917L..37C} {917, L37}

\bibitem[\protect\citeauthoryear{{Ciardi}, {Ferrara}, {Marri}  \&
  {Raimondo}}{{Ciardi} et~al.}{2001}]{ciardi2001}
{Ciardi} B.,  {Ferrara} A.,  {Marri} S.,   {Raimondo} G.,  2001, \mn@doi
  [\mnras] {10.1046/j.1365-8711.2001.04316.x}, \href
  {https://ui.adsabs.harvard.edu/abs/2001MNRAS.324..381C} {324, 381}

\bibitem[\protect\citeauthoryear{{Compostella}, {Cantalupo}  \&
  {Porciani}}{{Compostella} et~al.}{2013}]{compostella2013}
{Compostella} M.,  {Cantalupo} S.,   {Porciani} C.,  2013, \mn@doi [\mnras]
  {10.1093/mnras/stt1510}, \href
  {https://ui.adsabs.harvard.edu/abs/2013MNRAS.435.3169C} {435, 3169}

\bibitem[\protect\citeauthoryear{{Compostella}, {Cantalupo}  \&
  {Porciani}}{{Compostella} et~al.}{2014}]{compostella2014}
{Compostella} M.,  {Cantalupo} S.,   {Porciani} C.,  2014, \mn@doi [\mnras]
  {10.1093/mnras/stu2035}, \href
  {https://ui.adsabs.harvard.edu/abs/2014MNRAS.445.4186C} {445, 4186}

\bibitem[\protect\citeauthoryear{{Cristiani}, {Serrano}, {Fontanot}, {Vanzella}
   \& {Monaco}}{{Cristiani} et~al.}{2016}]{Christiani2016}
{Cristiani} S.,  {Serrano} L.~M.,  {Fontanot} F.,  {Vanzella} E.,   {Monaco}
  P.,  2016, \mn@doi [\mnras] {10.1093/mnras/stw1810}, \href
  {https://ui.adsabs.harvard.edu/abs/2016MNRAS.462.2478C} {462, 2478}

\bibitem[\protect\citeauthoryear{{Croft}, {Weinberg}, {Katz}  \&
  {Hernquist}}{{Croft} et~al.}{1997}]{Croft1997}
{Croft} R. A.~C.,  {Weinberg} D.~H.,  {Katz} N.,   {Hernquist} L.,  1997,
  \mn@doi [\apj] {10.1086/304723}, \href
  {https://ui.adsabs.harvard.edu/abs/1997ApJ...488..532C} {488, 532}

\bibitem[\protect\citeauthoryear{{Davies}, {Furlanetto}  \& {Dixon}}{{Davies}
  et~al.}{2017}]{davies2017}
{Davies} F.~B.,  {Furlanetto} S.~R.,   {Dixon} K.~L.,  2017, \mn@doi [\mnras]
  {10.1093/mnras/stw2868}, \href
  {https://ui.adsabs.harvard.edu/abs/2017MNRAS.465.2886D} {465, 2886}

\bibitem[\protect\citeauthoryear{{Dixon} \& {Furlanetto}}{{Dixon} \&
  {Furlanetto}}{2009}]{Dixon2009}
{Dixon} K.~L.,  {Furlanetto} S.~R.,  2009, \mn@doi [\apj]
  {10.1088/0004-637X/706/2/970}, \href
  {https://ui.adsabs.harvard.edu/abs/2009ApJ...706..970D} {706, 970}

\bibitem[\protect\citeauthoryear{{Eftekharzadeh} et~al.,}{{Eftekharzadeh}
  et~al.}{2015}]{Eftekharzadeh2015}
{Eftekharzadeh} S.,  et~al., 2015, \mn@doi [\mnras] {10.1093/mnras/stv1763},
  \href {https://ui.adsabs.harvard.edu/abs/2015MNRAS.453.2779E} {453, 2779}

\bibitem[\protect\citeauthoryear{Eide, Graziani, Ciardi, Feng, Kakiichi  \&
  Di~Matteo}{Eide et~al.}{2018}]{Marius2018}
Eide M.~B.,  Graziani L.,  Ciardi B.,  Feng Y.,  Kakiichi K.,   Di~Matteo T.,
  2018, \mn@doi [Monthly Notices of the Royal Astronomical Society]
  {10.1093/mnras/sty272}, 476, 1174–1190

\bibitem[\protect\citeauthoryear{{Eide}, {Ciardi}, {Graziani}, {Busch}, {Feng}
  \& {Di Matteo}}{{Eide} et~al.}{2020}]{Marius2020}
{Eide} M.~B.,  {Ciardi} B.,  {Graziani} L.,  {Busch} P.,  {Feng} Y.,   {Di
  Matteo} T.,  2020, \mn@doi [\mnras] {10.1093/mnras/staa2774}, \href
  {https://ui.adsabs.harvard.edu/abs/2020MNRAS.498.6083E} {498, 6083}

\bibitem[\protect\citeauthoryear{{Fan} et~al.,}{{Fan} et~al.}{2006}]{fan2006}
{Fan} X.,  et~al., 2006, \mn@doi [\aj] {10.1086/504836}, \href
  {https://ui.adsabs.harvard.edu/abs/2006AJ....132..117F} {132, 117}

\bibitem[\protect\citeauthoryear{{Fardal}, {Giroux}  \& {Shull}}{{Fardal}
  et~al.}{1998}]{Fardal1998}
{Fardal} M.~A.,  {Giroux} M.~L.,   {Shull} J.~M.,  1998, \mn@doi [\aj]
  {10.1086/300359}, \href
  {https://ui.adsabs.harvard.edu/abs/1998AJ....115.2206F} {115, 2206}

\bibitem[\protect\citeauthoryear{{Fudamoto}, {Inoue}  \& {Sugahara}}{{Fudamoto}
  et~al.}{2022}]{Fudamoto2022}
{Fudamoto} Y.,  {Inoue} A.~K.,   {Sugahara} Y.,  2022, \mn@doi [\apjl]
  {10.3847/2041-8213/ac982b}, \href
  {https://ui.adsabs.harvard.edu/abs/2022ApJ...938L..24F} {938, L24}

\bibitem[\protect\citeauthoryear{Furlanetto \& Dixon}{Furlanetto \&
  Dixon}{2010}]{Furlanetto2011}
Furlanetto S.~R.,  Dixon K.~L.,  2010, \mn@doi [The Astrophysical Journal]
  {10.1088/0004-637X/714/1/355}, 714, 355

\bibitem[\protect\citeauthoryear{{Furlanetto} \& {Lidz}}{{Furlanetto} \&
  {Lidz}}{2011}]{Furnaletto&Lidz2011}
{Furlanetto} S.~R.,  {Lidz} A.,  2011, \mn@doi [\apj]
  {10.1088/0004-637X/735/2/117}, \href
  {https://ui.adsabs.harvard.edu/abs/2011ApJ...735..117F} {735, 117}

\bibitem[\protect\citeauthoryear{{Gaikwad} et~al.,}{{Gaikwad}
  et~al.}{2020}]{Gaikwad2020}
{Gaikwad} P.,  et~al., 2020, \mn@doi [\mnras] {10.1093/mnras/staa907}, \href
  {https://ui.adsabs.harvard.edu/abs/2020MNRAS.494.5091G} {494, 5091}

\bibitem[\protect\citeauthoryear{{Gaikwad}, {Srianand}, {Haehnelt}  \&
  {Choudhury}}{{Gaikwad} et~al.}{2021}]{gaikwad2021}
{Gaikwad} P.,  {Srianand} R.,  {Haehnelt} M.~G.,   {Choudhury} T.~R.,  2021,
  \mn@doi [\mnras] {10.1093/mnras/stab2017}, \href
  {https://ui.adsabs.harvard.edu/abs/2021MNRAS.506.4389G} {506, 4389}

\bibitem[\protect\citeauthoryear{{Gallerani}, {Choudhury}  \&
  {Ferrara}}{{Gallerani} et~al.}{2006}]{gallerani2006}
{Gallerani} S.,  {Choudhury} T.~R.,   {Ferrara} A.,  2006, \mn@doi [\mnras]
  {10.1111/j.1365-2966.2006.10553.x}, \href
  {https://ui.adsabs.harvard.edu/abs/2006MNRAS.370.1401G} {370, 1401}

\bibitem[\protect\citeauthoryear{{Gallerani}, {Ferrara}, {Fan}  \&
  {Choudhury}}{{Gallerani} et~al.}{2008}]{gallerani2008}
{Gallerani} S.,  {Ferrara} A.,  {Fan} X.,   {Choudhury} T.~R.,  2008, \mn@doi
  [\mnras] {10.1111/j.1365-2966.2008.13029.x}, \href
  {https://ui.adsabs.harvard.edu/abs/2008MNRAS.386..359G} {386, 359}

\bibitem[\protect\citeauthoryear{{Garaldi}}{{Garaldi}}{2023}]{Garaldi2023}
{Garaldi} E.,  2023, \mn@doi [The Journal of Open Source Software]
  {10.21105/joss.05407}, \href
  {https://ui.adsabs.harvard.edu/abs/2023JOSS....8.5407G} {8, 5407}

\bibitem[\protect\citeauthoryear{{Garaldi}, {Compostella}  \&
  {Porciani}}{{Garaldi} et~al.}{2019a}]{garaldi2019}
{Garaldi} E.,  {Compostella} M.,   {Porciani} C.,  2019a, \mn@doi [\mnras]
  {10.1093/mnras/sty3414}, \href
  {https://ui.adsabs.harvard.edu/abs/2019MNRAS.483.5301G} {483, 5301}

\bibitem[\protect\citeauthoryear{{Garaldi}, {Gnedin}  \& {Madau}}{{Garaldi}
  et~al.}{2019b}]{Garaldi2019b}
{Garaldi} E.,  {Gnedin} N.~Y.,   {Madau} P.,  2019b, \mn@doi [\apj]
  {10.3847/1538-4357/ab12dc}, \href
  {https://ui.adsabs.harvard.edu/abs/2019ApJ...876...31G} {876, 31}

\bibitem[\protect\citeauthoryear{{Garaldi}, {Kannan}, {Smith}, {Springel},
  {Pakmor}, {Vogelsberger}  \& {Hernquist}}{{Garaldi}
  et~al.}{2022}]{garaldi2022}
{Garaldi} E.,  {Kannan} R.,  {Smith} A.,  {Springel} V.,  {Pakmor} R.,
  {Vogelsberger} M.,   {Hernquist} L.,  2022, \mn@doi [\mnras]
  {10.1093/mnras/stac257}, \href
  {https://ui.adsabs.harvard.edu/abs/2022MNRAS.512.4909G} {512, 4909}

\bibitem[\protect\citeauthoryear{{Garzilli}, {Bolton}, {Kim}, {Leach}  \&
  {Viel}}{{Garzilli} et~al.}{2012}]{Garzili2012}
{Garzilli} A.,  {Bolton} J.~S.,  {Kim} T.~S.,  {Leach} S.,   {Viel} M.,  2012,
  \mn@doi [\mnras] {10.1111/j.1365-2966.2012.21223.x}, \href
  {https://ui.adsabs.harvard.edu/abs/2012MNRAS.424.1723G} {424, 1723}

\bibitem[\protect\citeauthoryear{{Giroux}, {Fardal}  \& {Shull}}{{Giroux}
  et~al.}{1995}]{Giroux1995}
{Giroux} M.~L.,  {Fardal} M.~A.,   {Shull} J.~M.,  1995, \mn@doi [\apj]
  {10.1086/176236}, \href
  {https://ui.adsabs.harvard.edu/abs/1995ApJ...451..477G} {451, 477}

\bibitem[\protect\citeauthoryear{{Glatzle}, {Ciardi}  \& {Graziani}}{{Glatzle}
  et~al.}{2019}]{glatzle2019}
{Glatzle} M.,  {Ciardi} B.,   {Graziani} L.,  2019, \mn@doi [\mnras]
  {10.1093/mnras/sty2514}, \href
  {https://ui.adsabs.harvard.edu/abs/2019MNRAS.482..321G} {482, 321}

\bibitem[\protect\citeauthoryear{{Glatzle}, {Graziani}  \& {Ciardi}}{{Glatzle}
  et~al.}{2022}]{glatzle2022}
{Glatzle} M.,  {Graziani} L.,   {Ciardi} B.,  2022, \mn@doi [\mnras]
  {10.1093/mnras/stab3459}, \href
  {https://ui.adsabs.harvard.edu/abs/2022MNRAS.510.1068G} {510, 1068}

\bibitem[\protect\citeauthoryear{{Gleser}, {Nusser}, {Benson}, {Ohno}  \&
  {Sugiyama}}{{Gleser} et~al.}{2005}]{Gleser2005}
{Gleser} L.,  {Nusser} A.,  {Benson} A.~J.,  {Ohno} H.,   {Sugiyama} N.,  2005,
  \mn@doi [\mnras] {10.1111/j.1365-2966.2005.09276.x}, \href
  {https://ui.adsabs.harvard.edu/abs/2005MNRAS.361.1399G} {361, 1399}

\bibitem[\protect\citeauthoryear{{Gnedin}, {Becker}  \& {Fan}}{{Gnedin}
  et~al.}{2017}]{Gnedin2017}
{Gnedin} N.~Y.,  {Becker} G.~D.,   {Fan} X.,  2017, \mn@doi [\apj]
  {10.3847/1538-4357/aa6c24}, \href
  {https://ui.adsabs.harvard.edu/abs/2017ApJ...841...26G} {841, 26}

\bibitem[\protect\citeauthoryear{{Goulding} et~al.,}{{Goulding}
  et~al.}{2023}]{Goulding2023}
{Goulding} A.~D.,  et~al., 2023, \mn@doi [\apjl] {10.3847/2041-8213/acf7c5},
  \href {https://ui.adsabs.harvard.edu/abs/2023ApJ...955L..24G} {955, L24}

\bibitem[\protect\citeauthoryear{{Graziani}, {Maselli}  \& {Ciardi}}{{Graziani}
  et~al.}{2013}]{graziani2013}
{Graziani} L.,  {Maselli} A.,   {Ciardi} B.,  2013, \mn@doi [\mnras]
  {10.1093/mnras/stt206}, \href
  {https://ui.adsabs.harvard.edu/abs/2013MNRAS.431..722G} {431, 722}

\bibitem[\protect\citeauthoryear{{Graziani}, {Ciardi}  \& {Glatzle}}{{Graziani}
  et~al.}{2018}]{graziani2018}
{Graziani} L.,  {Ciardi} B.,   {Glatzle} M.,  2018, \mn@doi [\mnras]
  {10.1093/mnras/sty1367}, \href
  {https://ui.adsabs.harvard.edu/abs/2018MNRAS.479.4320G} {479, 4320}

\bibitem[\protect\citeauthoryear{{Greene} et~al.,}{{Greene}
  et~al.}{2023}]{Greene2023}
{Greene} J.~E.,  et~al., 2023, \mn@doi [arXiv e-prints]
  {10.48550/arXiv.2309.05714}, \href
  {https://ui.adsabs.harvard.edu/abs/2023arXiv230905714G} {p. arXiv:2309.05714}

\bibitem[\protect\citeauthoryear{{Greiner}, {Bolmer}, {Yates}, {Habouzit},
  {Ba{\~n}ados}, {Afonso}  \& {Schady}}{{Greiner} et~al.}{2021}]{Greiner2021}
{Greiner} J.,  {Bolmer} J.,  {Yates} R.~M.,  {Habouzit} M.,  {Ba{\~n}ados} E.,
  {Afonso} P.~M.~J.,   {Schady} P.,  2021, \mn@doi [\aap]
  {10.1051/0004-6361/202140790}, \href
  {https://ui.adsabs.harvard.edu/abs/2021A&A...654A..79G} {654, A79}

\bibitem[\protect\citeauthoryear{{Harikane} et~al.,}{{Harikane}
  et~al.}{2023}]{Harikane2023}
{Harikane} Y.,  et~al., 2023, \mn@doi [\apj] {10.3847/1538-4357/ad029e}, \href
  {https://ui.adsabs.harvard.edu/abs/2023ApJ...959...39H} {959, 39}

\bibitem[\protect\citeauthoryear{{Heap}, {Williger}, {Smette}, {Hubeny},
  {Sahu}, {Jenkins}, {Tripp}  \& {Winkler}}{{Heap} et~al.}{2000}]{Heap2000}
{Heap} S.~R.,  {Williger} G.~M.,  {Smette} A.,  {Hubeny} I.,  {Sahu} M.~S.,
  {Jenkins} E.~B.,  {Tripp} T.~M.,   {Winkler} J.~N.,  2000, \mn@doi [\apj]
  {10.1086/308719}, \href
  {https://ui.adsabs.harvard.edu/abs/2000ApJ...534...69H} {534, 69}

\bibitem[\protect\citeauthoryear{{Hiss}, {Walther}, {Hennawi}, {O{\~n}orbe},
  {O'Meara}, {Rorai}  \& {Luki{\'c}}}{{Hiss} et~al.}{2018}]{hiss2018}
{Hiss} H.,  {Walther} M.,  {Hennawi} J.~F.,  {O{\~n}orbe} J.,  {O'Meara} J.~M.,
   {Rorai} A.,   {Luki{\'c}} Z.,  2018, \mn@doi [\apj]
  {10.3847/1538-4357/aada86}, \href
  {https://ui.adsabs.harvard.edu/abs/2018ApJ...865...42H} {865, 42}

\bibitem[\protect\citeauthoryear{{Hui} \& {Gnedin}}{{Hui} \&
  {Gnedin}}{1997}]{hui&gnedin1997}
{Hui} L.,  {Gnedin} N.~Y.,  1997, \mn@doi [\mnras] {10.1093/mnras/292.1.27},
  \href {https://ui.adsabs.harvard.edu/abs/1997MNRAS.292...27H} {292, 27}

\bibitem[\protect\citeauthoryear{{Hunter}}{{Hunter}}{2007}]{Hunter2007}
{Hunter} J.~D.,  2007, \mn@doi [Computing in Science and Engineering]
  {10.1109/MCSE.2007.55}, \href
  {https://ui.adsabs.harvard.edu/abs/2007CSE.....9...90H} {9, 90}

\bibitem[\protect\citeauthoryear{Jones, Oliphant  \& Peterson}{Jones
  et~al.}{2001}]{Jones2001}
Jones E.,  Oliphant T.,   Peterson P.,  2001

\bibitem[\protect\citeauthoryear{{Juod{\v{z}}balis} et~al.,}{{Juod{\v{z}}balis}
  et~al.}{2023}]{Juodzbalis2023}
{Juod{\v{z}}balis} I.,  et~al., 2023, \mn@doi [\mnras]
  {10.1093/mnras/stad2396}, \href
  {https://ui.adsabs.harvard.edu/abs/2023MNRAS.525.1353J} {525, 1353}

\bibitem[\protect\citeauthoryear{{Kakiichi}, {Graziani}, {Ciardi}, {Meiksin},
  {Compostella}, {Eide}  \& {Zaroubi}}{{Kakiichi} et~al.}{2017}]{koki2017}
{Kakiichi} K.,  {Graziani} L.,  {Ciardi} B.,  {Meiksin} A.,  {Compostella} M.,
  {Eide} M.~B.,   {Zaroubi} S.,  2017, \mn@doi [\mnras] {10.1093/mnras/stx603},
  \href {https://ui.adsabs.harvard.edu/abs/2017MNRAS.468.3718K} {468, 3718}

\bibitem[\protect\citeauthoryear{{Kapahtia} \& {Choudhury}}{{Kapahtia} \&
  {Choudhury}}{2024}]{Kapahtia2024}
{Kapahtia} A.,  {Choudhury} T.~R.,  2024, \mn@doi [arXiv e-prints]
  {10.48550/arXiv.2402.03794}, \href
  {https://ui.adsabs.harvard.edu/abs/2024arXiv240203794K} {p. arXiv:2402.03794}

\bibitem[\protect\citeauthoryear{{Khrykin}, {Hennawi}, {McQuinn}  \&
  {Worseck}}{{Khrykin} et~al.}{2016}]{Khrykin2016}
{Khrykin} I.~S.,  {Hennawi} J.~F.,  {McQuinn} M.,   {Worseck} G.,  2016,
  \mn@doi [\apj] {10.3847/0004-637X/824/2/133}, \href
  {https://ui.adsabs.harvard.edu/abs/2016ApJ...824..133K} {824, 133}

\bibitem[\protect\citeauthoryear{{Khrykin}, {Hennawi}  \& {McQuinn}}{{Khrykin}
  et~al.}{2017}]{Khrykin2017}
{Khrykin} I.~S.,  {Hennawi} J.~F.,   {McQuinn} M.,  2017, \mn@doi [\apj]
  {10.3847/1538-4357/aa6621}, \href
  {https://ui.adsabs.harvard.edu/abs/2017ApJ...838...96K} {838, 96}

\bibitem[\protect\citeauthoryear{{Khrykin}, {Hennawi}  \& {Worseck}}{{Khrykin}
  et~al.}{2019}]{Khrykin2019}
{Khrykin} I.~S.,  {Hennawi} J.~F.,   {Worseck} G.,  2019, \mn@doi [\mnras]
  {10.1093/mnras/stz135}, \href
  {https://ui.adsabs.harvard.edu/abs/2019MNRAS.484.3897K} {484, 3897}

\bibitem[\protect\citeauthoryear{{Khrykin}, {Hennawi}, {Worseck}  \&
  {Davies}}{{Khrykin} et~al.}{2021}]{Khrykin2021}
{Khrykin} I.~S.,  {Hennawi} J.~F.,  {Worseck} G.,   {Davies} F.~B.,  2021,
  \mn@doi [\mnras] {10.1093/mnras/stab1288}, \href
  {https://ui.adsabs.harvard.edu/abs/2021MNRAS.505..649K} {505, 649}

\bibitem[\protect\citeauthoryear{{Krawczyk}, {Richards}, {Mehta}, {Vogeley},
  {Gallagher}, {Leighly}, {Ross}  \& {Schneider}}{{Krawczyk}
  et~al.}{2013}]{Krawczyk2013}
{Krawczyk} C.~M.,  {Richards} G.~T.,  {Mehta} S.~S.,  {Vogeley} M.~S.,
  {Gallagher} S.~C.,  {Leighly} K.~M.,  {Ross} N.~P.,   {Schneider} D.~P.,
  2013, \mn@doi [\apjs] {10.1088/0067-0049/206/1/4}, \href
  {https://ui.adsabs.harvard.edu/abs/2013ApJS..206....4K} {206, 4}

\bibitem[\protect\citeauthoryear{{Kulkarni}, {Worseck}  \&
  {Hennawi}}{{Kulkarni} et~al.}{2019}]{kulkarni2019}
{Kulkarni} G.,  {Worseck} G.,   {Hennawi} J.~F.,  2019, \mn@doi [\mnras]
  {10.1093/mnras/stz1493}, \href
  {https://ui.adsabs.harvard.edu/abs/2019MNRAS.488.1035K} {488, 1035}

\bibitem[\protect\citeauthoryear{{La Plante}, {Trac}, {Croft}  \& {Cen}}{{La
  Plante} et~al.}{2017}]{laplante2017}
{La Plante} P.,  {Trac} H.,  {Croft} R.,   {Cen} R.,  2017, \mn@doi [\apj]
  {10.3847/1538-4357/aa7136}, \href
  {https://ui.adsabs.harvard.edu/abs/2017ApJ...841...87L} {841, 87}

\bibitem[\protect\citeauthoryear{{Larson} et~al.,}{{Larson}
  et~al.}{2023}]{Larson2023}
{Larson} R.~L.,  et~al., 2023, \mn@doi [\apjl] {10.3847/2041-8213/ace619},
  \href {https://ui.adsabs.harvard.edu/abs/2023ApJ...953L..29L} {953, L29}

\bibitem[\protect\citeauthoryear{{Laurent} et~al.,}{{Laurent}
  et~al.}{2017}]{Laurent2017}
{Laurent} P.,  et~al., 2017, \mn@doi [\jcap] {10.1088/1475-7516/2017/07/017},
  \href {https://ui.adsabs.harvard.edu/abs/2017JCAP...07..017L} {2017, 017}

\bibitem[\protect\citeauthoryear{{Lidz}, {Faucher-Gigu{\`e}re}, {Dall'Aglio},
  {McQuinn}, {Fechner}, {Zaldarriaga}, {Hernquist}  \& {Dutta}}{{Lidz}
  et~al.}{2010}]{Lidz2010}
{Lidz} A.,  {Faucher-Gigu{\`e}re} C.-A.,  {Dall'Aglio} A.,  {McQuinn} M.,
  {Fechner} C.,  {Zaldarriaga} M.,  {Hernquist} L.,   {Dutta} S.,  2010,
  \mn@doi [\apj] {10.1088/0004-637X/718/1/199}, \href
  {https://ui.adsabs.harvard.edu/abs/2010ApJ...718..199L} {718, 199}

\bibitem[\protect\citeauthoryear{{Madau} \& {Haardt}}{{Madau} \&
  {Haardt}}{2015}]{madau2015}
{Madau} P.,  {Haardt} F.,  2015, \mn@doi [\apjl] {10.1088/2041-8205/813/1/L8},
  \href {https://ui.adsabs.harvard.edu/abs/2015ApJ...813L...8M} {813, L8}

\bibitem[\protect\citeauthoryear{{Maiolino} et~al.,}{{Maiolino}
  et~al.}{2023a}]{Maiolinoa2023}
{Maiolino} R.,  et~al., 2023a, \mn@doi [arXiv e-prints]
  {10.48550/arXiv.2305.12492}, \href
  {https://ui.adsabs.harvard.edu/abs/2023arXiv230512492M} {p. arXiv:2305.12492}

\bibitem[\protect\citeauthoryear{{Maiolino} et~al.,}{{Maiolino}
  et~al.}{2023b}]{Maiolino2023}
{Maiolino} R.,  et~al., 2023b, \mn@doi [arXiv e-prints]
  {10.48550/arXiv.2308.01230}, \href
  {https://ui.adsabs.harvard.edu/abs/2023arXiv230801230M} {p. arXiv:2308.01230}

\bibitem[\protect\citeauthoryear{{Makan}, {Worseck}, {Davies}, {Hennawi},
  {Prochaska}  \& {Richter}}{{Makan} et~al.}{2021}]{makan2021}
{Makan} K.,  {Worseck} G.,  {Davies} F.~B.,  {Hennawi} J.~F.,  {Prochaska}
  J.~X.,   {Richter} P.,  2021, \mn@doi [\apj] {10.3847/1538-4357/abee17},
  \href {https://ui.adsabs.harvard.edu/abs/2021ApJ...912...38M} {912, 38}

\bibitem[\protect\citeauthoryear{{Makan}, {Worseck}, {Davies}, {Hennawi},
  {Prochaska}  \& {Richter}}{{Makan} et~al.}{2022}]{makan2022}
{Makan} K.,  {Worseck} G.,  {Davies} F.~B.,  {Hennawi} J.~F.,  {Prochaska}
  J.~X.,   {Richter} P.,  2022, \mn@doi [\apj] {10.3847/1538-4357/ac524a},
  \href {https://ui.adsabs.harvard.edu/abs/2022ApJ...927..175M} {927, 175}

\bibitem[\protect\citeauthoryear{{Mao}, {Koda}, {Shapiro}, {Iliev}, {Mellema},
  {Park}, {Ahn}  \& {Bianco}}{{Mao} et~al.}{2020}]{Mao2020}
{Mao} Y.,  {Koda} J.,  {Shapiro} P.~R.,  {Iliev} I.~T.,  {Mellema} G.,  {Park}
  H.,  {Ahn} K.,   {Bianco} M.,  2020, \mn@doi [\mnras]
  {10.1093/mnras/stz2986}, \href
  {https://ui.adsabs.harvard.edu/abs/2020MNRAS.491.1600M} {491, 1600}

\bibitem[\protect\citeauthoryear{{Marinacci} et~al.,}{{Marinacci}
  et~al.}{2018}]{marinacci2018}
{Marinacci} F.,  et~al., 2018, \mn@doi [\mnras] {10.1093/mnras/sty2206}, \href
  {https://ui.adsabs.harvard.edu/abs/2018MNRAS.480.5113M} {480, 5113}

\bibitem[\protect\citeauthoryear{{Maselli} \& {Ferrara}}{{Maselli} \&
  {Ferrara}}{2005}]{Maselli2005}
{Maselli} A.,  {Ferrara} A.,  2005, \mn@doi [\mnras]
  {10.1111/j.1365-2966.2005.09682.x}, \href
  {https://ui.adsabs.harvard.edu/abs/2005MNRAS.364.1429M} {364, 1429}

\bibitem[\protect\citeauthoryear{{Maselli}, {Ferrara}  \& {Ciardi}}{{Maselli}
  et~al.}{2003}]{maselli2003}
{Maselli} A.,  {Ferrara} A.,   {Ciardi} B.,  2003, \mn@doi [\mnras]
  {10.1046/j.1365-8711.2003.06979.x}, \href
  {https://ui.adsabs.harvard.edu/abs/2003MNRAS.345..379M} {345, 379}

\bibitem[\protect\citeauthoryear{{Maselli}, {Ciardi}  \& {Kanekar}}{{Maselli}
  et~al.}{2009}]{maselli2009}
{Maselli} A.,  {Ciardi} B.,   {Kanekar} A.,  2009, \mn@doi [\mnras]
  {10.1111/j.1365-2966.2008.14197.x}, \href
  {https://ui.adsabs.harvard.edu/abs/2009MNRAS.393..171M} {393, 171}

\bibitem[\protect\citeauthoryear{{Masters} et~al.,}{{Masters}
  et~al.}{2012}]{Masters2012}
{Masters} D.,  et~al., 2012, \mn@doi [\apj] {10.1088/0004-637X/755/2/169},
  \href {https://ui.adsabs.harvard.edu/abs/2012ApJ...755..169M} {755, 169}

\bibitem[\protect\citeauthoryear{{McGreer}, {Mesinger}  \& {Fan}}{{McGreer}
  et~al.}{2011}]{Mcgreer2011}
{McGreer} I.~D.,  {Mesinger} A.,   {Fan} X.,  2011, \mn@doi [\mnras]
  {10.1111/j.1365-2966.2011.18935.x}, \href
  {https://ui.adsabs.harvard.edu/abs/2011MNRAS.415.3237M} {415, 3237}

\bibitem[\protect\citeauthoryear{{McGreer} et~al.,}{{McGreer}
  et~al.}{2013}]{Mcgreer2013}
{McGreer} I.~D.,  et~al., 2013, \mn@doi [\apj] {10.1088/0004-637X/768/2/105},
  \href {https://ui.adsabs.harvard.edu/abs/2013ApJ...768..105M} {768, 105}

\bibitem[\protect\citeauthoryear{{McGreer}, {Mesinger}  \&
  {D'Odorico}}{{McGreer} et~al.}{2015}]{Mcgreer2015}
{McGreer} I.~D.,  {Mesinger} A.,   {D'Odorico} V.,  2015, \mn@doi [\mnras]
  {10.1093/mnras/stu2449}, \href
  {https://ui.adsabs.harvard.edu/abs/2015MNRAS.447..499M} {447, 499}

\bibitem[\protect\citeauthoryear{{McQuinn}, {Lidz}, {Zaldarriaga}, {Hernquist},
  {Hopkins}, {Dutta}  \& {Faucher-Gigu{\`e}re}}{{McQuinn}
  et~al.}{2009}]{McQuinn2009}
{McQuinn} M.,  {Lidz} A.,  {Zaldarriaga} M.,  {Hernquist} L.,  {Hopkins} P.~F.,
   {Dutta} S.,   {Faucher-Gigu{\`e}re} C.-A.,  2009, \mn@doi [\apj]
  {10.1088/0004-637X/694/2/842}, \href
  {https://ui.adsabs.harvard.edu/abs/2009ApJ...694..842M} {694, 842}

\bibitem[\protect\citeauthoryear{{Meiksin} \& {Tittley}}{{Meiksin} \&
  {Tittley}}{2012}]{Meiksin2012}
{Meiksin} A.,  {Tittley} E.~R.,  2012, \mn@doi [\mnras]
  {10.1111/j.1365-2966.2011.20380.x}, \href
  {https://ui.adsabs.harvard.edu/abs/2012MNRAS.423....7M} {423, 7}

\bibitem[\protect\citeauthoryear{{Mesinger}}{{Mesinger}}{2010}]{Mesinger2010}
{Mesinger} A.,  2010, \mn@doi [\mnras] {10.1111/j.1365-2966.2010.16995.x},
  \href {https://ui.adsabs.harvard.edu/abs/2010MNRAS.407.1328M} {407, 1328}

\bibitem[\protect\citeauthoryear{{Miralda-Escude}}{{Miralda-Escude}}{1993}]{Miralda1993}
{Miralda-Escude} J.,  1993, \mn@doi [\mnras] {10.1093/mnras/262.1.273}, \href
  {https://ui.adsabs.harvard.edu/abs/1993MNRAS.262..273M} {262, 273}

\bibitem[\protect\citeauthoryear{{Morey}, {Eilers}, {Davies}, {Hennawi}  \&
  {Simcoe}}{{Morey} et~al.}{2021}]{Morey2021}
{Morey} K.~A.,  {Eilers} A.-C.,  {Davies} F.~B.,  {Hennawi} J.~F.,   {Simcoe}
  R.~A.,  2021, \mn@doi [\apj] {10.3847/1538-4357/ac1c70}, \href
  {https://ui.adsabs.harvard.edu/abs/2021ApJ...921...88M} {921, 88}

\bibitem[\protect\citeauthoryear{{Naiman} et~al.,}{{Naiman}
  et~al.}{2018}]{naiman2018}
{Naiman} J.~P.,  et~al., 2018, \mn@doi [\mnras] {10.1093/mnras/sty618}, \href
  {https://ui.adsabs.harvard.edu/abs/2018MNRAS.477.1206N} {477, 1206}

\bibitem[\protect\citeauthoryear{{Nelson} et~al.,}{{Nelson}
  et~al.}{2018}]{nelson2018}
{Nelson} D.,  et~al., 2018, \mn@doi [\mnras] {10.1093/mnras/stx3040}, \href
  {https://ui.adsabs.harvard.edu/abs/2018MNRAS.475..624N} {475, 624}

\bibitem[\protect\citeauthoryear{{Oogi}, {Enoki}, {Ishiyama}, {Kobayashi},
  {Makiya}  \& {Nagashima}}{{Oogi} et~al.}{2016}]{Oogi2016}
{Oogi} T.,  {Enoki} M.,  {Ishiyama} T.,  {Kobayashi} M. A.~R.,  {Makiya} R.,
  {Nagashima} M.,  2016, \mn@doi [\mnras] {10.1093/mnrasl/slv169}, \href
  {https://ui.adsabs.harvard.edu/abs/2016MNRAS.456L..30O} {456, L30}

\bibitem[\protect\citeauthoryear{{Pakmor}, {Bauer}  \& {Springel}}{{Pakmor}
  et~al.}{2011}]{pakmor2011}
{Pakmor} R.,  {Bauer} A.,   {Springel} V.,  2011, \mn@doi [\mnras]
  {10.1111/j.1365-2966.2011.19591.x}, \href
  {https://ui.adsabs.harvard.edu/abs/2011MNRAS.418.1392P} {418, 1392}

\bibitem[\protect\citeauthoryear{{Palanque-Delabrouille}
  et~al.,}{{Palanque-Delabrouille} et~al.}{2013}]{Palanque2013}
{Palanque-Delabrouille} N.,  et~al., 2013, \mn@doi [\aap]
  {10.1051/0004-6361/201220379}, \href
  {https://ui.adsabs.harvard.edu/abs/2013A&A...551A..29P} {551, A29}

\bibitem[\protect\citeauthoryear{{Pan}, {Jiang}, {Fan}, {Wu}  \& {Yang}}{{Pan}
  et~al.}{2022}]{Pan2022}
{Pan} Z.,  {Jiang} L.,  {Fan} X.,  {Wu} J.,   {Yang} J.,  2022, \mn@doi [\apj]
  {10.3847/1538-4357/ac5aab}, \href
  {https://ui.adsabs.harvard.edu/abs/2022ApJ...928..172P} {928, 172}

\bibitem[\protect\citeauthoryear{{Partl}, {Maselli}, {Ciardi}, {Ferrara}  \&
  {M{\"u}ller}}{{Partl} et~al.}{2011}]{partl2011}
{Partl} A.~M.,  {Maselli} A.,  {Ciardi} B.,  {Ferrara} A.,   {M{\"u}ller} V.,
  2011, \mn@doi [\mnras] {10.1111/j.1365-2966.2011.18401.x}, \href
  {https://ui.adsabs.harvard.edu/abs/2011MNRAS.414..428P} {414, 428}

\bibitem[\protect\citeauthoryear{{Paschos} \& {Norman}}{{Paschos} \&
  {Norman}}{2005}]{paschos2005}
{Paschos} P.,  {Norman} M.~L.,  2005, \mn@doi [\apj] {10.1086/431787}, \href
  {https://ui.adsabs.harvard.edu/abs/2005ApJ...631...59P} {631, 59}

\bibitem[\protect\citeauthoryear{{Pillepich} et~al.,}{{Pillepich}
  et~al.}{2018}]{pillepich2018}
{Pillepich} A.,  et~al., 2018, \mn@doi [\mnras] {10.1093/mnras/stx2656}, \href
  {https://ui.adsabs.harvard.edu/abs/2018MNRAS.473.4077P} {473, 4077}

\bibitem[\protect\citeauthoryear{{Planck Collaboration} et~al.,}{{Planck
  Collaboration} et~al.}{2016}]{planck2016}
{Planck Collaboration} et~al., 2016, \mn@doi [\aap]
  {10.1051/0004-6361/201527101}, \href
  {https://ui.adsabs.harvard.edu/abs/2016A&A...594A...1P} {594, A1}

\bibitem[\protect\citeauthoryear{{Porciani}, {Magliocchetti}  \&
  {Norberg}}{{Porciani} et~al.}{2004}]{Porciani2004}
{Porciani} C.,  {Magliocchetti} M.,   {Norberg} P.,  2004, \mn@doi [\mnras]
  {10.1111/j.1365-2966.2004.08408.x}, \href
  {https://ui.adsabs.harvard.edu/abs/2004MNRAS.355.1010P} {355, 1010}

\bibitem[\protect\citeauthoryear{{Puchwein}, {Haardt}, {Haehnelt}  \&
  {Madau}}{{Puchwein} et~al.}{2019}]{puchwein2019}
{Puchwein} E.,  {Haardt} F.,  {Haehnelt} M.~G.,   {Madau} P.,  2019, \mn@doi
  [\mnras] {10.1093/mnras/stz222}, \href
  {https://ui.adsabs.harvard.edu/abs/2019MNRAS.485...47P} {485, 47}

\bibitem[\protect\citeauthoryear{{Reed} et~al.,}{{Reed}
  et~al.}{2015}]{Reed2015}
{Reed} S.~L.,  et~al., 2015, \mn@doi [\mnras] {10.1093/mnras/stv2031}, \href
  {https://ui.adsabs.harvard.edu/abs/2015MNRAS.454.3952R} {454, 3952}

\bibitem[\protect\citeauthoryear{{Reimers}, {Fechner}, {Hagen}, {Jakobsen},
  {Tytler}  \& {Kirkman}}{{Reimers} et~al.}{2005}]{Reimers2005}
{Reimers} D.,  {Fechner} C.,  {Hagen} H.~J.,  {Jakobsen} P.,  {Tytler} D.,
  {Kirkman} D.,  2005, \mn@doi [\aap] {10.1051/0004-6361:20053365}, \href
  {https://ui.adsabs.harvard.edu/abs/2005A&A...442...63R} {442, 63}

\bibitem[\protect\citeauthoryear{{Richards} et~al.,}{{Richards}
  et~al.}{2005}]{Richards2005}
{Richards} G.~T.,  et~al., 2005, \mn@doi [\mnras]
  {10.1111/j.1365-2966.2005.09096.x}, \href
  {https://ui.adsabs.harvard.edu/abs/2005MNRAS.360..839R} {360, 839}

\bibitem[\protect\citeauthoryear{{Richards} et~al.,}{{Richards}
  et~al.}{2006}]{Richards2006}
{Richards} G.~T.,  et~al., 2006, \mn@doi [\aj] {10.1086/503559}, \href
  {https://ui.adsabs.harvard.edu/abs/2006AJ....131.2766R} {131, 2766}

\bibitem[\protect\citeauthoryear{{Ricotti}, {Gnedin}  \& {Shull}}{{Ricotti}
  et~al.}{2000}]{Riccoti2000}
{Ricotti} M.,  {Gnedin} N.~Y.,   {Shull} J.~M.,  2000, \mn@doi [\apj]
  {10.1086/308733}, \href
  {https://ui.adsabs.harvard.edu/abs/2000ApJ...534...41R} {534, 41}

\bibitem[\protect\citeauthoryear{{Rodr{\'\i}guez-Torres}
  et~al.,}{{Rodr{\'\i}guez-Torres} et~al.}{2017}]{rodrigues2017}
{Rodr{\'\i}guez-Torres} S.~A.,  et~al., 2017, \mn@doi [\mnras]
  {10.1093/mnras/stx454}, \href
  {https://ui.adsabs.harvard.edu/abs/2017MNRAS.468..728R} {468, 728}

\bibitem[\protect\citeauthoryear{{Rorai}, {Carswell}, {Haehnelt}, {Becker},
  {Bolton}  \& {Murphy}}{{Rorai} et~al.}{2018}]{rorai2018}
{Rorai} A.,  {Carswell} R.~F.,  {Haehnelt} M.~G.,  {Becker} G.~D.,  {Bolton}
  J.~S.,   {Murphy} M.~T.,  2018, \mn@doi [\mnras] {10.1093/mnras/stx2862},
  \href {https://ui.adsabs.harvard.edu/abs/2018MNRAS.474.2871R} {474, 2871}

\bibitem[\protect\citeauthoryear{{Ross} et~al.,}{{Ross}
  et~al.}{2013}]{Ross2013}
{Ross} N.~P.,  et~al., 2013, \mn@doi [\apj] {10.1088/0004-637X/773/1/14}, \href
  {https://ui.adsabs.harvard.edu/abs/2013ApJ...773...14R} {773, 14}

\bibitem[\protect\citeauthoryear{{Schaye}, {Theuns}, {Leonard}  \&
  {Efstathiou}}{{Schaye} et~al.}{2000a}]{Schaye1999}
{Schaye} J.,  {Theuns} T.,  {Leonard} A.,   {Efstathiou} G.,  2000a, in
  {Hammer} F.,  {Thuan} T.~X.,  {Cayatte} V.,  {Guiderdoni} B.,   {Thanh Van}
  J.~T.,  eds, Building Galaxies; from the Primordial Universe to the Present.
  p.~455 (\mn@eprint {arXiv} {astro-ph/9905364}),
  \mn@doi{10.48550/arXiv.astro-ph/9905364}

\bibitem[\protect\citeauthoryear{{Schaye}, {Theuns}, {Rauch}, {Efstathiou}  \&
  {Sargent}}{{Schaye} et~al.}{2000b}]{Schaye2000}
{Schaye} J.,  {Theuns} T.,  {Rauch} M.,  {Efstathiou} G.,   {Sargent} W. L.~W.,
   2000b, \mn@doi [\mnras] {10.1046/j.1365-8711.2000.03815.x}, \href
  {https://ui.adsabs.harvard.edu/abs/2000MNRAS.318..817S} {318, 817}

\bibitem[\protect\citeauthoryear{{Schmidt} \& {Green}}{{Schmidt} \&
  {Green}}{1983}]{Schmidt1983}
{Schmidt} M.,  {Green} R.~F.,  1983, \mn@doi [\apj] {10.1086/161048}, \href
  {https://ui.adsabs.harvard.edu/abs/1983ApJ...269..352S} {269, 352}

\bibitem[\protect\citeauthoryear{{Schneider} et~al.,}{{Schneider}
  et~al.}{2010}]{Schneider2010}
{Schneider} D.~P.,  et~al., 2010, \mn@doi [\aj] {10.1088/0004-6256/139/6/2360},
  \href {https://ui.adsabs.harvard.edu/abs/2010AJ....139.2360S} {139, 2360}

\bibitem[\protect\citeauthoryear{{Shen} et~al.,}{{Shen}
  et~al.}{2009}]{Shen2009}
{Shen} Y.,  et~al., 2009, \mn@doi [\apj] {10.1088/0004-637X/697/2/1656}, \href
  {https://ui.adsabs.harvard.edu/abs/2009ApJ...697.1656S} {697, 1656}

\bibitem[\protect\citeauthoryear{{Shen}, {Hopkins}, {Faucher-Gigu{\`e}re},
  {Alexander}, {Richards}, {Ross}  \& {Hickox}}{{Shen} et~al.}{2020}]{Shen2020}
{Shen} X.,  {Hopkins} P.~F.,  {Faucher-Gigu{\`e}re} C.-A.,  {Alexander} D.~M.,
  {Richards} G.~T.,  {Ross} N.~P.,   {Hickox} R.~C.,  2020, \mn@doi [\mnras]
  {10.1093/mnras/staa1381}, \href
  {https://ui.adsabs.harvard.edu/abs/2020MNRAS.495.3252S} {495, 3252}

\bibitem[\protect\citeauthoryear{{Smette}, {Heap}, {Williger}, {Tripp},
  {Jenkins}  \& {Songaila}}{{Smette} et~al.}{2002}]{Smette2002}
{Smette} A.,  {Heap} S.~R.,  {Williger} G.~M.,  {Tripp} T.~M.,  {Jenkins}
  E.~B.,   {Songaila} A.,  2002, \mn@doi [\apj] {10.1086/324397}, \href
  {https://ui.adsabs.harvard.edu/abs/2002ApJ...564..542S} {564, 542}

\bibitem[\protect\citeauthoryear{{Sokasian}, {Abel}  \& {Hernquist}}{{Sokasian}
  et~al.}{2002}]{Sokasian2002}
{Sokasian} A.,  {Abel} T.,   {Hernquist} L.,  2002, \mn@doi [\mnras]
  {10.1046/j.1365-8711.2002.05291.x}, \href
  {https://ui.adsabs.harvard.edu/abs/2002MNRAS.332..601S} {332, 601}

\bibitem[\protect\citeauthoryear{{Springel}}{{Springel}}{2010}]{springel2010}
{Springel} V.,  2010, \mn@doi [\mnras] {10.1111/j.1365-2966.2009.15715.x},
  \href {https://ui.adsabs.harvard.edu/abs/2010MNRAS.401..791S} {401, 791}

\bibitem[\protect\citeauthoryear{{Springel} \& {Hernquist}}{{Springel} \&
  {Hernquist}}{2003}]{springel2003}
{Springel} V.,  {Hernquist} L.,  2003, \mn@doi [\mnras]
  {10.1046/j.1365-8711.2003.06206.x}, \href
  {https://ui.adsabs.harvard.edu/abs/2003MNRAS.339..289S} {339, 289}

\bibitem[\protect\citeauthoryear{{Springel} et~al.,}{{Springel}
  et~al.}{2018}]{volker2018}
{Springel} V.,  et~al., 2018, \mn@doi [\mnras] {10.1093/mnras/stx3304}, \href
  {https://ui.adsabs.harvard.edu/abs/2018MNRAS.475..676S} {475, 676}

\bibitem[\protect\citeauthoryear{{Tepper-Garc{\'\i}a}}{{Tepper-Garc{\'\i}a}}{2006}]{tepper2006}
{Tepper-Garc{\'\i}a} T.,  2006, \mn@doi [\mnras]
  {10.1111/j.1365-2966.2006.10450.x}, \href
  {https://ui.adsabs.harvard.edu/abs/2006MNRAS.369.2025T} {369, 2025}

\bibitem[\protect\citeauthoryear{{Theuns}, {Schaye}  \& {Haehnelt}}{{Theuns}
  et~al.}{2000}]{Tom2000}
{Theuns} T.,  {Schaye} J.,   {Haehnelt} M.~G.,  2000, \mn@doi [\mnras]
  {10.1046/j.1365-8711.2000.03423.x}, \href
  {https://ui.adsabs.harvard.edu/abs/2000MNRAS.315..600T} {315, 600}

\bibitem[\protect\citeauthoryear{{Theuns}, {Schaye}, {Zaroubi}, {Kim},
  {Tzanavaris}  \& {Carswell}}{{Theuns} et~al.}{2002}]{Theuns2002}
{Theuns} T.,  {Schaye} J.,  {Zaroubi} S.,  {Kim} T.-S.,  {Tzanavaris} P.,
  {Carswell} B.,  2002, \mn@doi [\apjl] {10.1086/339998}, \href
  {https://ui.adsabs.harvard.edu/abs/2002ApJ...567L.103T} {567, L103}

\bibitem[\protect\citeauthoryear{{Timlin} et~al.,}{{Timlin}
  et~al.}{2018}]{Timlin2018}
{Timlin} J.~D.,  et~al., 2018, \mn@doi [\apj] {10.3847/1538-4357/aab9ac}, \href
  {https://ui.adsabs.harvard.edu/abs/2018ApJ...859...20T} {859, 20}

\bibitem[\protect\citeauthoryear{{Upton Sanderbeck} \& {Bird}}{{Upton
  Sanderbeck} \& {Bird}}{2020}]{upton2020}
{Upton Sanderbeck} P.,  {Bird} S.,  2020, \mn@doi [\mnras]
  {10.1093/mnras/staa1850}, \href
  {https://ui.adsabs.harvard.edu/abs/2020MNRAS.496.4372U} {496, 4372}

\bibitem[\protect\citeauthoryear{{Walther}, {O{\~n}orbe}, {Hennawi}  \&
  {Luki{\'c}}}{{Walther} et~al.}{2019}]{walther2019}
{Walther} M.,  {O{\~n}orbe} J.,  {Hennawi} J.~F.,   {Luki{\'c}} Z.,  2019,
  \mn@doi [\apj] {10.3847/1538-4357/aafad1}, \href
  {https://ui.adsabs.harvard.edu/abs/2019ApJ...872...13W} {872, 13}

\bibitem[\protect\citeauthoryear{{Weinberger} et~al.,}{{Weinberger}
  et~al.}{2017}]{weinberger2017}
{Weinberger} R.,  et~al., 2017, \mn@doi [\mnras] {10.1093/mnras/stw2944}, \href
  {https://ui.adsabs.harvard.edu/abs/2017MNRAS.465.3291W} {465, 3291}

\bibitem[\protect\citeauthoryear{{Weinberger} et~al.,}{{Weinberger}
  et~al.}{2018}]{weinberger2018}
{Weinberger} R.,  et~al., 2018, \mn@doi [\mnras] {10.1093/mnras/sty1733}, \href
  {https://ui.adsabs.harvard.edu/abs/2018MNRAS.479.4056W} {479, 4056}

\bibitem[\protect\citeauthoryear{{White} et~al.,}{{White}
  et~al.}{2012}]{White2012}
{White} M.,  et~al., 2012, \mn@doi [\mnras] {10.1111/j.1365-2966.2012.21251.x},
  \href {https://ui.adsabs.harvard.edu/abs/2012MNRAS.424..933W} {424, 933}

\bibitem[\protect\citeauthoryear{{Worseck}, {Prochaska}, {Hennawi}  \&
  {McQuinn}}{{Worseck} et~al.}{2016}]{worseck2016}
{Worseck} G.,  {Prochaska} J.~X.,  {Hennawi} J.~F.,   {McQuinn} M.,  2016,
  \mn@doi [\apj] {10.3847/0004-637X/825/2/144}, \href
  {https://ui.adsabs.harvard.edu/abs/2016ApJ...825..144W} {825, 144}

\bibitem[\protect\citeauthoryear{{Worseck}, {Davies}, {Hennawi}  \&
  {Prochaska}}{{Worseck} et~al.}{2019}]{worseck2019}
{Worseck} G.,  {Davies} F.~B.,  {Hennawi} J.~F.,   {Prochaska} J.~X.,  2019,
  \mn@doi [\apj] {10.3847/1538-4357/ab0fa1}, \href
  {https://ui.adsabs.harvard.edu/abs/2019ApJ...875..111W} {875, 111}

\bibitem[\protect\citeauthoryear{{{\v{S}}oltinsk{\'y}}, {Bolton}, {Molaro},
  {Hatch}, {Haehnelt}, {Keating}, {Kulkarni}  \&
  {Puchwein}}{{{\v{S}}oltinsk{\'y}} et~al.}{2023}]{Soltinsky2023}
{{\v{S}}oltinsk{\'y}} T.,  {Bolton} J.~S.,  {Molaro} M.,  {Hatch} N.,
  {Haehnelt} M.~G.,  {Keating} L.~C.,  {Kulkarni} G.,   {Puchwein} E.,  2023,
  \mn@doi [\mnras] {10.1093/mnras/stac3710}, \href
  {https://ui.adsabs.harvard.edu/abs/2023MNRAS.519.3027S} {519, 3027}

\bibitem[\protect\citeauthoryear{{van der Walt}, {Colbert}  \&
  {Varoquaux}}{{van der Walt} et~al.}{2011}]{vander2011}
{van der Walt} S.,  {Colbert} S.~C.,   {Varoquaux} G.,  2011, \mn@doi
  [Computing in Science and Engineering] {10.1109/MCSE.2011.37}, \href
  {https://ui.adsabs.harvard.edu/abs/2011CSE....13b..22V} {13, 22}

\makeatother
\end{thebibliography}




\appendix

\section{Convergence tests}
\label{appendix:convergence}

In this section, we describe a series of convergence tests we performed for the simulations presented in this paper. Starting from our fiducial run, we systematically vary individual numerical parameters and assess their impact on the simulated IGM properties. We refer the readers to Table \ref{tab:table-sim} for a detailed overview of all simulation parameters.

\begin{figure}
\includegraphics[width=\columnwidth]{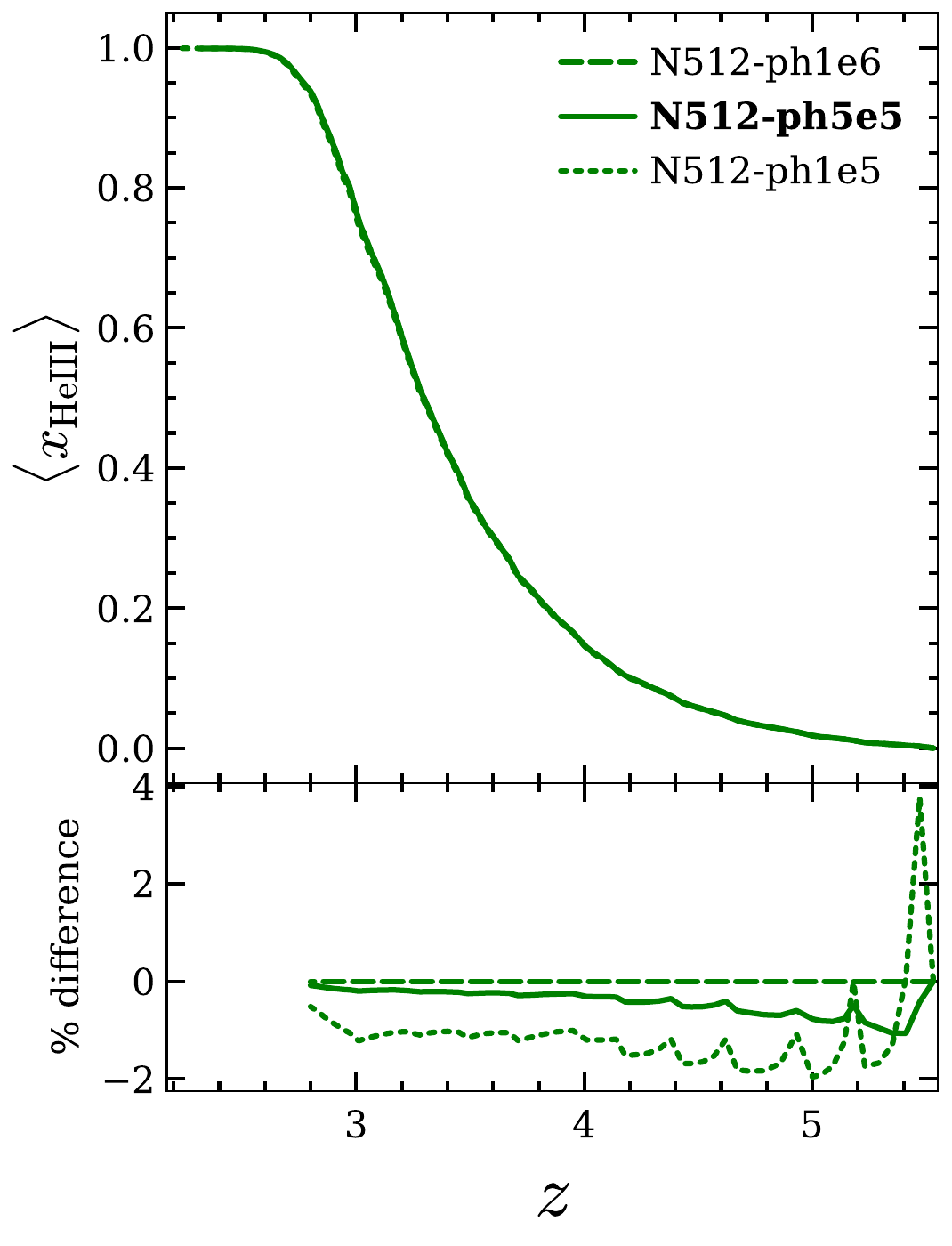}
\caption{{\it Top panel}: evolution of the volume-averaged He~{\sc iii} fraction in simulations with  $N_{\gamma}$ = $10^6$ (dashed curve), $5 \times 10^5$ (solid) and $10^5$ (dotted). {\it Bottom panel}: relative differences with respect to the highest resolution simulation \texttt{N512-ph1e6}. The fiducial simulation is marked in bold font.}
\label{fig:photon_pckt_convergence}
\end{figure}

\subsection{Convergence with photon sampling}
\label{appendix:convergence_Ngamma}

We start by investigating the numerical convergence with respect to the number of photon packets emitted by each source at each timestep of the simulation ($N_{\gamma}$). We explore $N_{\gamma}$ = $10^5$, $5 \times 10^5$ and $10^6$, corresponding to the runs labeled as \texttt{N512-ph1e5}, \texttt{N512-ph5e5} and \texttt{N512-ph1e6}, respectively. We present the  evolution of volume-averaged He~{\sc iii} fraction in these three models in the top panel of figure \ref{fig:photon_pckt_convergence}. Our reference simulation \texttt{N512-ph5e5} (solid line) shows an excellent convergence (within $2\%$) with the highest resolution simulation (\texttt{N512-ph1e6}, dashed line) in the entire simulated redshift range, with a relative difference (bottom panel) below 1\% towards the end of reionization. The lowest resolution simulation \texttt{N512-ph1e5} (dotted line) is also converged within 3\% throughout the entire reionization history. 

\begin{figure}
\includegraphics[width=\columnwidth]{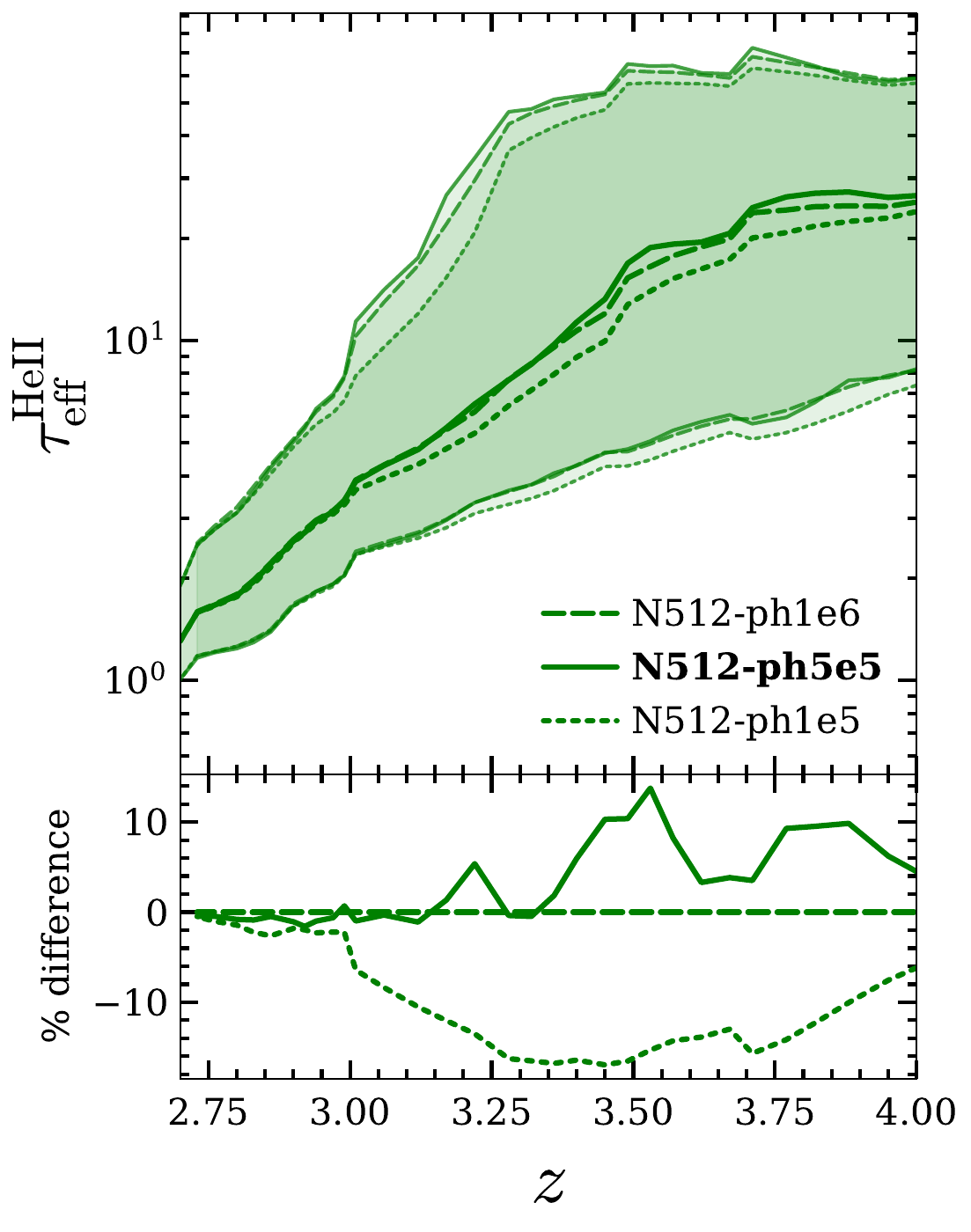}
\caption{{\it Top panel}: evolution of the He~{\sc ii} optical depth in simulations with  $N_{\gamma}$ = $10^6$ (dashed curve), $5 \times 10^5$ (solid) and $10^5$ (dotted). The shaded regions denote the 68$\%$ confidence intervals. {\it Bottom panel}: relative differences with respect to the highest resolution simulation \texttt{N512-ph1e6}. The fiducial simulation is marked in bold font.}
\label{fig:photon_pckt_convergence_tau}
\end{figure}

The convergence in global quantities like the reionization history is  however not indicative of convergence in local observed quantities. For this reason, in figure \ref{fig:photon_pckt_convergence_tau} we show the redshift evolution of the He~{\sc ii} effective optical depth for the three values of $N_{\gamma}$ employed. Also in this case \texttt{N512-ph5e5} and \texttt{N512-ph1e6} exhibit an excellent convergence (within 2\%) for \textit{z} $\lesssim$ 3.1. We note that this holds true not only for the median effective optical depth, but also for the entire distribution, as shown by the overlapping shaded regions (corresponding to the central 68\% of the data) in the figure. Unlike the reionization history case, the effective optical depth in \texttt{N512-ph1e5} is significantly different than in the runs with a larger $N_{\gamma}$, except for the tail end of reionization, when the entire volume is fully ionize and therefore the role of photon sampling is significantly reduced.

This analysis demonstrates that \texttt{N512-ph5e5} achieves a very good convergence in terms of photon sampling.

\begin{figure} 
    \includegraphics[width=\columnwidth]{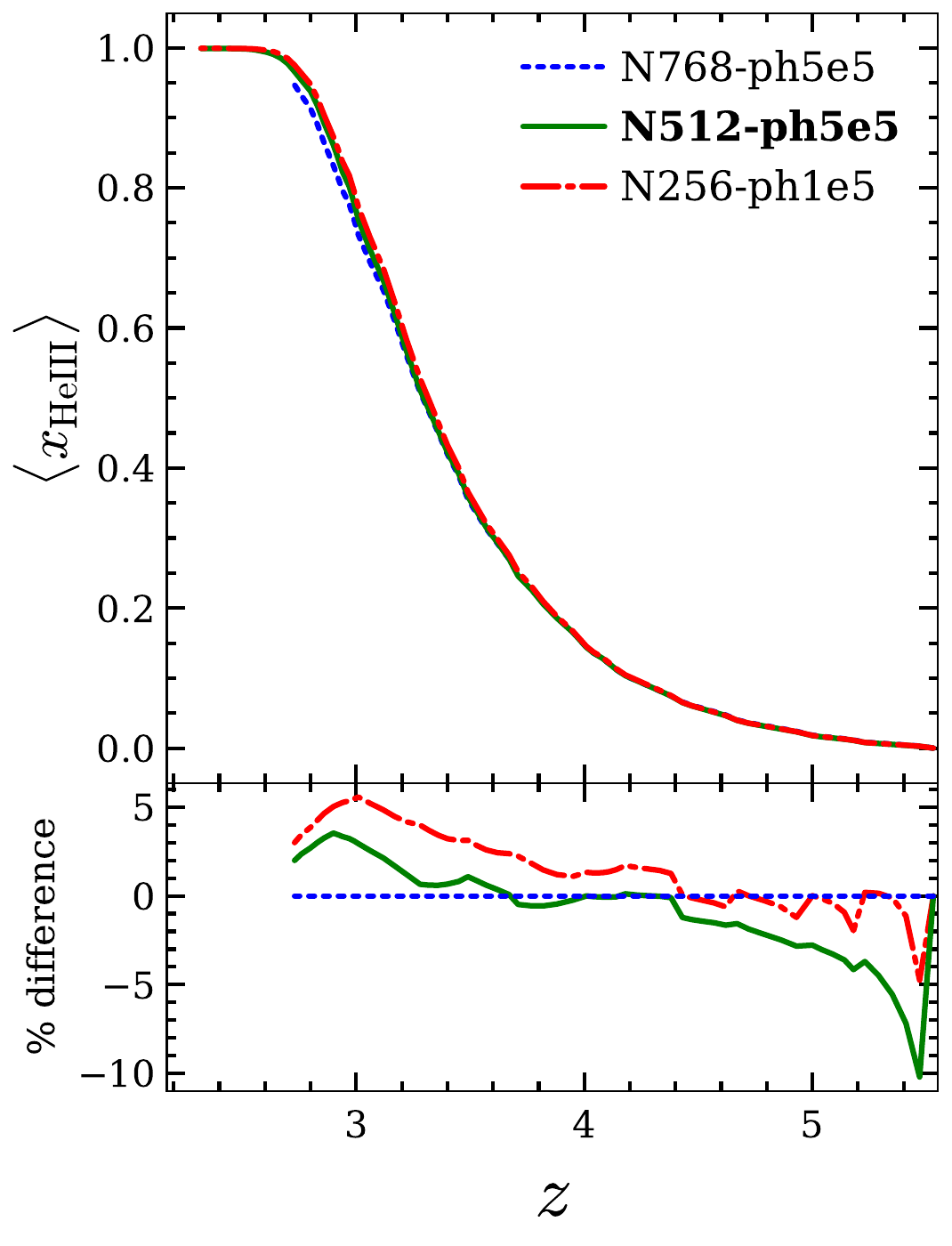}
    \caption{{\it Top panel}: evolution of the volume averaged He~{\sc iii} fraction in simulations with $N_{\rm{grid}}=768^{3}$ (blue dotted curve), $512^{3}$ (green solid) and $256^{3}$ (red dash-dotted). {\it Bottom panel}: corresponding 
    relative difference  with respect to \texttt{N768-ph5e5}. The fiducial simulation is marked in bold font.}
    \label{fig:Ngrid_convergence}
\end{figure}

\subsection{Convergence with grid dimension}
\label{appendix:convergence_Ngrid}

In order to test the convergence of our fiducial simulation with respect to the physical resolution of the baryonic component, we have run a set of simulations systematically varying the number of grid points used to discretize the simulated volume, namely $N_{\rm{grid}}$= $256^{3}$, $512^{3}$, $768^{3}$. Figure \ref{fig:Ngrid_convergence} shows the volume-averaged He~{\sc iii} fraction in these simulations (top panel) and their differences relative to \texttt{N768-ph5e5} (bottom panel). Note that these runs employ a value of $N_{\gamma}$ that ensures convergence in the radiation sampling (see the previous section). 
The most remarkable feature is that $N_{\rm{grid}}$ guides the \textit{speed} of reionization, with lower resolution runs starting slower (i.e. with negative values in the bottom panel) but proceeding faster and eventually completing reionization earlier. This is a consequence of the fact that higher resolution simulations better resolve density contrast. The low-density channels enable faster photons escape in the early phases of reionization, but high-density regions are more resilient to reionization, slowing down this process. Nevertheless, the relative difference between our fiducial model and the higher-resolution one remains below $\approx$3\% at all redshift relevant for helium reionization, showing an excellent convergence. 

\begin{figure} 
    \includegraphics[width=\columnwidth]{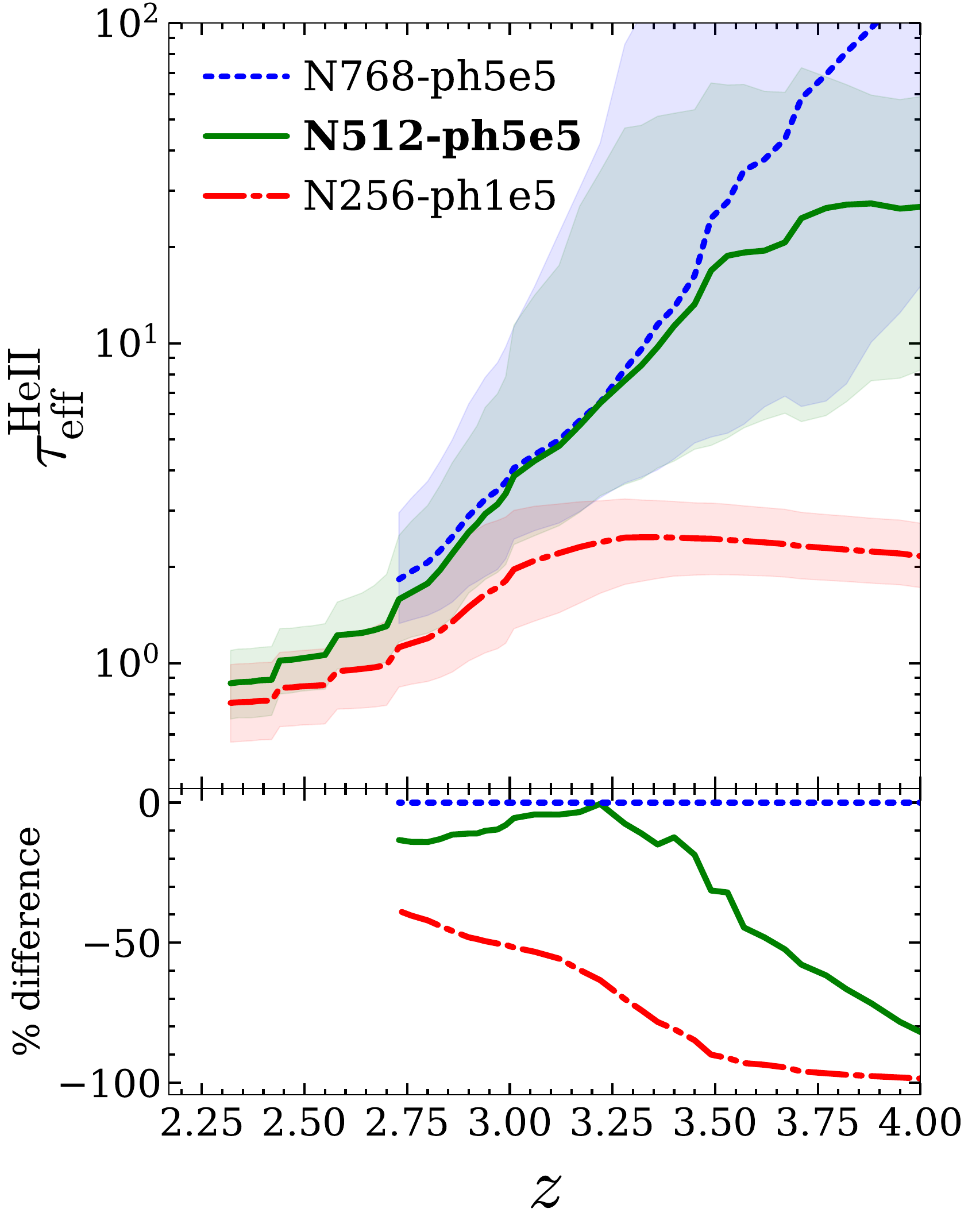}
    \caption{{\it Top panel}: evolution of the He~{\sc ii} optical depth in simulations with $N_{\rm{grid}}=768^{3}$ (blue dotted curve), $512^{3}$ (green solid) and $256^{3}$ (red dash-dotted). The shaded regions denote the 68$\%$ confidence intervals. {\it Bottom panel}: relative differences with respect to \texttt{N768-ph5e5}. The fiducial simulation is marked in bold font.}
    \label{fig:Ngrid_convergence_tau}
\end{figure}

In figure \ref{fig:Ngrid_convergence_tau} we show the evolution of the He~{\sc ii} effective optical depth in these different models (top panel, with solid lines indicating the median value and shaded regions marking the central 68\% of the data) and the relative difference of their medians (bottom panel). For this observable, our fiducial simulation is converged within 10\% at $z \lesssim 3.3$, while at earlier times it significantly under-predicts $\tau_\mathrm{eff}^\mathrm{HeII}$. Remarkably, however, the low-$\tau_\mathrm{eff}^\mathrm{HeII}$ part of the distribution is in very good agreement all the way to $z\lesssim 3.8$, since in the second half of reionization the increased gas resolution mostly affects the high-density (and therefore high-$\tau_\mathrm{eff}^\mathrm{HeII}$) regions, as described above. Since at $z\gtrsim3$ observations are only sensitive to the low-$\tau_\mathrm{eff}^\mathrm{HeII}$ part of the intrinsic distribution (see figure \ref{fig:taueff} and relative discussion), we deem this an acceptable convergence level. Finally, \texttt{N256-ph1e5} displays very poor convergence throughout the entire simulation evolution.

\subsection{Convergence with periodic boundary condition}
\label{appendix:convergence_PBC}

\begin{figure} 
    \includegraphics[width=\columnwidth]{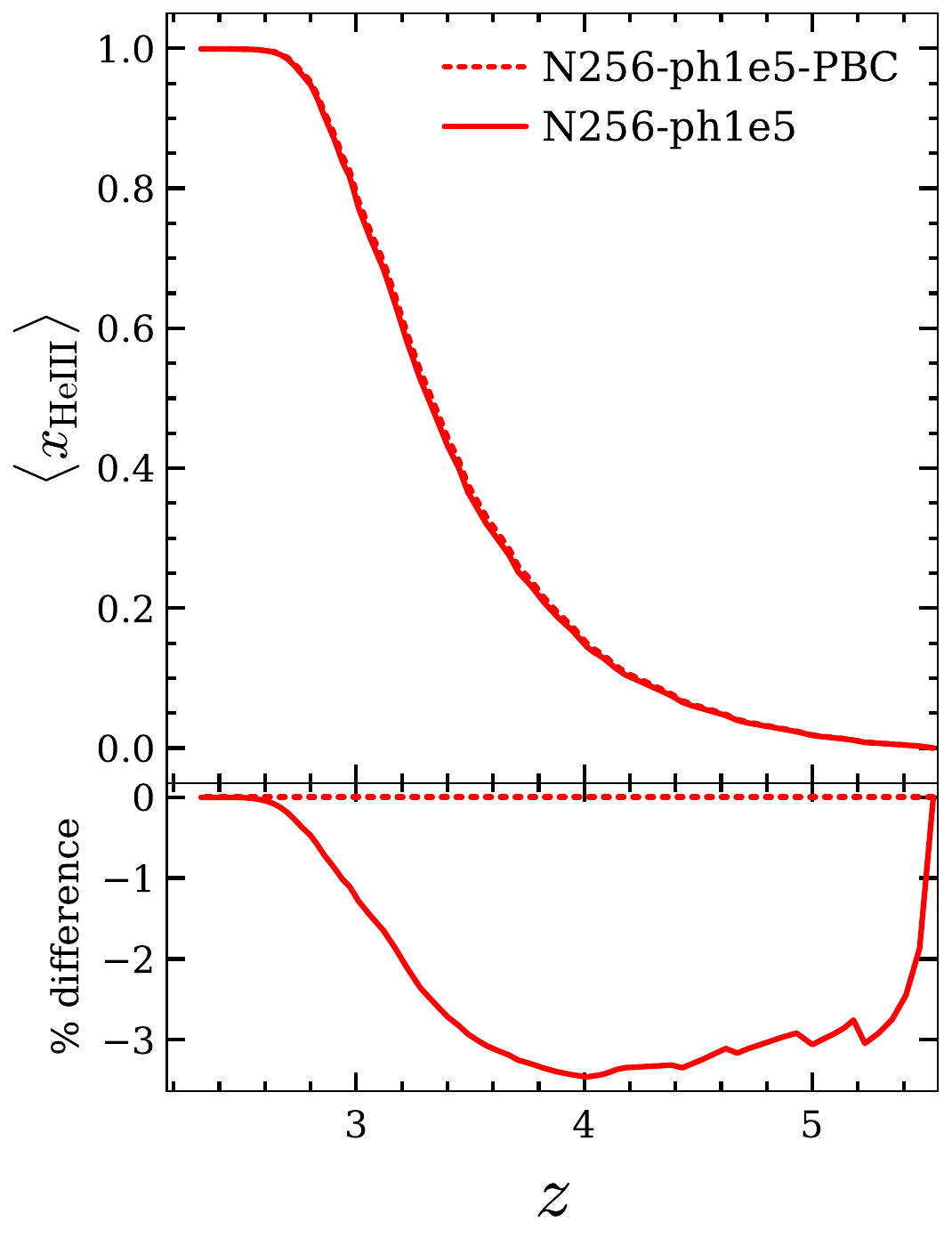}
    \caption{{\it Top panel}: evolution of the volume averaged He~{\sc iii} fraction with (dotted line) and without (solid) periodic boundary condition for simulations run with $N_{\rm{grid}}$= $256^{3}$. {\it Bottom panel}: relative difference with respect to \texttt{N256-ph1e5-PBC}. }
    \label{fig:PBC_convergence}
\end{figure}

\begin{figure} 
    \includegraphics[width=\columnwidth]{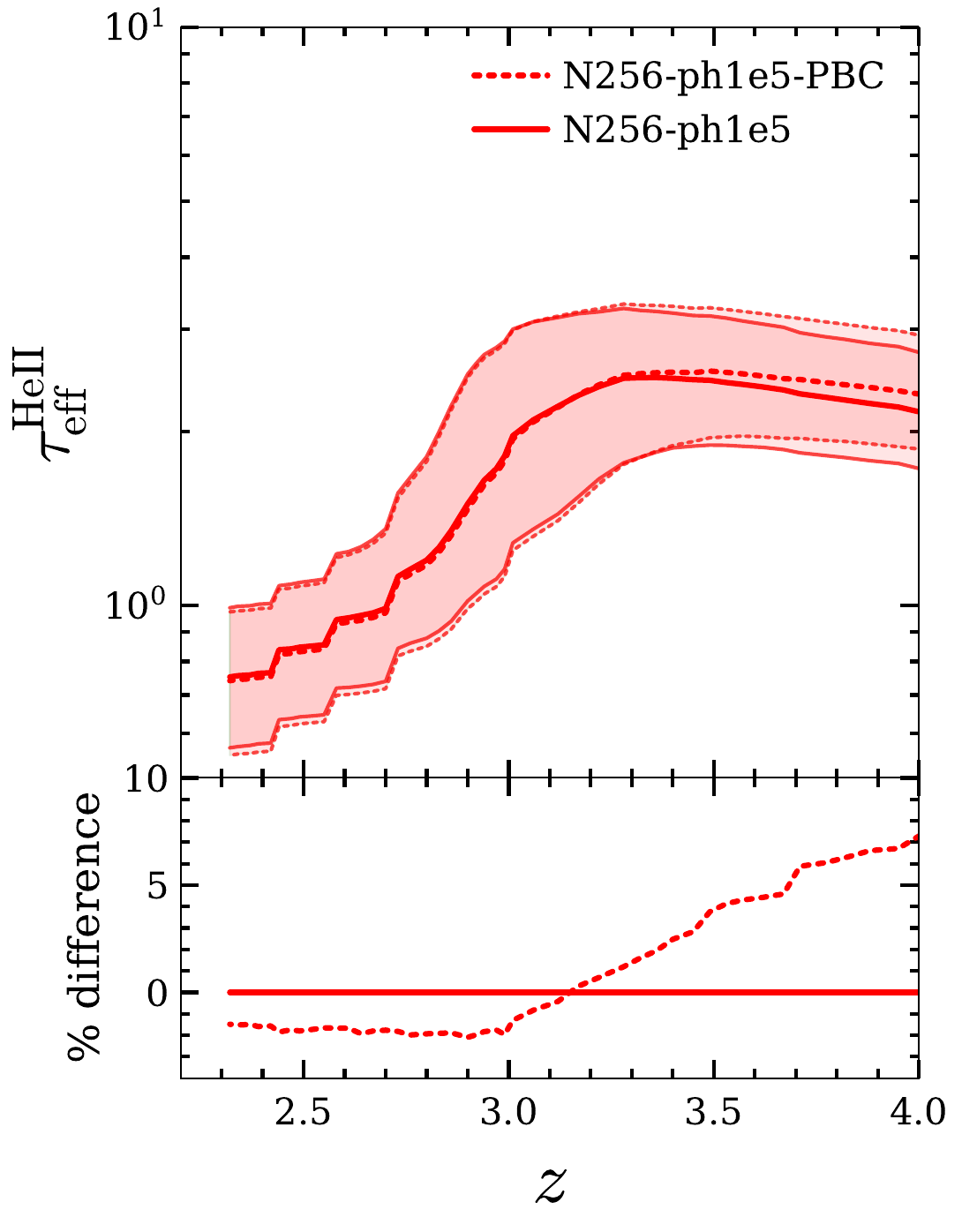}
    \caption{{\it Top panel}: evolution of the He~{\sc ii} optical depth with (dotted line) and without (solid) periodic boundary condition for simulations run with $N_{\rm{grid}}$= $256^{3}$. The shaded regions denote the 68 $\%$ confidence intervals. {\it Bottom panel}: relative differences with respect  to \texttt{N256-ph1e5-PBC}.}
    \label{fig:PBC_convergence_tau}
\end{figure}

Finally, we analyze the impact of periodic boundary conditions (PBC) in our simulations. Since we are purely interested in a comparative analysis, the actual convergence of their physical predictions is of little importance here. Therefore we have employed low-resolution runs that are computationally cheaper. These simulations have been run only until $z_\mathrm{final}=2.73$, as by then perfect convergence has been already established. 
The inclusion of PBC has a small impact on both the reionization history ($\lesssim$3\%) and $\tau_\mathrm{eff}^\mathrm{HeII}$ ($\lesssim$10\%) in the initial stages. Such difference is progressively reduced until perfect ($\lesssim$1\%) convergence is reached at $z\approx3$. 
In figure \ref{fig:PBC_convergence} we show the evolution of the volume averaged He~{\sc iii} fraction (top panel) and their relative difference (bottom panel) for simulations with (\texttt{N256-ph1e5-PBC}, dotted line) and without (\texttt{N256-ph1e5}, solid) PBC. In figure \ref{fig:PBC_convergence_tau} we report the He~{\sc ii} effective optical depth evolution and the relative difference for the same two simulations. Note that we have conducted the same test for \texttt{N512-ph5e5} (i.e. employing the same $N_{\gamma}$ as in our fiducial run) until $z_\mathrm{final} \sim 4$, finding similar results.

\begin{figure}
    \includegraphics[width=\columnwidth]{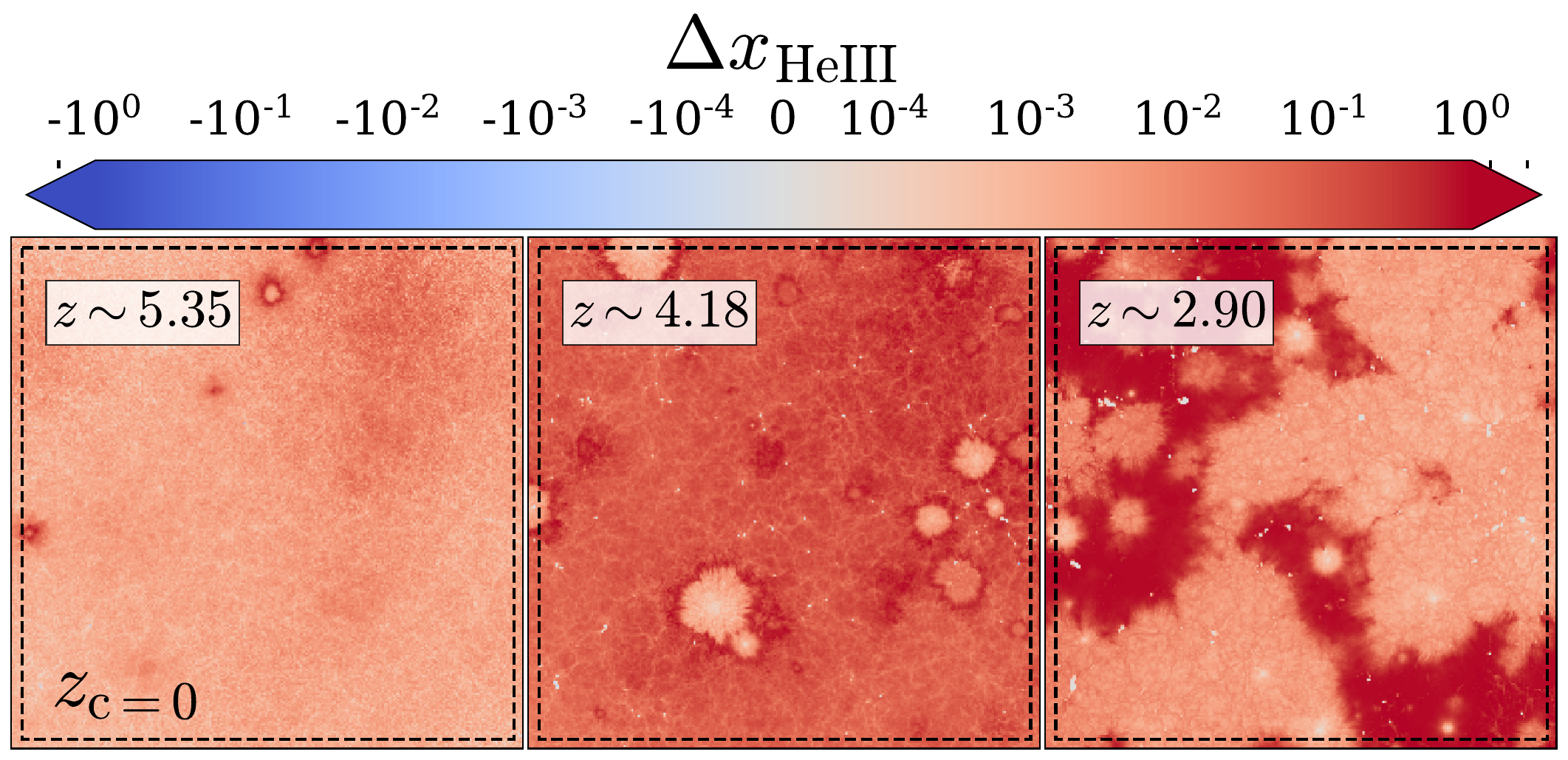}
    \includegraphics[width=\columnwidth]{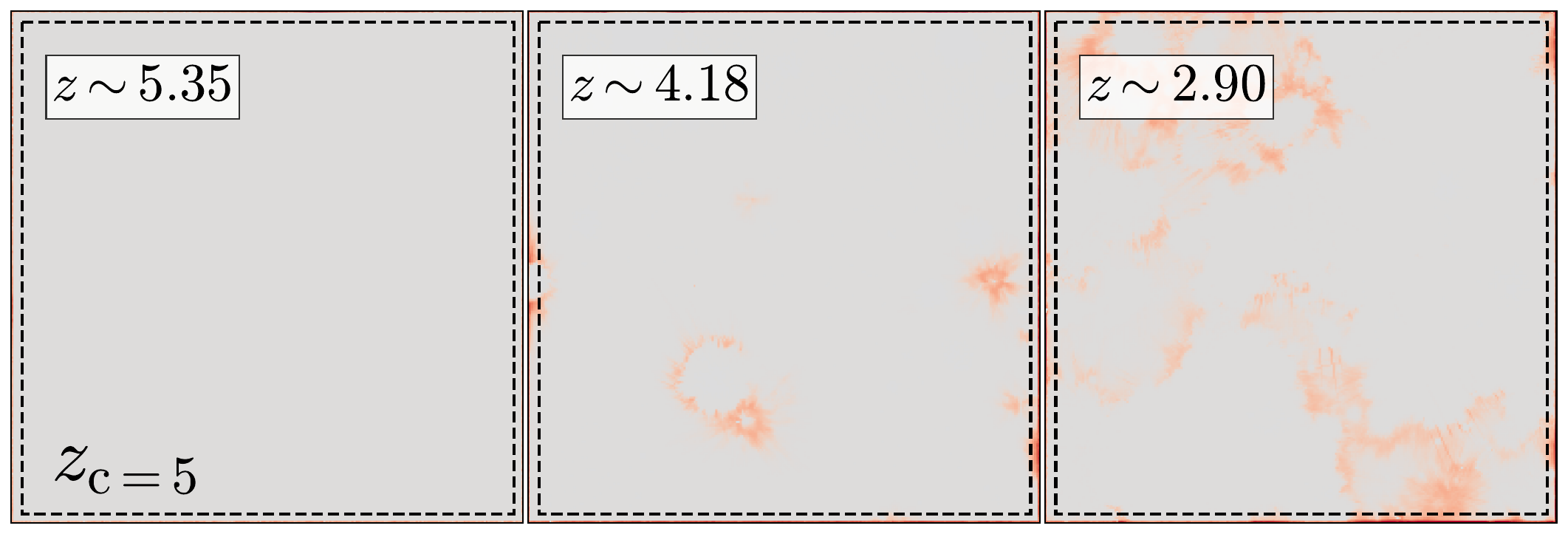}
    \includegraphics[width=\columnwidth]{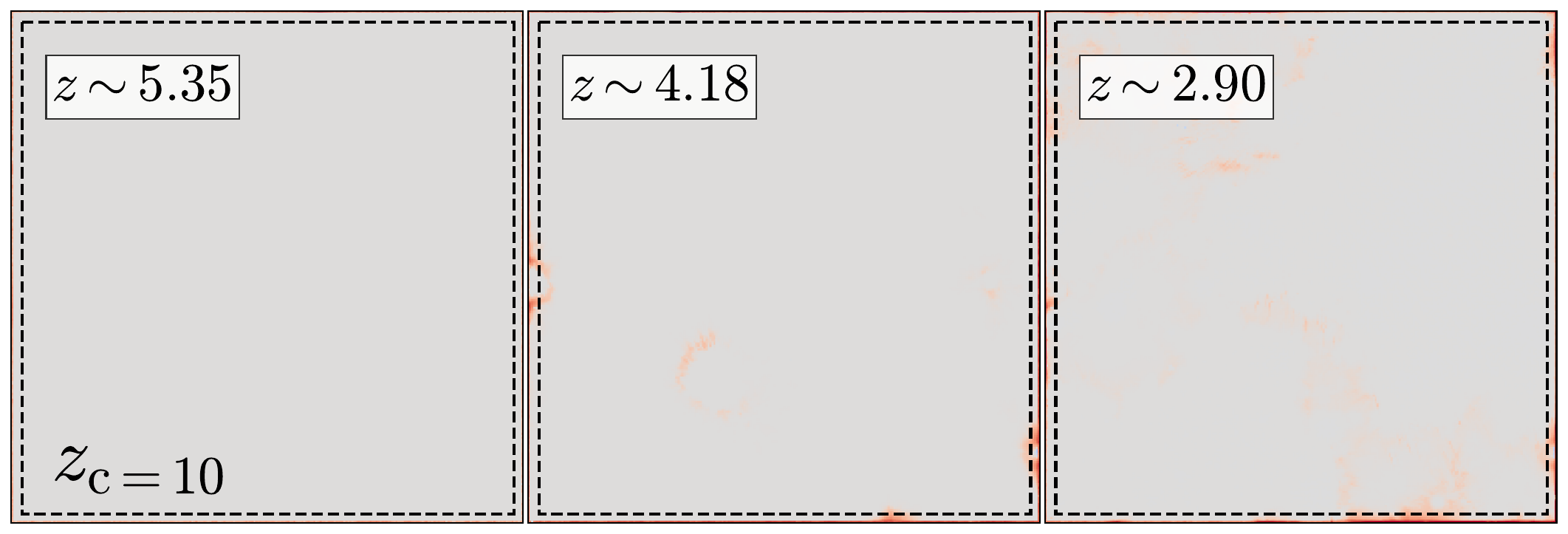}
    
    \caption{Maps showing the difference in He~{\sc iii} fraction between \texttt{N256-ph1e5-PBC} and \texttt{N256-ph1e5} at \textit{z} =5.35, 4.18, and 2.90 (from left to right). The rows refer to slices of thickness 800 $h^{-1}$~ckpc located at various distances along the $z$-direction (marked in grid unit) from the edge of the box. The outside region of the black dashed rectangle shows the parts of the edges of all slices which are finally removed from the analysis.}
    \label{fig:reion_history_pbc}
\end{figure}

As mentioned in section \ref{method}, we have removed from our analysis the 5 layers of cells closest to each edge of the simulation box. The reason is shown in figure \ref{fig:reion_history_pbc}, where we display the difference in He~{\sc iii} fraction between \texttt{N256-ph1e5-PBC} and \texttt{N256-ph1e5} in three slices (spanning the entire simulation box in two dimensions) placed at the edge of the simulation box (top row), 5 cells away (middle row) and 10 cells away (bottom row). Already in the middle panel, the effect of the PBC are negligible. Notice that the edge of each slice will also be removed as it is close to one of the edges of the simulation box, as indicated by black dashed rectangles in the slices shown.

\section{Dependence on BH-to-halo relation}
\label{appendix:BH_to_halo}

\begin{figure} 
    \includegraphics[width=\columnwidth]{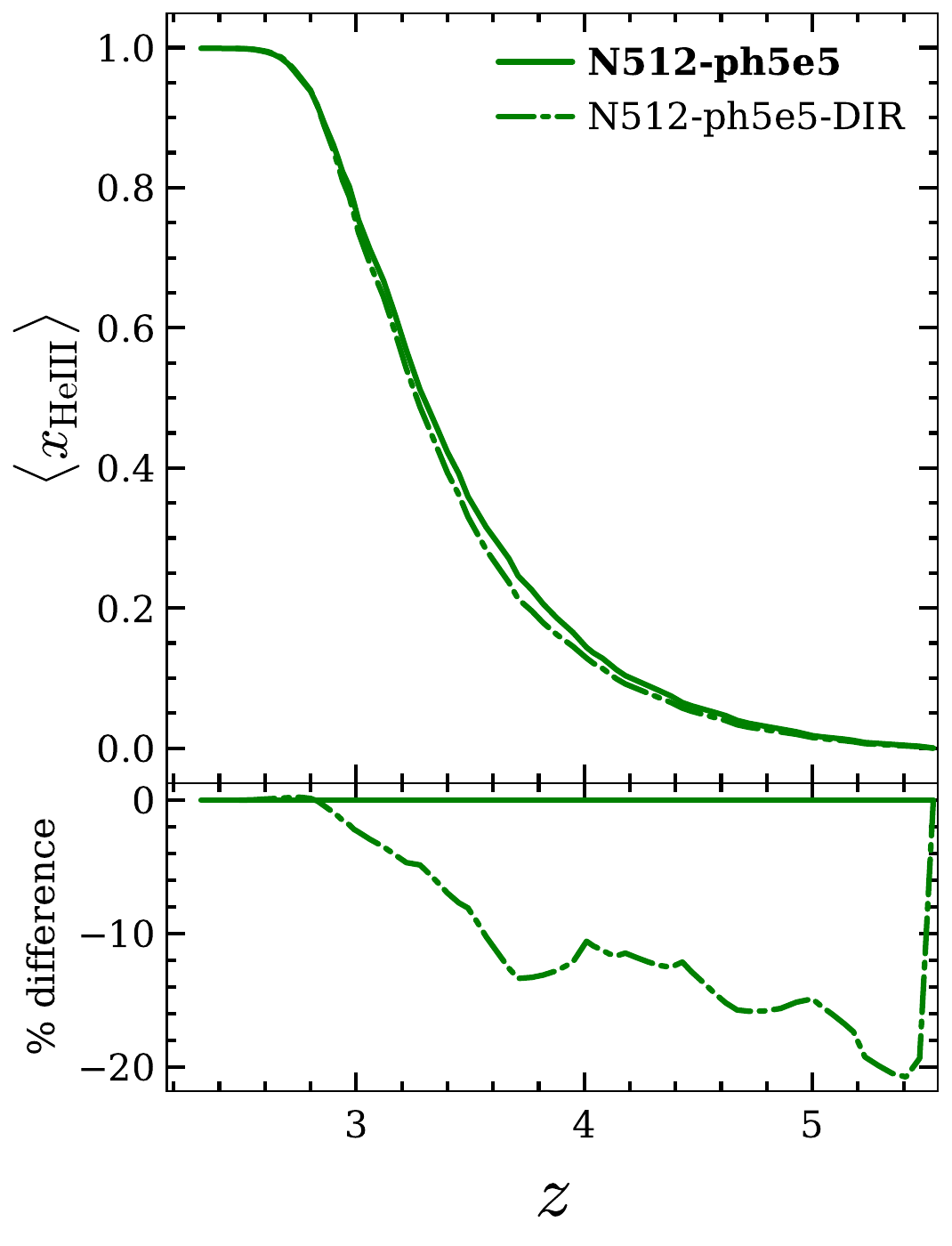}
    \caption{{\it Top panel:} Volume averaged He~{\sc iii} fraction for two different methods to assign quasars to the DM haloes: abundance matching with scatter (\texttt{N512-ph5e5}, solid curve) and  abundance matching without any scatter (\texttt{N512-ph5e5-DIR}, dashed curve). {\it Bottom panel}: percentage difference between the curves in the top panel. The fiducial simulation is marked in bold font.}
    \label{fig:two_approaches}
\end{figure}

\begin{figure} 
    \includegraphics[width=\columnwidth]{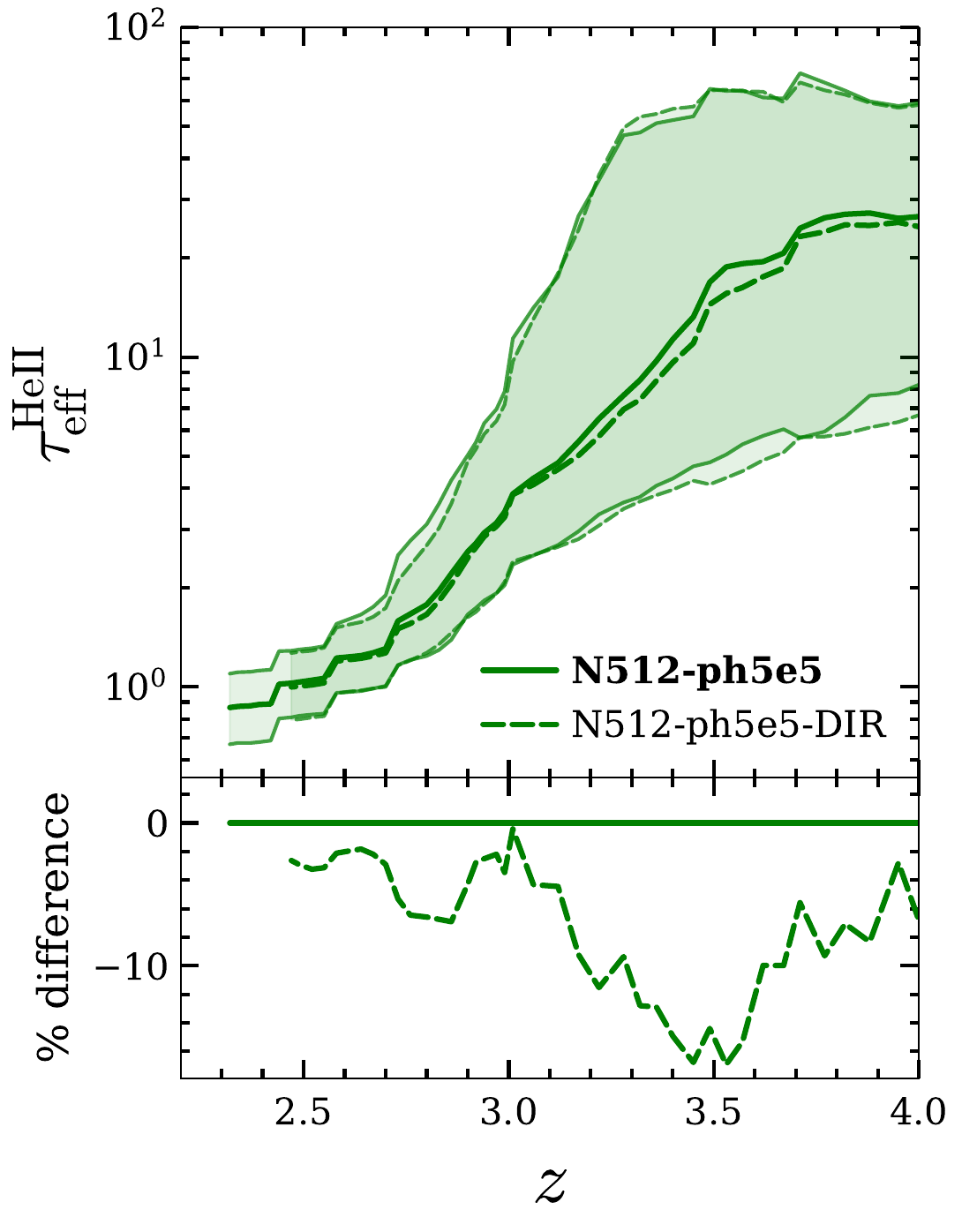}
    \caption{{\it Top Panel}: He~{\sc ii} effective optical depth for two different methods to assign quasars to the DM haloes: abundance matching with scatter (\texttt{N512-ph5e5}, solid curve) and  abundance matching without any scatter (\texttt{N512-ph5e5-DIR}, dashed curve). 68 $\%$ confidence intervals are denoted by the shaded regions. {\it Bottom panel}: percentage difference between the curves in the top panel. The fiducial simulation is marked in bold font.}
    \label{fig:two_approaches_tau}
\end{figure}

In this Appendix we investigate the impact on the results of our two BHs positioning schemes (dubbed `fiducial' and `direct' in Section \ref{tng+crash}). To do so, we run an additional simulation, \texttt{N512-ph5e5-DIR}, that differs from our fiducial run \texttt{N512-ph5e5} just in the fact that QSOs are placed in the simulation using our `direct' method. 
In figure \ref{fig:two_approaches} we compare the evolution of the volume-averaged He~{\sc iii} fraction obtained using these two approaches. 
AThe reionization process progresses slower in the `direct' method, as QSOs are placed in more massive haloes, and therefore --~on average~-- in higher density regions. Consequently, the higher recombination slows down the advancements of reionization fronts, and therefore of the IGM ionization. By $z\approx3$, helium reionization is completed and therefore these differences vanish. The situation is similar for the He~{\sc ii} effective optical depth shown in figure \ref{fig:two_approaches_tau}. The difference always remain within 20\% and decreases with time, until it settles on $\approx 5\%$ at $z\leq 3$. This is in line with the differences in reionization histories. Additionally, at $z\leq 3$ the residual difference is due to the fact that photons emitted by QSOs placed through the `direct' method are more strongly absorbed before reaching the IGM, because of the higher local gas densities, and hence residual neutral fractions.

\bsp	
\label{lastpage}
\end{document}